\documentclass[showpacs,onecolumn,preprintnumbers,amsmath,amsfonts,amssymb,floatfix,aps,superscriptaddress]{revtex4}

\usepackage{graphicx}
\usepackage{epsfig}
\usepackage{amsmath,mathtools}
\usepackage[hang,nooneline]{subfigure}
\usepackage[normalem]{ulem}
\usepackage{color}
\usepackage{hyperref}
\usepackage{bm}
\usepackage{times}
\usepackage{makecell}
\usepackage{array}
\usepackage{diagbox}
\DeclareSymbolFont{matha}{OML}{txmi}{m}{it}
\DeclareMathSymbol{\varv}{\mathord}{matha}{118}

\newcounter{fig}

\DeclareMathOperator{\sech}{sech}

\newcommand{\ii}{\mathrm{i}}

\begin{document}

\title{Breathers in lattices with alternating strain-hardening and strain-softening interactions}

\author{M.~M. Lee}
\email[Email: ]{mlee195@calpoly.edu}
\affiliation{Mathematics Department,
California Polytechnic State University,
San Luis Obispo, CA 93407-0403, USA}

\author{E.~G. Charalampidis}
\email[Email: ]{echarala@calpoly.edu}
\affiliation{Mathematics Department,
California Polytechnic State University,
San Luis Obispo, CA 93407-0403, USA}

\author{S. Xing}
\email[Email: ]{sixing@calpoly.edu}
\affiliation{Department of Mechanical Engineering, 
California Polytechnic State University,
San Luis Obispo, CA 93407-0403, USA}

\author{C. Chong}
\email[Email: ]{cchong@bowdoin.edu}
\affiliation{Department of Mathematics, 
Bowdoin College, 
Brunswick, ME 04011, USA}

\author{P.~G. Kevrekidis}
\email[Email: ]{kevrekid@math.umass.edu}
\affiliation{Department of Mathematics and
Statistics, University of Massachusetts
Amherst, Amherst, MA 01003-4515, USA}

\begin{abstract}
This work focuses on the study of time-periodic solutions, including breathers, 
in a nonlinear lattice consisting of elements whose contacts alternate between 
strain-hardening and strain-softening. The existence, stability, and bifurcation 
structure of such solutions, as well as the system dynamics in the presence of 
damping and driving are studied systematically. It is found that the linear resonant 
peaks in the system bend toward the frequency gap in the presence of nonlinearity. 
The time-periodic solutions that lie within the frequency gap compare well to 
Hamiltonian breathers if the damping and driving are small. In the Hamiltonian 
limit of the problem, we use a multiple scale analysis to derive a Nonlinear Schr\"odinger 
(NLS) equation to construct both acoustic and optical breathers. The latter compare 
very well with the numerically obtained breathers in the Hamiltonian limit.
\end{abstract}

\date{\today}


\maketitle

\section{Introduction}

Nonlinear lattices play a pivotal role in the modeling of a wide range of physical, 
biological, chemical, and electrical systems~\cite{FPUreview,pgk:2011,moti,Morsch,sievers}. 
One of the prototypical examples, and most relevant for the present study, is the 
Fermi-Pasta-Ulam-Tsingou (FPUT) chain, which is simply a mass-spring system with nonlinear 
springs~\cite{FPU55}. While the types of behavior that are possible in FPUT chains
span a wide range, the structure we will focus on here is the so-called discrete 
breather, which is a solution that is localized in space and periodic in time. 
Discrete breathers (or just breathers) have been studied in a variety of physical 
systems, including, optical waveguide arrays and photorefractive crystals~\cite{moti}, 
micromechanical oscillator arrays~\cite{sievers2}, Josephson-junction ladders~\cite{alex,alex2}, 
electrical lattices with nonlinear elements~\cite{remoissenet}, layered antiferromagnetic 
crystals~\cite{lars3,lars4}, halide-bridged transition metal complexes~\cite{swanson}, 
dynamical models of the DNA double strand~\cite{Peybi}, Bose--Einstein condensates in 
optical lattices~\cite{Morsch}, and magnetic lattices~\cite{moleron,Mehrem2017,Marc2017,magBreathers,Chong_2021,cantilevers}
among others.  

One mechanical realization of an FPUT lattice that has attracted significant attention
is the so-called granular chain, in which case the nonlinear interaction is described 
by a power-law with nonlinearity exponent $p = 3/2>1$. Granular chains consist of closely 
packed arrays of particles that interact elastically upon compression~\cite{Nester2001,granularBook,yuli_book,Lindenberg_review,gc_review,sen08}. 
The contact force can be tuned to yield responses that range from near-linear to purely 
nonlinear, and the effective stiffness properties can also be easily changed by modifying 
the material, geometry, or contact angle of the elements in contact~\cite{Nester2001}, although 
the stiffness will be of the strain-hardening type  (implying an increase of elastic modulus 
with strain). Due to the strain-hardening nature of the granular chain, the monomer one (where 
all particles are identical) does not have genuine breather solutions~\cite{JAMES201339} 
(although so-called dark breathers are possible~\cite{dark,dark2}). Breathers are possible in 
granular chains with defects~\cite{Theocharis2009,Nature11} or in mass-dimer chains~\cite{Boechler2010,Theocharis10,hooge12}
and have been studied in great detail.
 
More recently, strain-softening materials (implying a decrease of elastic modulus with strain) 
have been emerging as a new playground for the formation of nonlinear waves~\cite{Deng2017, Raney},
and the dynamics here will be distinct from their strain-hardening counterparts. For instance, 
rarefaction waves form in lattices with strain-softening interactions~\cite{shock_trans_granular,yasuda,HEC_DSW} 
instead of the classic solitary waves in the case of hardening lattices~\cite{Nester2001}. 
Examples of lattices that have been argued to exhibit strain-softening interaction include 
tensegrity~\cite{carpe}, origami~\cite{origami,origami2}, stainless steel cylinders separated 
by polymer foams~\cite{RarefactionNester}, and hollow elliptical cylinders~\cite{HEC_DSW}. 
These strain-softening lattices can be modeled with FPUT type lattices with a power-law interaction 
with nonlinear exponent $0<p<1$. Breathers in such strain-softening lattices are far less studied
than their strain-hardening counterparts.

More novel still, to the best of our knowledge, is the study of breathers in lattices that feature 
both strain-hardening and strain-softening interactions, the focus of the present study. Such a 
lattice could be possible in an origami setting where each origami unit cell is tailored to have 
strain-hardening or strain-softening behavior. This is possible, in principle, given the highly 
tunable nature of origami by virtue of changes in crease-angle (see~\cite{Bistable_origami} for an 
example of an origami unit cell that can be tuned to exhibit both softening and hardening behavior,
depending, e.g., on the folding angle for Tachi-Miura polyhedra~\cite{origami}). Another possible 
realization is a chain of stainless steel cylinders in which only a subset of contacts is separated 
with a polymer foam in order to induce the strain-softening behavior.

The paper is structured as follows. The model equations and basic linear analysis of such a periodically 
alternating nonlinearity are given in Sec.~\ref{sec:theory}. Our main aim herein is to explore this model 
and especially its nonlinear time-periodic solutions. The numerical study of such breather (exponentially 
localized in space and periodic in time) waveforms of the damped-driven model (being more realistic for 
the potential experimental realization of such structures) is detailed in Sec.~\ref{sec:damped-driven}. 
The ideal case of the Hamiltonian lattice is considered in Sec.~\ref{sec:NLS}, where the Nonlinear 
Schr\"odinger (NLS) equation is derived using a multiple-scale analysis in order to construct an analytical 
prediction of both acoustic and optical breathers. Section~\ref{sec:theend} concludes the paper and offers 
suggestions for future studies that build on the present one.

\section{Theoretical Setup} \label{sec:theory}

\subsection{The model} 
A heterogeneous FPUT lattice with a power-law nonlinearity is given by \cite{Nester2001,FPU55}
\begin{align}
m_n \ddot{u}_{n} = A_n \left[d_n + u_{n-1} - u_{n}\right]^{p_{n}}_{+} - %
A_{n+1} \left[d_n + u_{n} - u_{n+1}\right]^{p_{n+1}}_{+} - \gamma\dot{u}_{n}, %
\quad n=1,\dots, N,
\label{gc_dimer_dim}              
\end{align}
where the over-dot represents differentiation with respect to time $t$, $u_n=u_n(t)\in\mathbb{R}$ 
is the displacement of the $n$-th particle from its equilibrium position, $\gamma$ is a parameter 
accounting for the dissipation, and $N$ stands for the total number of particles in the chain. The 
bracket $[\cdot]_{+}$ is defined via $[x]_{+}=\max(0,x)$, and accounts for the (absence of force 
under potential) loss of contact between adjacent nodes.  While general lattices with alternating 
stiffness will have mass ($m_n$), equilibrium position ($d_n$), and elastic properties ($A_n$) depend 
on the lattice location, we consider here a prototypical setting where we normalize them all to unity 
($m_n=A_n = d_n = 1$) for simplicity, except for the nonlinear exponent $p_n$. This will allow us to 
elucidate differences that arise specifically from variations in stiffness as opposed to other
variations. We consider nonlinear exponents that alternate in the following fashion
\begin{align}
p_{n}\coloneqq p+(-1)^{n}\delta  > 0,
\label{nonlin_pow}
\end{align}
where $p$ and $\delta$ are parameters, i.e., we explore a form of a ``nonlinearity dimer''. An important 
special case is that  of $p_n = p= \mathrm{const} = 3/2$ (i.e., $\delta=0$), which corresponds to the case 
of the monomer granular crystal chain~\cite{Nester2001}, in which case all particles are spheres. Other particle 
geometries can lead to other values of $p$~\cite{Nester2001}. In general, for $p>1$ the contact is strain-hardening. 
The practical interpretation of ``strain-hardening" is that the ``spring" connecting two nodes becomes harder
to deform the larger the applied strain (this can be easily discerned by simply plotting the force relation
$F(x) = [1 + x]^p$). Another special case is $p_n = p= \mathrm{const}  < 1$, which, as mentioned in the introduction, 
corresponds to tensegrity~\cite{carpe}, origami lattices~\cite{origami,origami2}, stainless steel cylinders separated 
by polymer foams~\cite{RarefactionNester}, and hollow elliptical cylinder lattices~\cite{HEC_DSW}. This is the 
``strain-softening" case, in which scenario the ``spring" connecting two nodes becomes easier to deform the larger 
the applied strain. For non-constant ($n$-dependent) nonlinear exponents of Eq.~\eqref{nonlin_pow}, the model 
[cf. Eq.~\eqref{gc_dimer_dim}] that we propose considers softening and hardening behavior in an alternating fashion 
with $p_{2n-1}<1$ and $p_{2n}>1$.  Since the case of alternating stiffness type is what we are interested in here,
we will mostly consider $p=1$ and $0 < \delta < p$. In this sense, our chain is, more concretely, a ``stiffness dimer", 
as opposed to the more commonly studied ``mass-dimer"~\cite{granularBook}.

The chain [cf. Eq.~\eqref{gc_dimer_dim}] is harmonically driven at the left boundary according to
\begin{align}
u_{0}=a_{d}\cos{(2\pi f_{d}t)},
\end{align}
with $a_{d}$ and $f_{d}$ being the amplitude and frequency of the (external) drive,
and thus rendering the model a damped-driven one. At the right boundary of the 
chain, we assume a fixed wall which is modeled by $u_{N+1}=0$.

It is worth mentioning that (the normalized version of) Eq.~\eqref{gc_dimer_dim} 
with $\gamma=0=a_{d}$ stems from the Euler-Lagrange equations:
\begin{equation}
\frac{\mathrm{d}}{\mathrm{d}t}\left(\frac{\partial \mathcal{L}_{n}}{\partial \dot{u}_{n}}\right)%
-\frac{\partial \mathcal{L}_{n}}{\partial u_{n}}=0,
\label{el_eqs}
\end{equation}
with discrete Lagrangian density $\mathcal{L}_{n}$ given by:
\begin{equation}
\mathcal{L}_{n}=\frac{1}{2}\dot{u}_{n}^{2}-%
\frac{1}{2}\left[V_{n-1}\!\left(u_{n}-u_{n-1}\right)+V_{n}\!\left(u_{n+1}-u_{n}\right)\right],
\label{lagr_dens}
\end{equation}
where
\begin{equation}
V_{n}\!\left(u_{n+1}-u_{n}\right)=\frac{1}{p_{n+1}+1}\left[1+u_{n}-u_{n+1}\right]_{+}^{p_{n+1}+1}%
+u_{n+1}-u_{n}-\frac{1}{p_{n+1}+1}.
\label{Vpot}
\end{equation}
For this special case with $\gamma=0=a_{d}$, the Hamiltonian
\begin{align}
H=\sum_{n=1}^{N}\frac{1}{2}\dot{u}_{n}^{2}+V_{n}(u_{n+1}-u_{n})
\label{ham}
\end{align}
is conserved when free boundary conditions are employed at the left and 
right ends of the chain, i.e., $u_{0}=u_{1}$ and $u_{N+1}=u_{N}$, or in
case the lattice is infinite in length, i.e., $N\mapsto\infty$.

\subsection{Linear Analysis}

To obtain the dispersion relation and the eigenfrequency spectrum for the
stiffness-dimer problem, we study in this section the linearized problem associated 
with Eq.~\eqref{gc_dimer_dim}. At first, if
\begin{align}
|u_{n}-u_{n+1}|\ll 1,
\label{lin_limit}
\end{align}
then Eq.~\eqref{gc_dimer_dim} reduces to
\begin{align}
\ddot{u}_{n}=p_{n}\left(u_{n-1}-u_{n}\right)%
-p_{n+1}\left(u_{n}-u_{n+1}\right)-\gamma\dot{u}_{n},
\label{gc_dimer_lin_dim}
\end{align}
which, upon introducing $v_{n}\coloneqq u_{2n}$ and $w_{n}\coloneqq u_{2n+1}$, i.e., even 
and odd displacements, respectively, can be written as
\begin{subequations}
\begin{align}
\ddot{v}_{n}&=p_{2}\left(w_{n-1}-v_{n}\right)%
-p_{1}\left(v_{n}-w_{n}\right)%
-\gamma\dot{v}_{n}
\label{gc_dimer_lin_system_1_dim}
\\
\ddot{w}_{n}&=p_{1}\left(v_{n}-w_{n}\right)%
-p_{2}\left(w_{n}-v_{n+1}\right)%
-\gamma\dot{w}_{n}, 
\label{gc_dimer_lin_system_2_dim}
\end{align}
\end{subequations}
with solutions given by the Bloch ansatz:
\begin{subequations}
\begin{align}
v_{n}&=\eta_{1}\,e^{\ii\left(\kappa\alpha j + \omega t\right)},
\label{ba_1_dim}\\
w_{n}&=\eta_{2}\,e^{\ii\left(\kappa\alpha j + \omega t\right)}.
\label{ba_2_dim}
\end{align}
\end{subequations}
Upon substituting Eqs.~\eqref{ba_1_dim}-\eqref{ba_2_dim} into
Eqs.~\eqref{gc_dimer_lin_system_1_dim}~\eqref{gc_dimer_lin_system_2_dim},
we arrive at the following linear system for the amplitudes $\eta_{1,2}$:
\begin{align}
\begin{bmatrix}
\omega^{2} - 2p-\ii\omega\gamma   &    p_{1}+p_{2}e^{-\ii \kappa \alpha} \\
p_{1} + p_{2}e^{\ii \kappa \alpha}                      & \omega^{2}-2p-\ii\omega\gamma \\
\end{bmatrix}
\begin{bmatrix}
\eta_{1} \\ \eta_{2}
\end{bmatrix}
=\begin{bmatrix}
  0 \\0 
\end{bmatrix},
\label{disp_mat_sys_dim}
\end{align}
where $2p=p_1+p_2$ therein. A non-trivial solution to the above system exists if the determinant 
of the matrix (containing the coefficients in Eq.~\eqref{disp_mat_sys_dim}) is $0$. This condition 
results in the following dispersion relation:
\begin{align}
\omega^{4}-2\ii \gamma\omega^{3}-\left[\gamma^{2}+4p\right]\omega^{2}%
+4\ii \gamma p\omega+4p_{1}p_{2}\sin^{2}{\left(\frac{\kappa\alpha}{2}\right)}=0,
\label{disp_eq_dim}
\end{align}
whose solutions read
\begin{subequations}
\begin{align}
\omega_{1}^{\left(\pm\right)}&=\pm%
\sqrt{2p-\sqrt{p_{1}^{2}+p_{2}^{2}+2p_{1}p_{2}\cos{\left(\kappa\alpha\right)}}
-\frac{\gamma^{2}}{4}}+\ii\frac{\gamma}{2},
\label{sol_omega1_dim}\\
\omega_{2}^{\left(\pm\right)}&=\pm%
\sqrt{2p+\sqrt{p_{1}^{2}+p_{2}^{2}+2p_{1}p_{2}\cos{\left(\kappa\alpha\right)}}
-\frac{\gamma^{2}}{4}}+\ii\frac{\gamma}{2},
\label{sol_omega2_dim}
\end{align}
\end{subequations}
and define the cut-off frequencies (at $\kappa=0$ and $\kappa=\pi/\alpha$):
\begin{subequations}
\begin{align}
f_{c,1}^{(+)}(\kappa=0)&=\ii\frac{\gamma}{2\pi},\label{cut_off_1_dim}\\ 
f_{c,1}^{(-)}(\kappa=0)&=0,\label{cut_off_2_dim} \\
f_{c,1}^{(\pm)}(\kappa=\pi/\alpha)&=\pm \frac{1}{2\pi}%
\sqrt{2\left(p+\delta\right)-\frac{\gamma^{2}}{4}}+\ii \frac{\gamma}{4\pi},\label{cut_off_3_dim}\\
f_{c,2}^{(\pm)}(\kappa=0)&=\pm\frac{1}{\pi}\sqrt{p-\frac{\gamma^{2}}{16}}+\ii\frac{\gamma}{4\pi},\label{cut_off_4_dim}\\
f_{c,2}^{(\pm)}(\kappa=\pi/\alpha)&=\pm %
\frac{1}{2\pi}\sqrt{2\left(p-\delta\right)-\frac{\gamma^{2}}{4}}+\ii\frac{\gamma}{4\pi}.\label{cut_off_5_dim}
\end{align}
\end{subequations}
Here, it is relevant to note that the relevant frequencies are defined according to 
$f=\omega/(2 \pi)$ from the results of the dispersion relation. Moreover, the imaginary 
nature of the frequencies reflects the presence of loss, since $\ii\omega$ entering the 
expressions of Eqs.~\eqref{ba_1_dim}-\eqref{ba_2_dim} leads to a negative real part, 
reflecting the lossy part of the relevant exponential term. The top left panel of Fig.~\ref{fig1} 
shows the dispersion curves and the cut-off frequencies (see, also Eqs.~\eqref{sol_omega1_dim}-\eqref{sol_omega2_dim}) 
as functions of the wavenumber $\kappa\alpha$.  The existence of a frequency gap can be discerned from the figure.

In the case of homogeneous Dirichlet boundary conditions, i.e., $u_{0}\equiv u_{N+1}=0$, and
for a finite domain, the above analytical solutions are no longer available, and the relevant 
eigenvalue problem is solved numerically (see, e.g., Ref.~\cite{Stathis} for details on the 
relevant calculation). Indeed, in the top right panel of Fig.~\ref{fig1}, the frequencies obtained 
numerically for $N=42$ are shown with filled circles together with the cut-off frequencies 
(obtained analytically from Eqs.~\eqref{cut_off_1_dim}-\eqref{cut_off_5_dim}) with dashed-dotted 
black lines. A perfect agreement can be discerned from this panel suggesting the consistency of 
our (linear) analysis. Example eigenvectors of the acoustic and optical bands are shown
in the bottom left and right panels, respectively. It is interesting to observe that in the
former (acoustic) case, the nodes bearing a softening nonlinearity are
excited, while in the latter (optical) one, it is instead the
hardening interaction nodes that are largely excited, while the
softening ones are, for the most part, suppressed.

\section{Time-periodic solutions in the damped-driven chain} \label{sec:damped-driven}
%
\begin{figure}[pt!]
\begin{center}
\includegraphics[height=.19\textheight, angle =0]{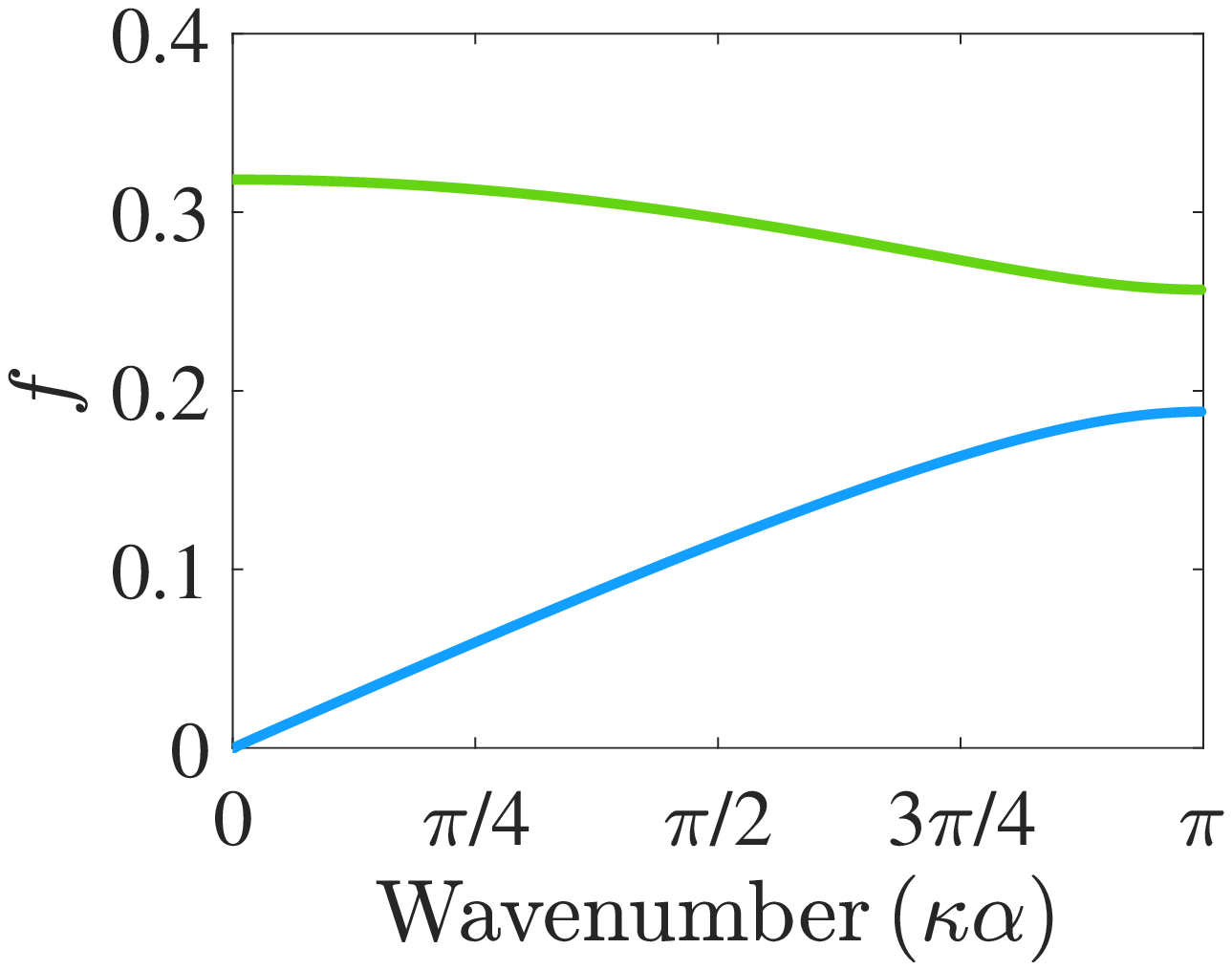}
\includegraphics[height=.19\textheight, angle =0]{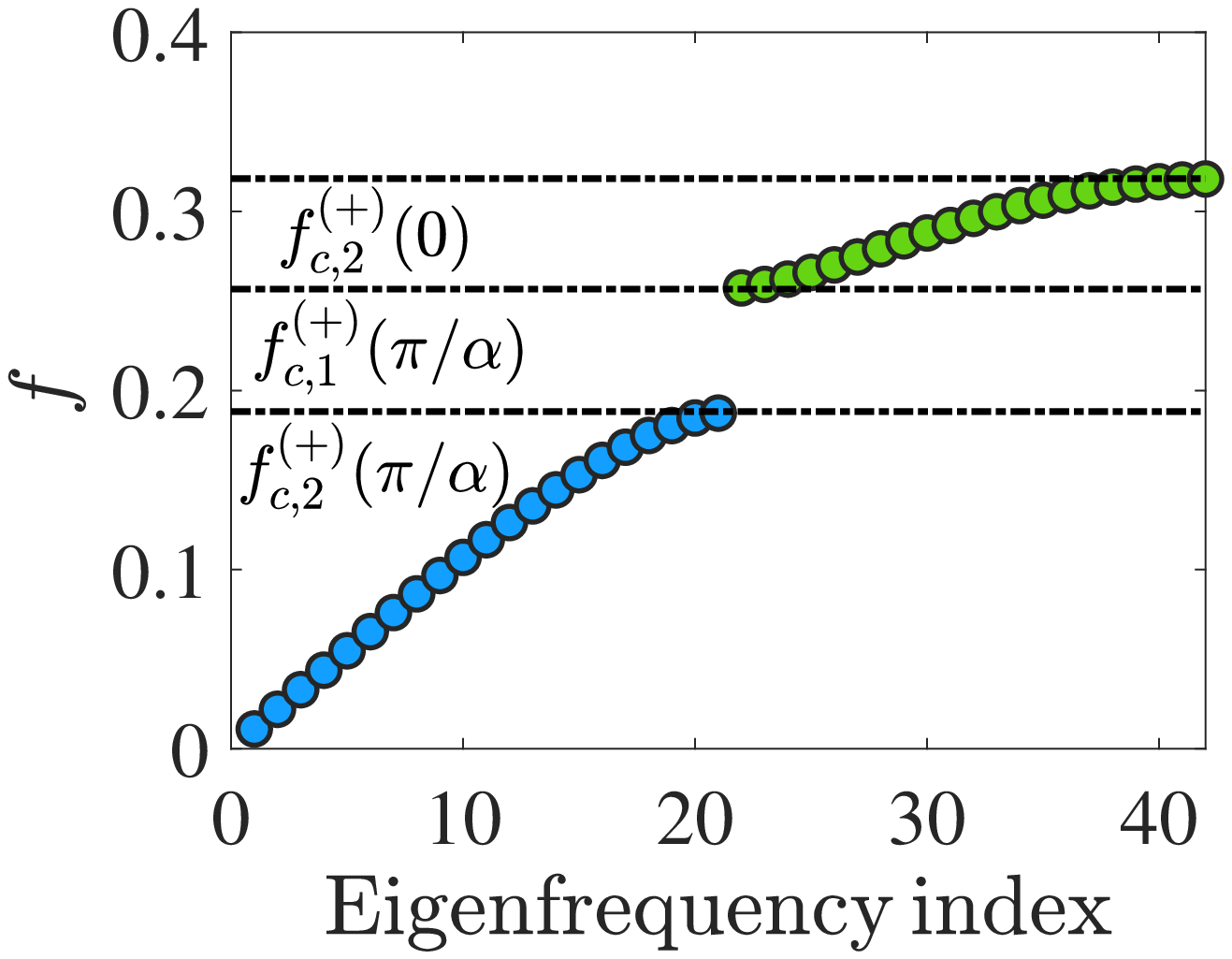}
\includegraphics[height=.19\textheight, angle =0]{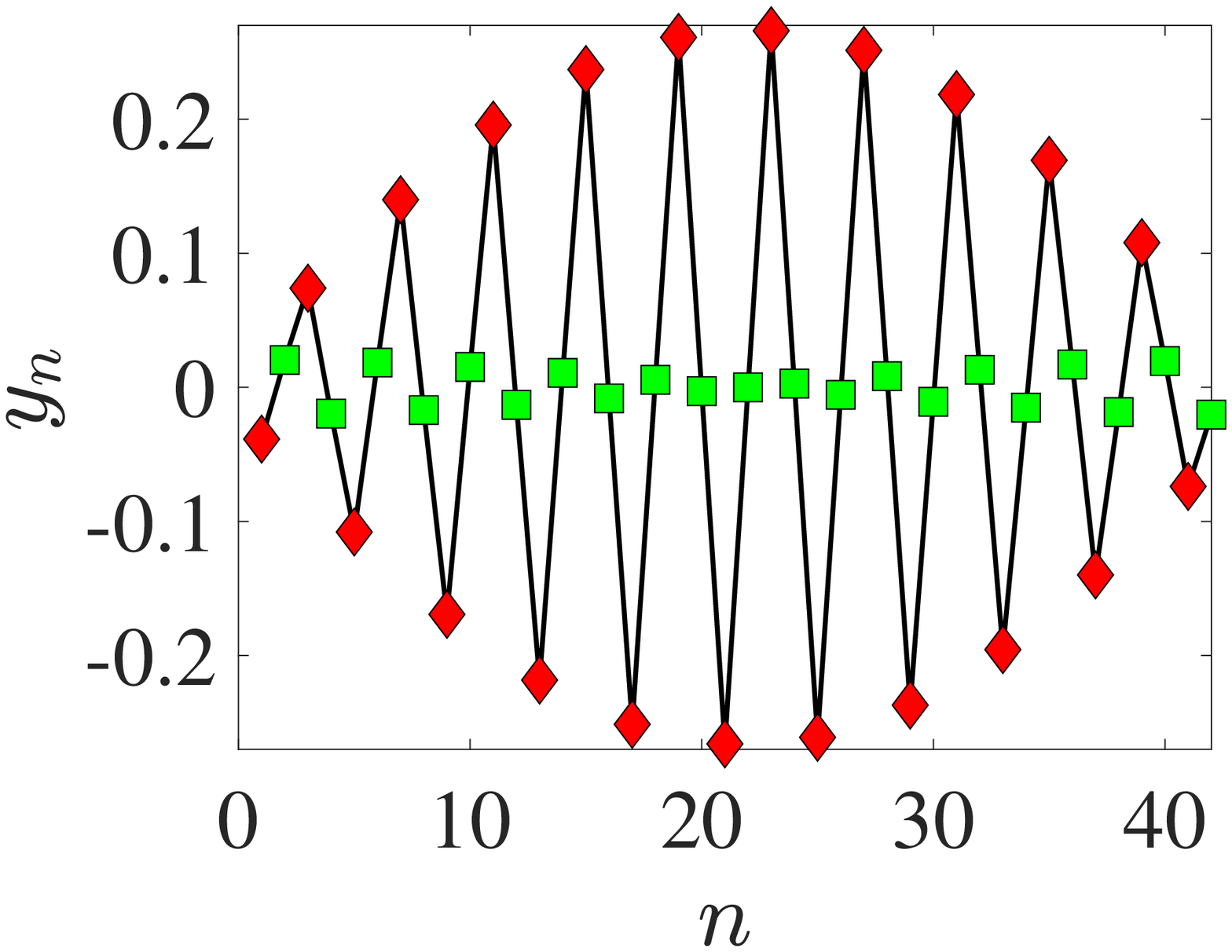}
\includegraphics[height=.19\textheight, angle =0]{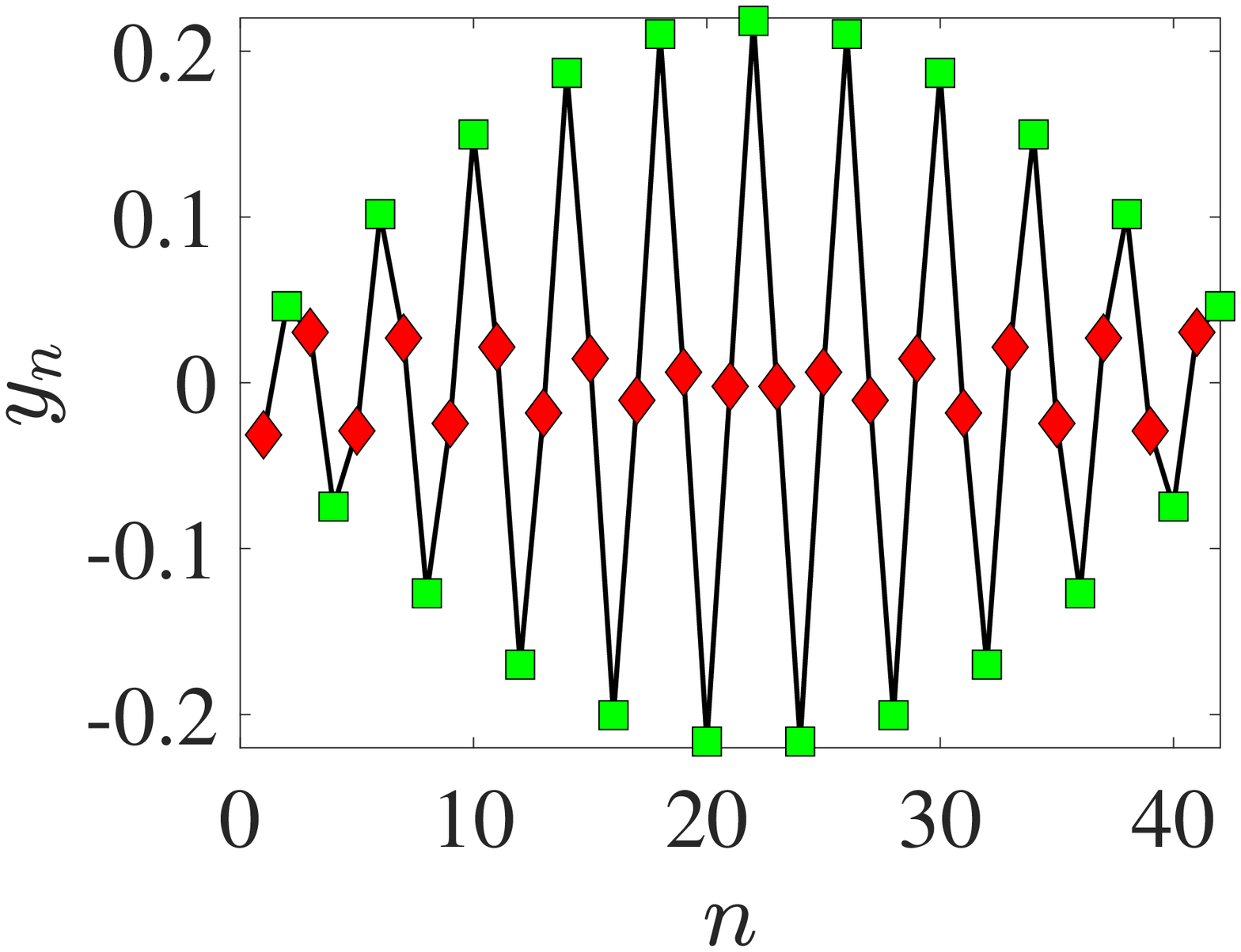}
\end{center}
\caption{(Color online)
Linear analysis of the stiffness-dimer model [cf. Eq.~\eqref{gc_dimer_dim}]
with $N=42$ particles, and for parameter values of $p=1$, $\delta=0.3$, 
and $\gamma=8\times 10^{-3}$. The top left panel presents the frequencies of the infinite 
lattice. Note that the blue and green colors stand for the acoustic and 
optical bands, respectively. The top right panel of the figure demonstrates 
the numerically obtained frequencies with filled blue and green circles. 
The horizontal dashed-dotted black lines in the panel highlight the cut-off 
frequencies: $f_{c,2}^{(+)}(\kappa=\pi/\alpha)\approx 0.1883$, $f_{c,1}^{(+)}(\kappa=\pi/\alpha)\approx 0.2566$, 
and $f_{c,2}^{(+)}(\kappa=0)\approx 0.3183$, respectively. It is important to
add here that, while the relevant frequencies bear an imaginary part 
(reflecting the lossy nature of the chain), only the real part of thereof is
represented in this figure. The bottom left panel is a plot of the strain $y_n = u_{n-1}-u_n$
of an eigenvector within the acoustic band (index 20). The red diamond corresponds to a softening interaction 
and the green square corresponds to a hardening interaction.  The bottom right panel
is the same as the bottom left, but for an eigenvector in the optical band (index 21).
}
\label{fig1}
\end{figure}

We now turn to the study of time-periodic solutions to the nonlinear problem with damping and driving.
We will use primarily numerical computations for their study, but we discuss an analytical
approximation in the next section using a reduction to the NLS equation in the Hamiltonian limit
of the problem.
 
We compute time-periodic orbits, and their spectral stability, of Eq.~\eqref{gc_dimer_dim} with 
period $T_d = 1/f_d$ with high precision using a standard Newton-type procedure, see Appendix A 
for details. To investigate the dynamical stability of the obtained 
states, a Floquet analysis is used to compute the multipliers associated with the solutions. If 
a solution has all Floquet multipliers on or inside the unit circle, the solution is called (spectrally) 
stable. An instability that results from a multiplier on the (positive) real line is called a real
(exponential in nature) instability. However, there can also be  oscillatory instabilities, which 
correspond to complex-conjugate pairs of Floquet multipliers  lying outside  the unit circle (in the 
complex plane). In the bifurcation diagrams that are to follow in this paper, solid blue segments will 
correspond to stable parametric regions while the dashed-dotted red segments correspond to real unstable 
regions and green dashed-dotted segments correspond to oscillatorily unstable regions. The bifurcation diagrams 
are obtained using pseudo-arclength continuation as it is implemented in the software package AUTO~\cite{Doedel}. 
In all of the bifurcation diagrams that follow below, the monitored quantity that will be used is the average 
energy over one period $T_{d}$ defined as:
\begin{align}
\overline{E}\coloneqq\frac{1}{T_{d}}\int_{0}^{T_{d}}E\,dt,
\label{avg_E}
\end{align}
where
\begin{align}
E=\sum_{n=1}^{N}e_{n}
\label{tot_E}
\end{align}
is the total energy, and 
\begin{align}
e_{n}=\frac{1}{2}\dot{u}_{n}^{2}+%
\frac{1}{2}\left[V_{n-1}\left(u_{n}-u_{n-1}\right)+V_{n}\left(u_{n+1}-u_{n}\right)\right],
\label{energ_dens}
\end{align}
is the (discrete) energy density. Note that $V_{n}$ in Eq.~\eqref{energ_dens} 
is given by Eq.~\eqref{Vpot}. 

\begin{figure}[pt!]
\begin{center}
\includegraphics[height=.17\textheight, angle =0]{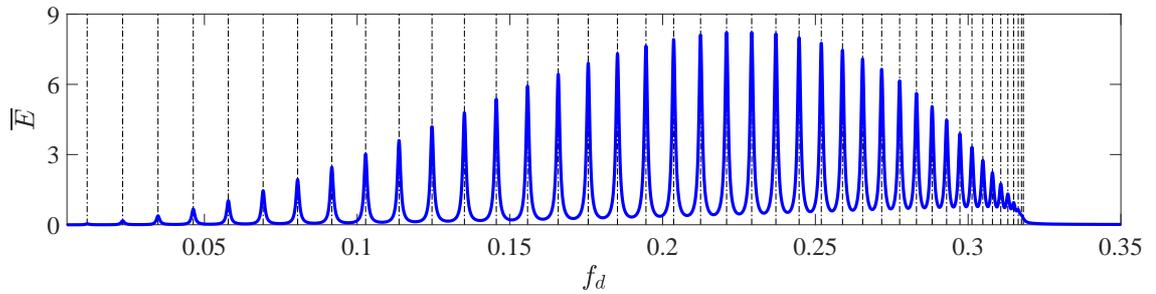}
\end{center}
\caption{(Color online)
The dependence of the average energy density over one period $\overline{E}$ 
[cf. Eq.~\eqref{avg_E}] as a function of the driving frequency $f_{d}$ for a value 
of the driving amplitude of $a_{d}=0.15$. The vertical dashed-dotted black lines 
correspond to the eigenmodes of the respective eigenvalue problem, for comparison. Note that in this case 
$p_{n}\equiv 1\,\forall n$, i.e., the lattice is linear. 
\label{fig2}
}
\end{figure}

\begin{figure}[pt!]
\begin{center}
\includegraphics[height=.17\textheight, angle =0]{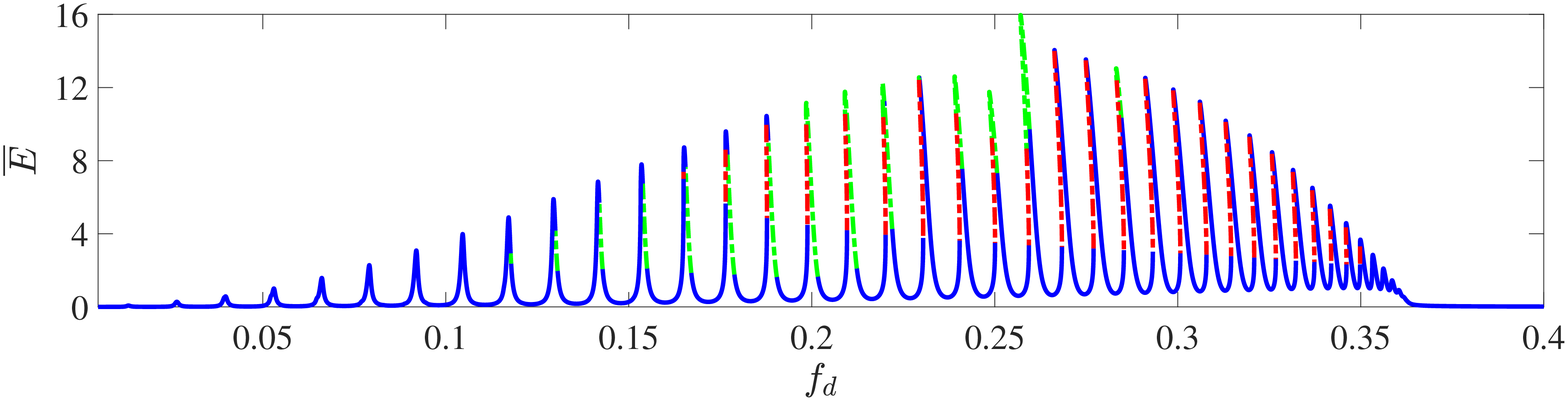}
\includegraphics[height=.17\textheight, angle =0]{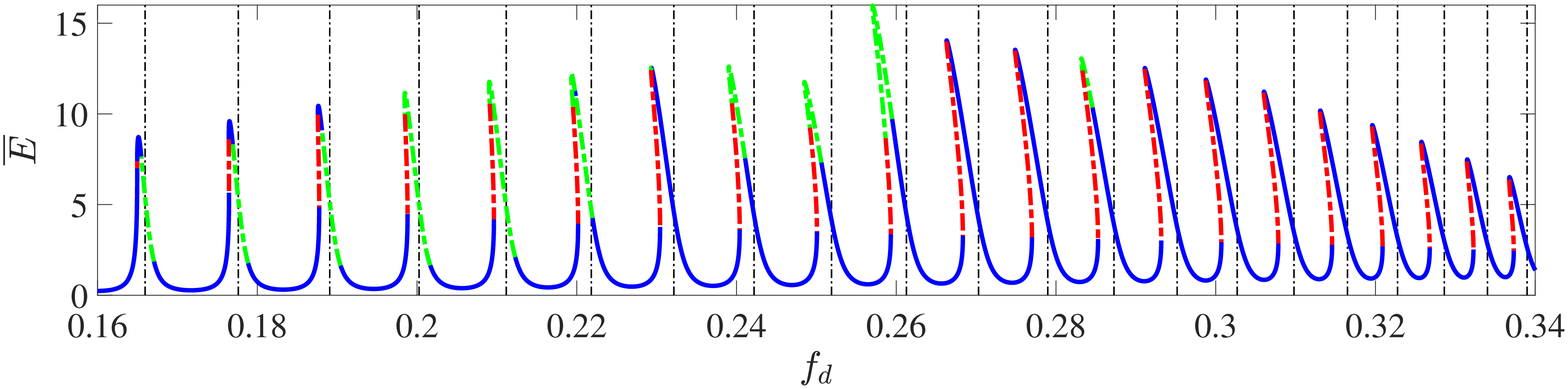}
\end{center}
\caption{(Color online)
Same as Fig.~\ref{fig2} but with $p_{n}\equiv p+\delta\,\forall n$.
Both top and bottom panels correspond to a value of the driving 
amplitude of $a_{d}=0.15$. The bottom panel is a zoom-in of the 
top one where the dashed-dotted black lines correspond to the 
eigenmodes of the linear problem. Note that the curves bend to 
the left (see the bottom panel).
\label{fig3}
}
\end{figure}

\begin{figure}[pt!]
\begin{center}
\includegraphics[height=.17\textheight, angle =0]{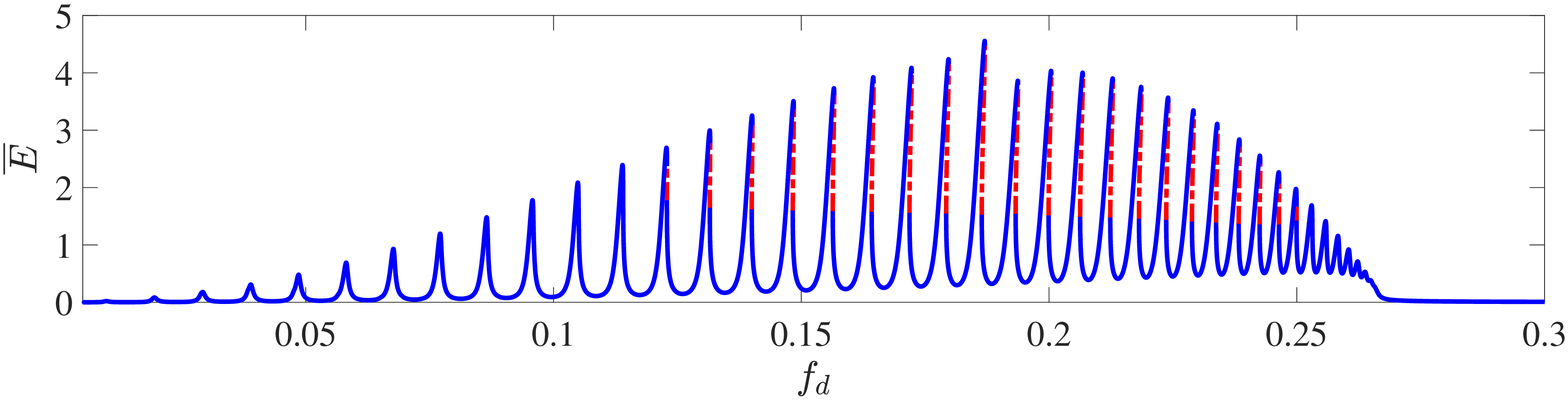}
\includegraphics[height=.17\textheight, angle =0]{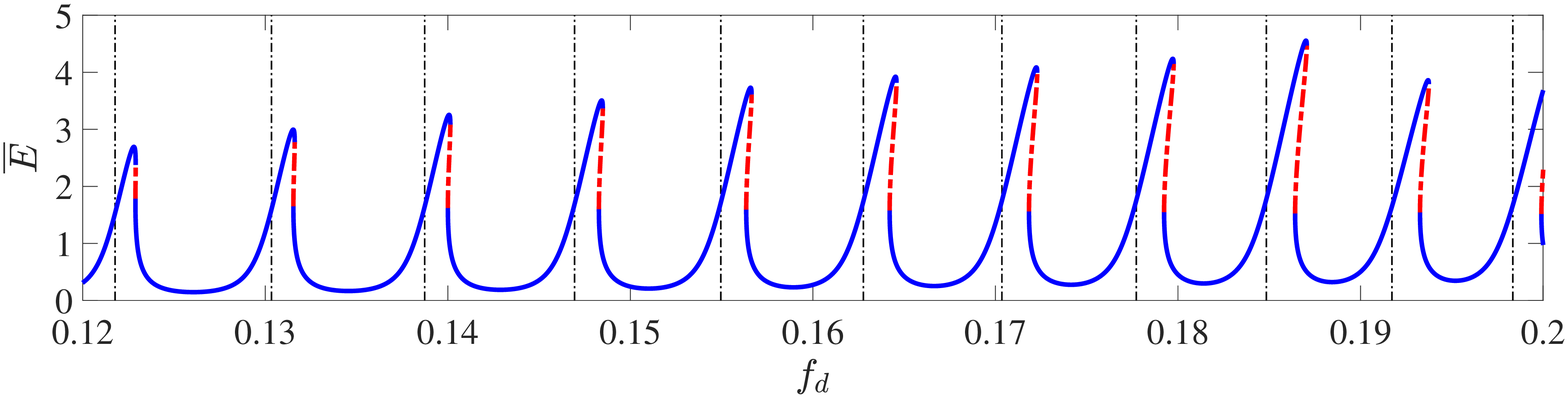}
\end{center}
\caption{(Color online)
Same as Fig.~\ref{fig2} but with $p_{n}\equiv p-\delta\,\forall n$.
Both top and bottom panels correspond to a value of the driving 
amplitude of $a_{d}=0.15$. Similar to the bottom panel of Fig.~\ref{fig3},
the bottom panel in the present figure is a zoom-in of the top one 
with the dashed-dotted black lines corresponding to the eigenmodes 
of the linear problem. In this case, the curves bend to the right.
\label{fig4}
}
\end{figure}

\begin{figure}[pt!]
\begin{center}
\includegraphics[height=.17\textheight, angle =0]{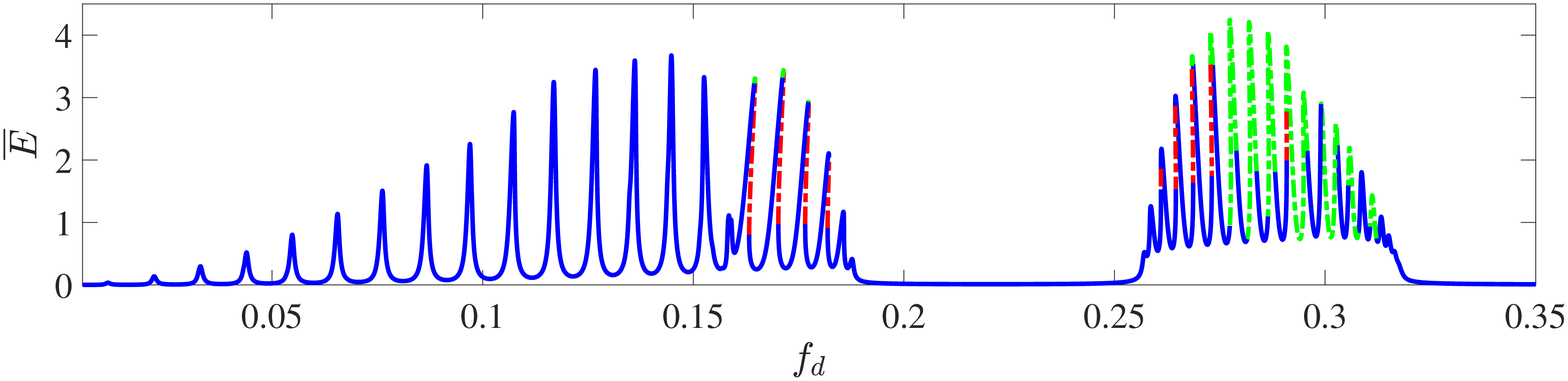}
\includegraphics[height=.17\textheight, angle =0]{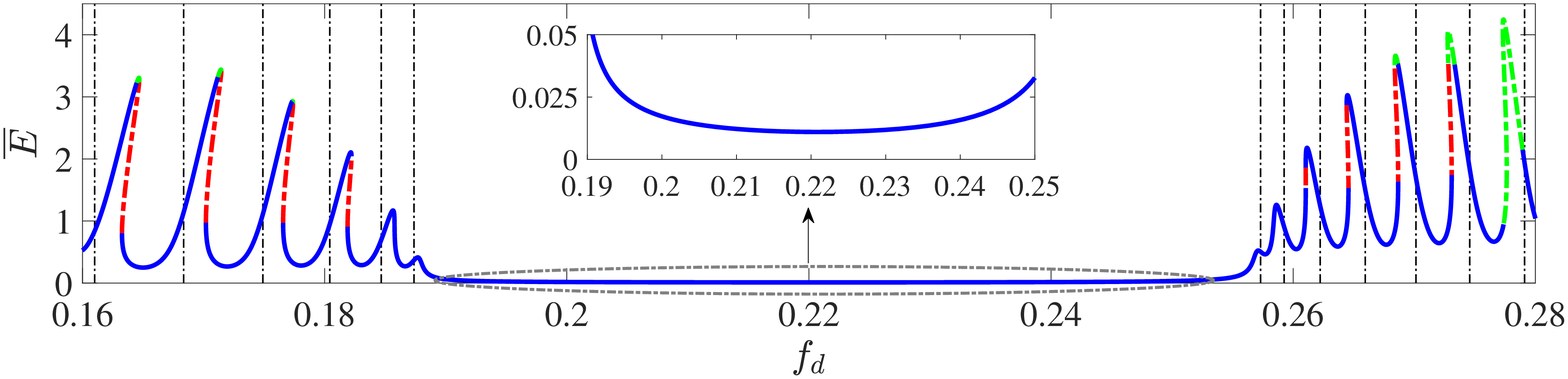}
\end{center}
\caption{(Color online)
Same as Fig.~\ref{fig2} but with $p_{n}$ given by Eq.~\eqref{nonlin_pow}
(and $a_{d}=0.15$). The format of this figure is the same as the 
one of Figs.~\ref{fig3} and~\ref{fig4}. On the left and right of the 
frequency gap (see also the inset in the bottom panel), the curves bend 
to the right and left, respectively, i.e., towards the frequency gap.
\label{fig5}
}
\end{figure}

For the numerical computations, we consider a chain with $N=42$ particles,
and set $p=1$, $\delta=0.3$, and $\gamma=8\times 10^{-3}$ (unless explicitly 
stated otherwise). The drive amplitude $a_d$ and frequency $f_d$ will be used 
as continuation parameters. Let us start our presentation by considering 
Figs.~\ref{fig2}-\ref{fig5}, which correspond to frequency continuations. 
In particular, we consider the following cases: linear monomer chain ($p_{n}\equiv p$) 
in Fig.~\ref{fig2}, a strain-hardening monomer chain ($p_{n}=p+\delta$) in 
Fig.~\ref{fig3}, a strain-softening monomer chain ($p_{n}=p-\delta$) in 
Fig.~\ref{fig4}, and a dimer chain with alternating stiffness ($p_{n}$ given by 
Eq.~\eqref{nonlin_pow}) in Fig.~\ref{fig5}. Figures~\ref{fig2}-\ref{fig5} showcase 
the average energy [cf. Eq.~\eqref{avg_E}] as a function of the driving frequency $f_{d}$.
When $p_{n}\equiv 1\,\forall n=1,\dots,N$, i.e., for the case corresponding to a linear lattice, 
it can be discerned from Fig.~\ref{fig2} that the resonant peaks match perfectly with the resonant 
frequencies of the respective linear problem shown with vertical dashed-dotted black lines 
(and solved numerically) in this case, as expected. Adding nonlinearity to the system will 
cause these linear resonant peaks to deform, or ``bend". For example, supposing that all 
contacts are strain-hardening, the resonant peaks will bend toward smaller frequencies, as 
shown in Fig.~\ref{fig3} where $p_{n}=p+\delta > 1$. If all contacts are strain-softening, 
the resonant peaks will bend toward larger frequencies as shown in Fig.~\ref{fig4} where $p_{n}=p-\delta < 1$.
The observations in Figs.~\ref{fig3} and~\ref{fig4} are expected, and are inline with 
the established behavior for oscillators with hardening or softening nonlinearities~\cite{Mook}. 
In the bottom panels of Figs.~\ref{fig3} and~\ref{fig4}, the resonant frequencies are depicted
by vertical dashed-dotted lines (similar to Fig.~\ref{fig2}) in order to ease the visualization
of the bending of the resonant curves (to the left or right).

Having investigated the baseline cases of pure hardening and pure softening stiffness, we now 
turn to the aspect of alternating stiffness, i.e., Eq.~\eqref{gc_dimer_dim} with nonlinearity 
exponents $p_{n}$ given by Eq.~\eqref{nonlin_pow}. The frequency continuation for this case is 
shown in Fig.~\ref{fig5} for $a_{d}=0.15$. Notice the resonant peaks close to the top edge of the 
acoustic band and the bottom edge of the optical band all bend toward the gap due to the interplay 
of softening and hardening. This finding is shown in detail in the bottom panel of Fig.~\ref{fig5}, 
which is a zoom of the top panel. Note that the eigenfrequency modes (shown with vertical dashed-dotted
lines) help highlight this effect as well as the existence of a frequency gap. Note that in traditional
mass-dimer lattices, resonant peaks from either the top of the acoustic band or the bottom of the 
optical band will bend into the frequency gap, but not both, like in the present case \cite{granularBook}.
The nature of the bending in the lattice with alternating stiffness can be understood in the small 
amplitude limit by inspection of the strain of the eigenvectors of the linear problem. In particular, within 
the acoustic band, the larger amplitude strains occur for odd indices, see the bottom left panel of Fig.~\ref{fig1}. 
According to Eq.~\eqref{nonlin_pow}, the odd numbered interactions are of the softening type. Hence, one would 
expect softening-type dynamics, involving resonant peaks bending toward the right, as seen in Fig.~\ref{fig5}. 
The situation in the optical band is reversed. The even numbered interactions have higher amplitude in the strain 
(see the bottom right panel of Fig.~\ref{fig1}) indicating that hardening-type behavior should be observed, involving 
resonant peaks bending to the left, like in Fig.~\ref{fig5}.

A discussion about the stability characteristics of the pertinent branches of Figs.~\ref{fig2}-\ref{fig5} is 
now in order. In the linear lattice of Fig.~\ref{fig2}, we observe that the entire  branch is spectrally stable. 
In the nonlinear cases of Figs.~\ref{fig3}-\ref{fig5}, the branches are mostly stable, with  exceptions occurring 
typically between two turning points. These segments, in line with what is known for classical oscillator 
systems~\cite{Mook}, are found to be unstable. For example, in Fig.~\ref{fig3} and for $f_{d}\approx[0.29,0.35]$, 
the alternation of stable and unstable segments between turning points can be discerned. The instability in this 
case corresponds to an exponential one (and is shown with dashed-dotted red segments). A similar observation 
is found in Fig.~\ref{fig4} for drive frequencies $f_{d}\approx[0.12,0.25]$. Interestingly, in some portions of 
the relevant branches, and in particular those of Figs.~\ref{fig3} and~\ref{fig5} (see the bottom panels therein), 
we find oscillatory instabilities. In these cases, the branches have been colored green to reflect this modified 
instability feature and the associated expected dynamics.

\begin{figure}[pt!]
\begin{center}
\includegraphics[height=.17\textheight, angle =0]{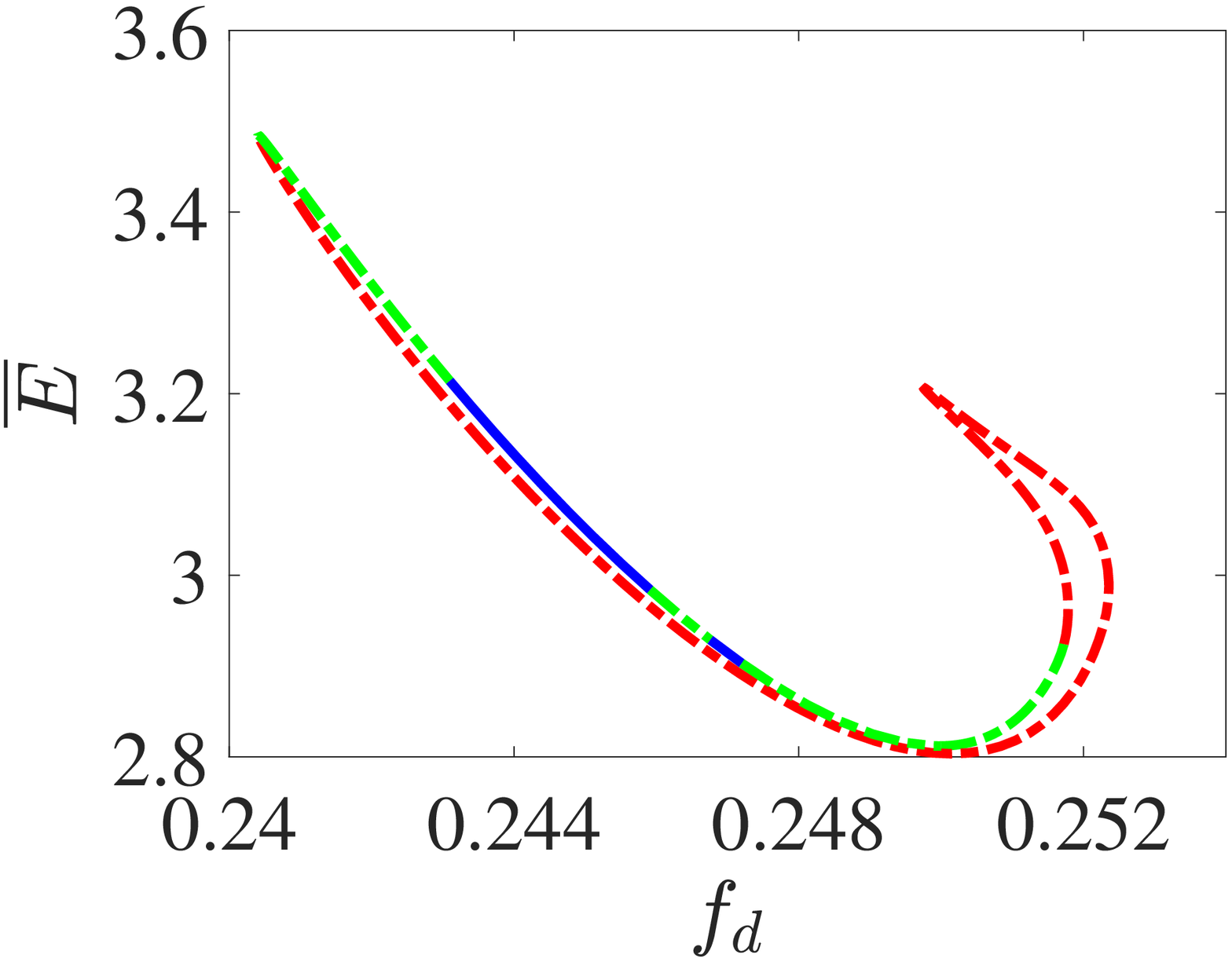}
\end{center}
\caption{(Color online)
Example of an isola for the same parameters given in Fig.~\ref{fig5}. 
\label{fig5b}
}
\end{figure}

Besides the solutions that live on the main resonant curve shown in Fig.~\ref{fig5}, there are also ``isolas" of 
solutions, namely closed curves of solutions in the bifurcation diagram~\cite{Isolas}; see Fig.~\ref{fig5b} for 
example (and~\cite{isola_strategy} on how it was obtained). This isola lives entirely within the spectral gap. 
Periodic solutions with frequency that lie within a spectral gap are interesting since they are (necessarily) 
spatially localized. Such solutions are called nonlinear localized modes, or breathers~\cite{Flach2007}. The case 
of damped-driven lattices where there is a driving source has attracted considerable attention in the present 
context. This is because of the intriguing interplay of localization due to the ``defect'' (i.e., the source) and 
that of the bulk breathers (or traveling waves) potentially available in the medium. This interplay can be used to 
produce hysteresis, metastability, resonances/antiresonances and related hallmarks of nonlinear dynamics that have 
been explored in a number of publications~\cite{Nature11,hooge12,PhysRevE.92.063203}. It is interesting to highlight 
that there is a portion of the relevant isola branch that is found to be spectrally stable. The starting point of our 
subsequent study in this work will be to explore the time-periodic solutions of this system that have a frequency that 
lies in the spectral gap in the presence of damping and driving. Having examined those, we will subsequently turn to 
the case of the breathers in the Hamiltonian lattice. 

\begin{figure}[pt!]
\begin{center}
\includegraphics[height=.17\textheight, angle =0]{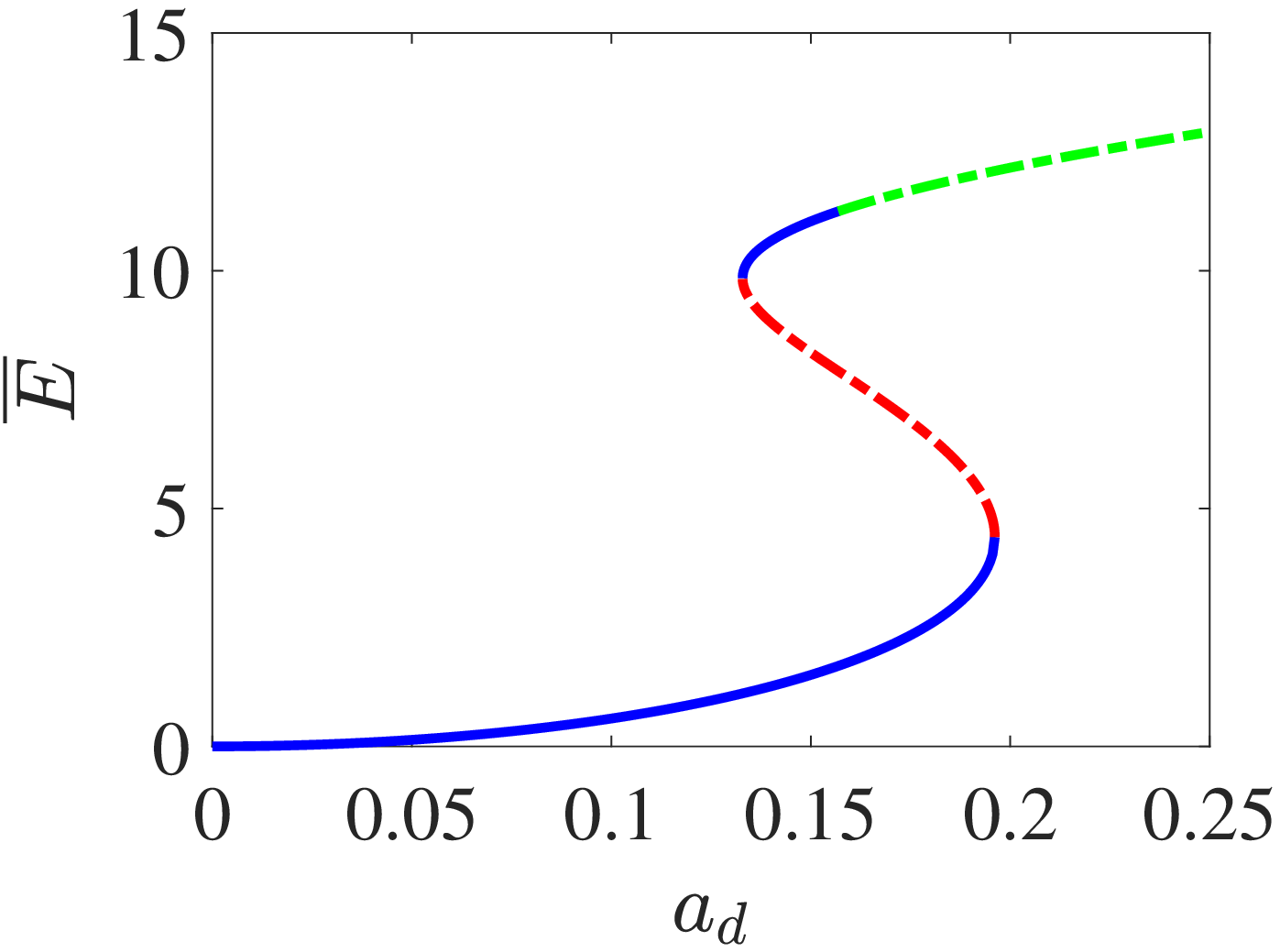}
\includegraphics[height=.17\textheight, angle =0]{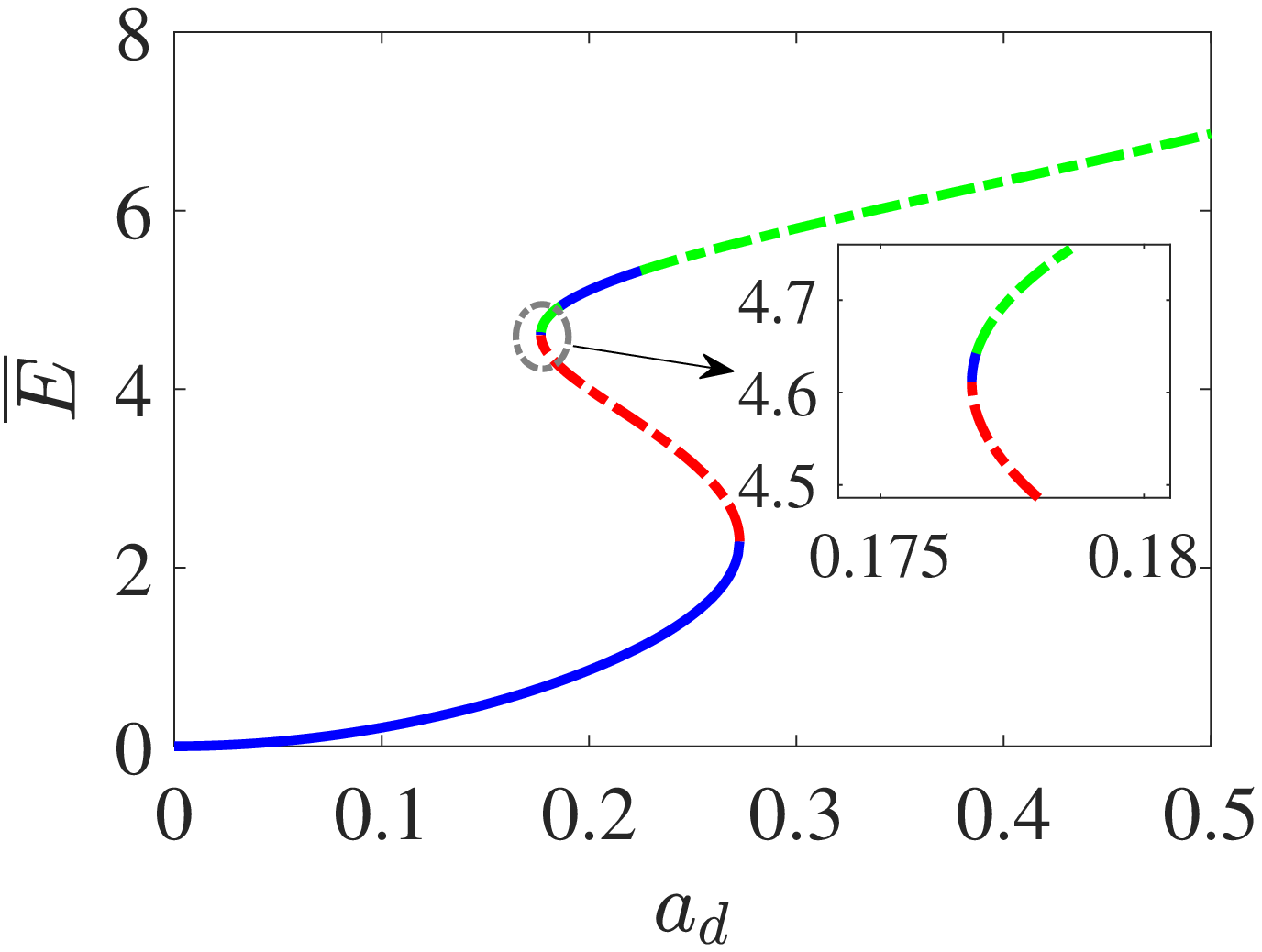}
\includegraphics[height=.17\textheight, angle =0]{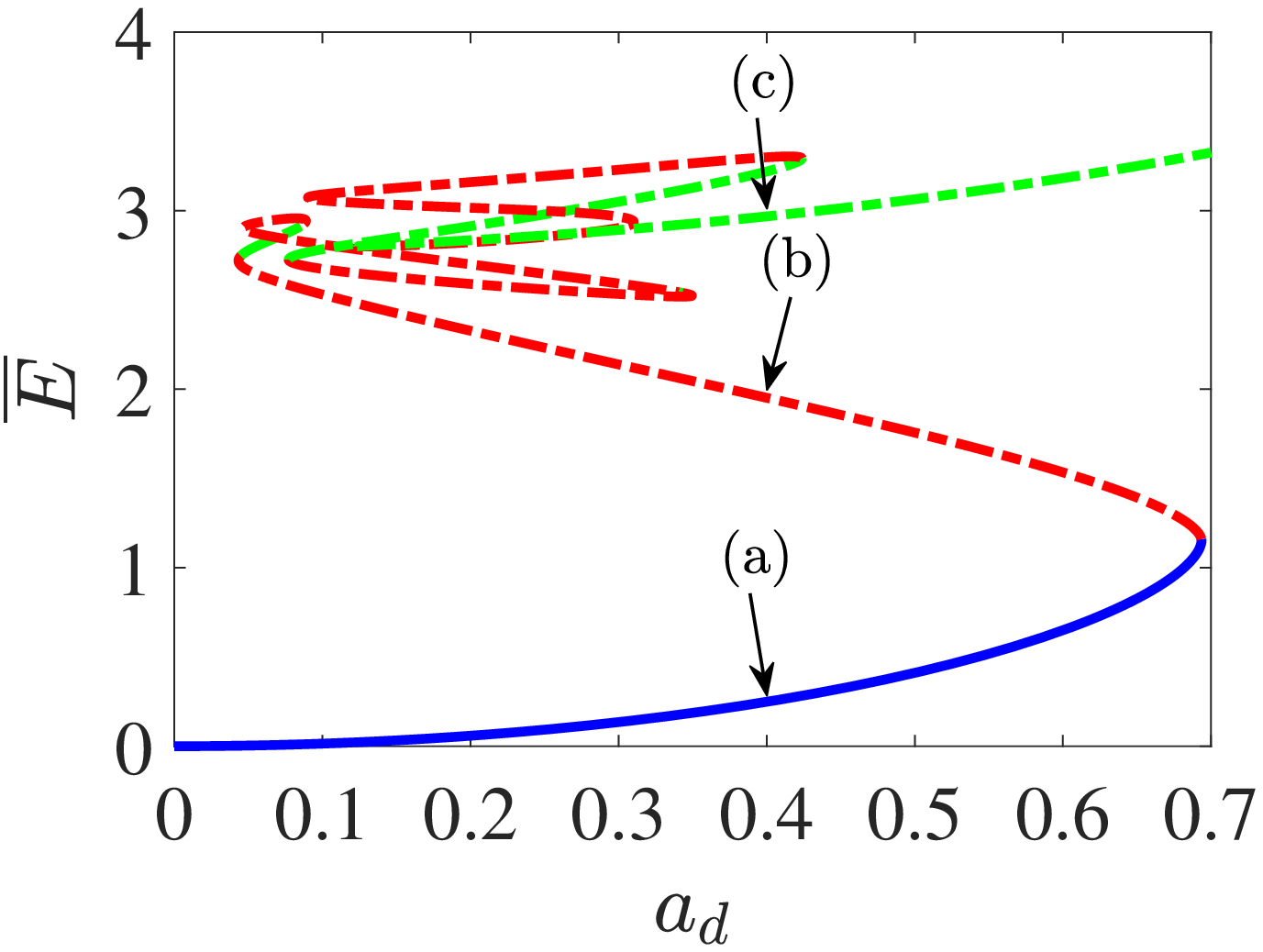}
\end{center}
\caption{(Color online)
The dependence of the average energy density over one period $\overline{E}$ 
[cf. Eq.~\eqref{avg_E}] as a function of $a_{d}$. The left and middle panels 
correspond to a value of driving frequency of $f_{d}=0.23$ with $p_{n}\equiv p+\delta\,\forall n$
(left) and $p_{n}\equiv p-\delta\,\forall n$ (right), respectively.
The right panel corresponds to the case with alternating nonlinearity 
and for a value of the driving frequency of $f_{d}=0.25$ (i.e., 
being inside the frequency gap). The labels in the panel are 
connected to the breather solutions depicted in Fig.~\ref{fig:profiles}.
\label{fig6}
}
\end{figure}

\begin{figure}[pt!]
\begin{center}
\includegraphics[height=.17\textheight, angle =0]{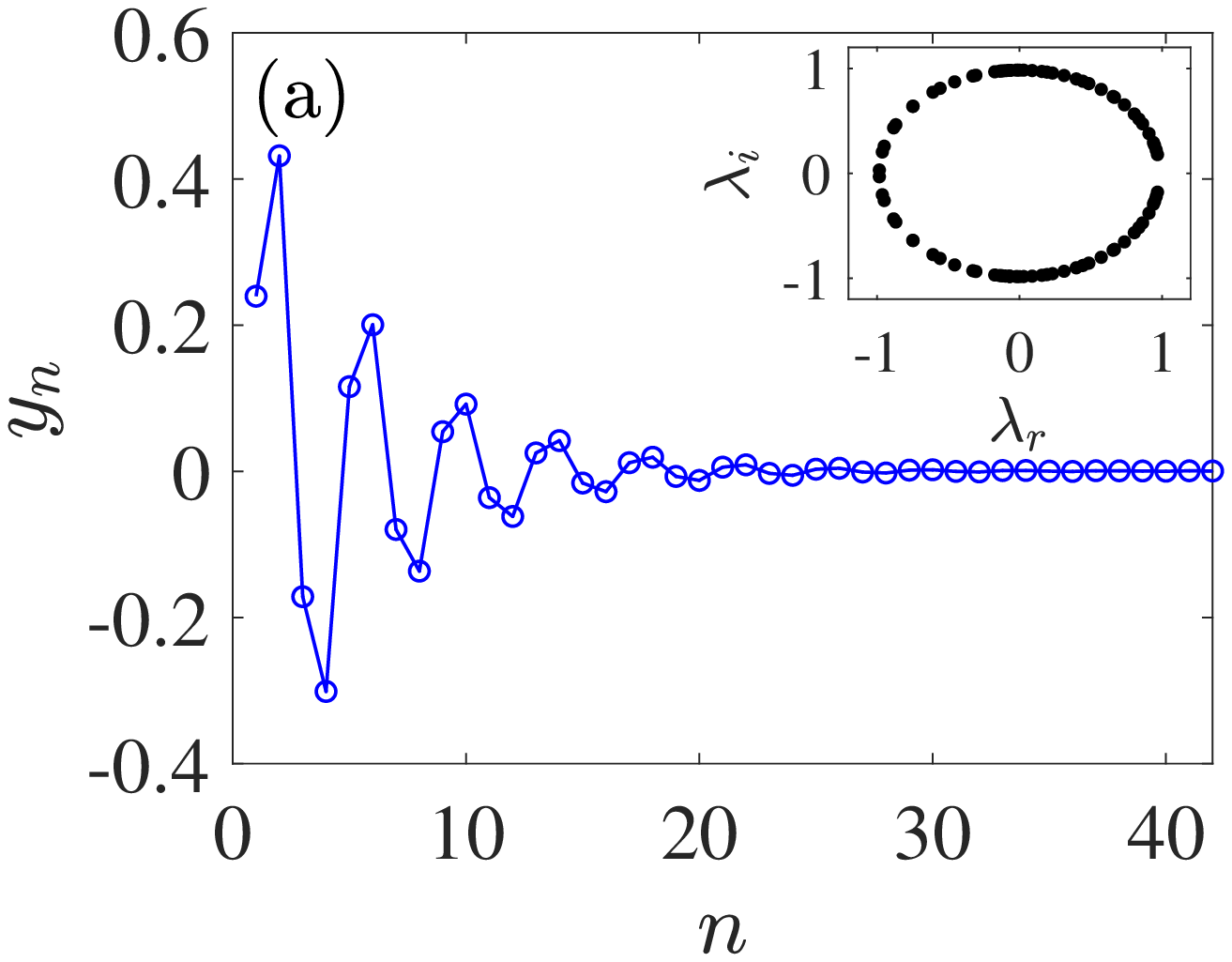}
\includegraphics[height=.17\textheight, angle =0]{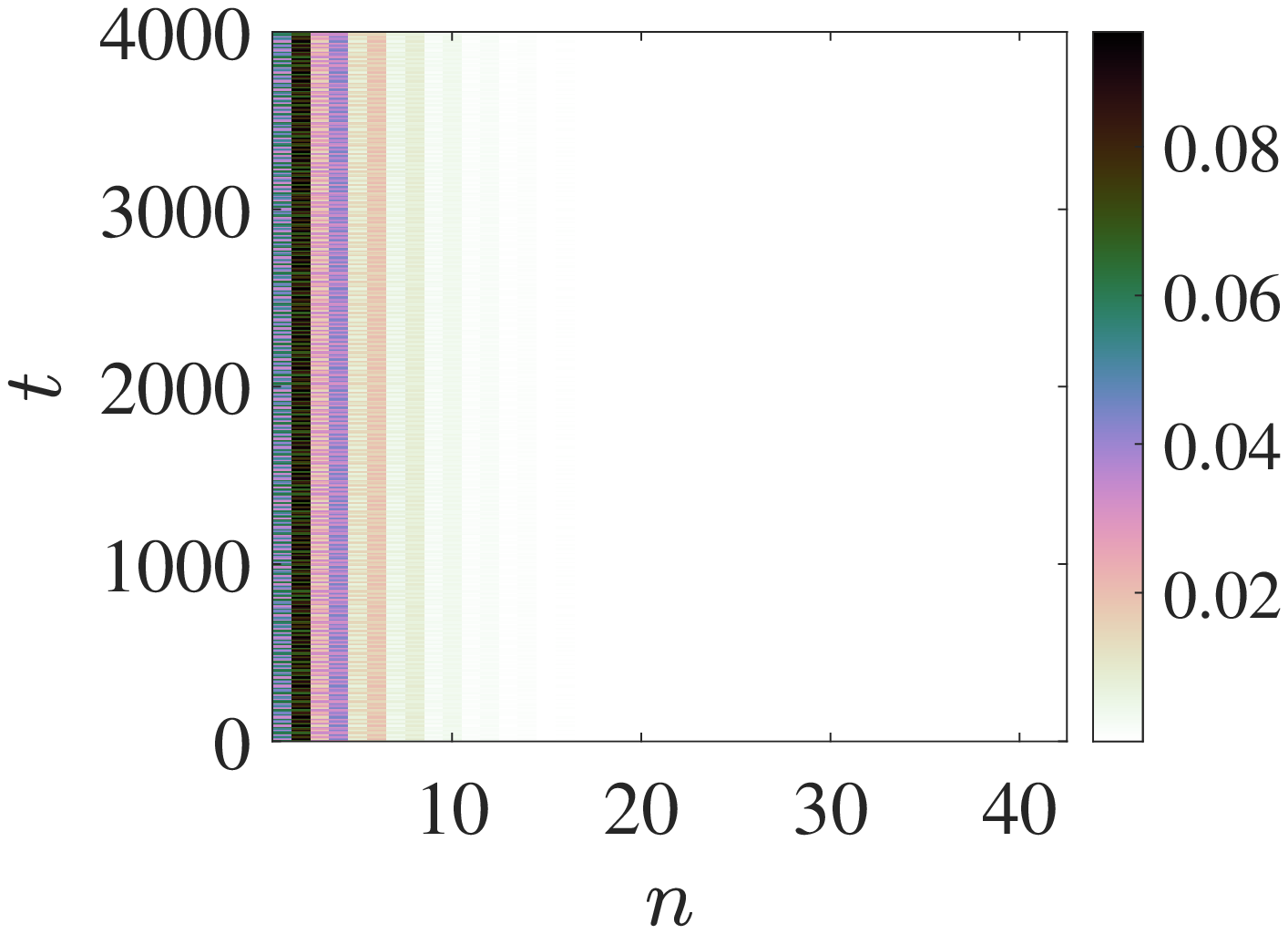}
\includegraphics[height=.17\textheight, angle =0]{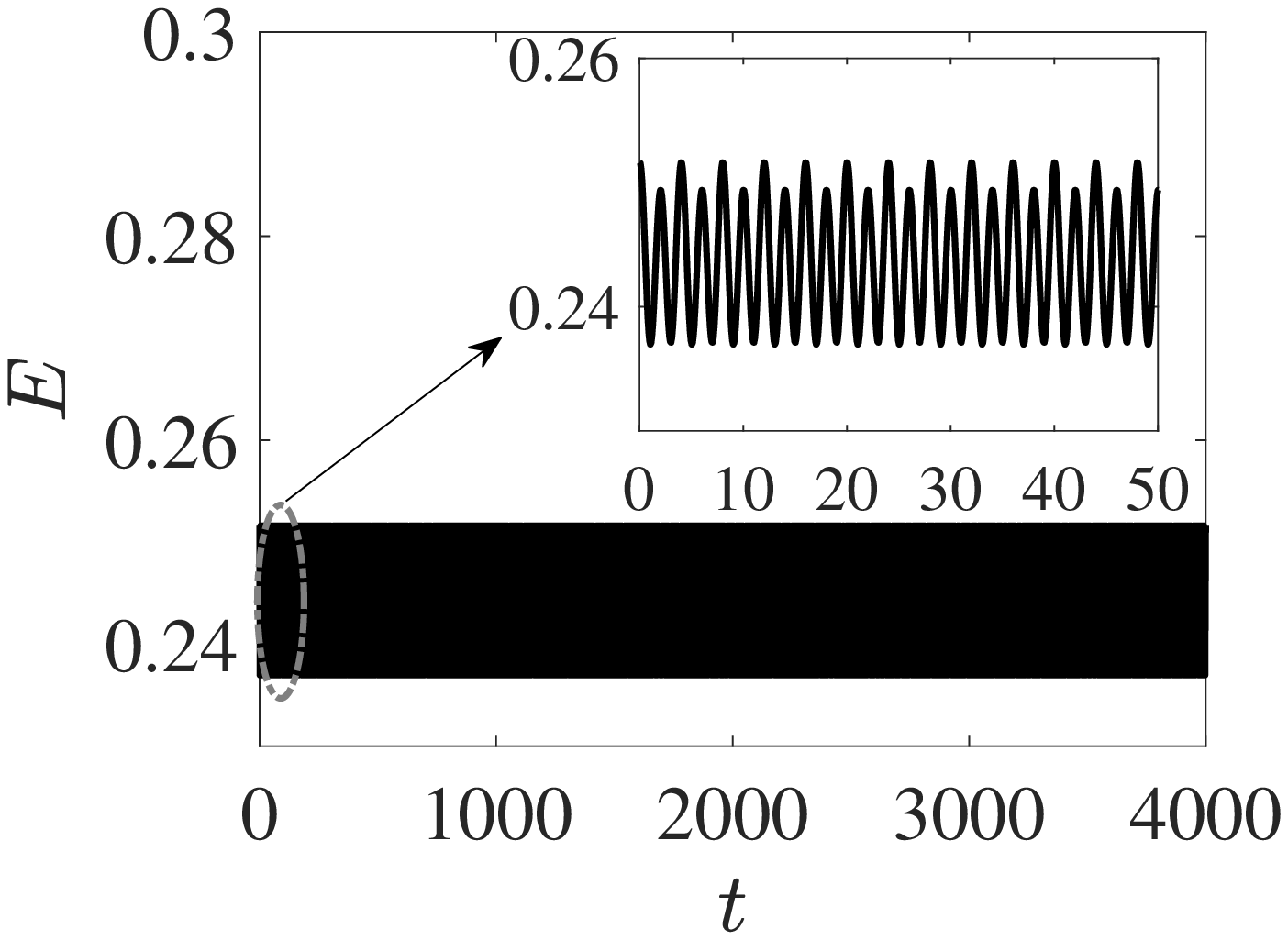}\\
\includegraphics[height=.17\textheight, angle =0]{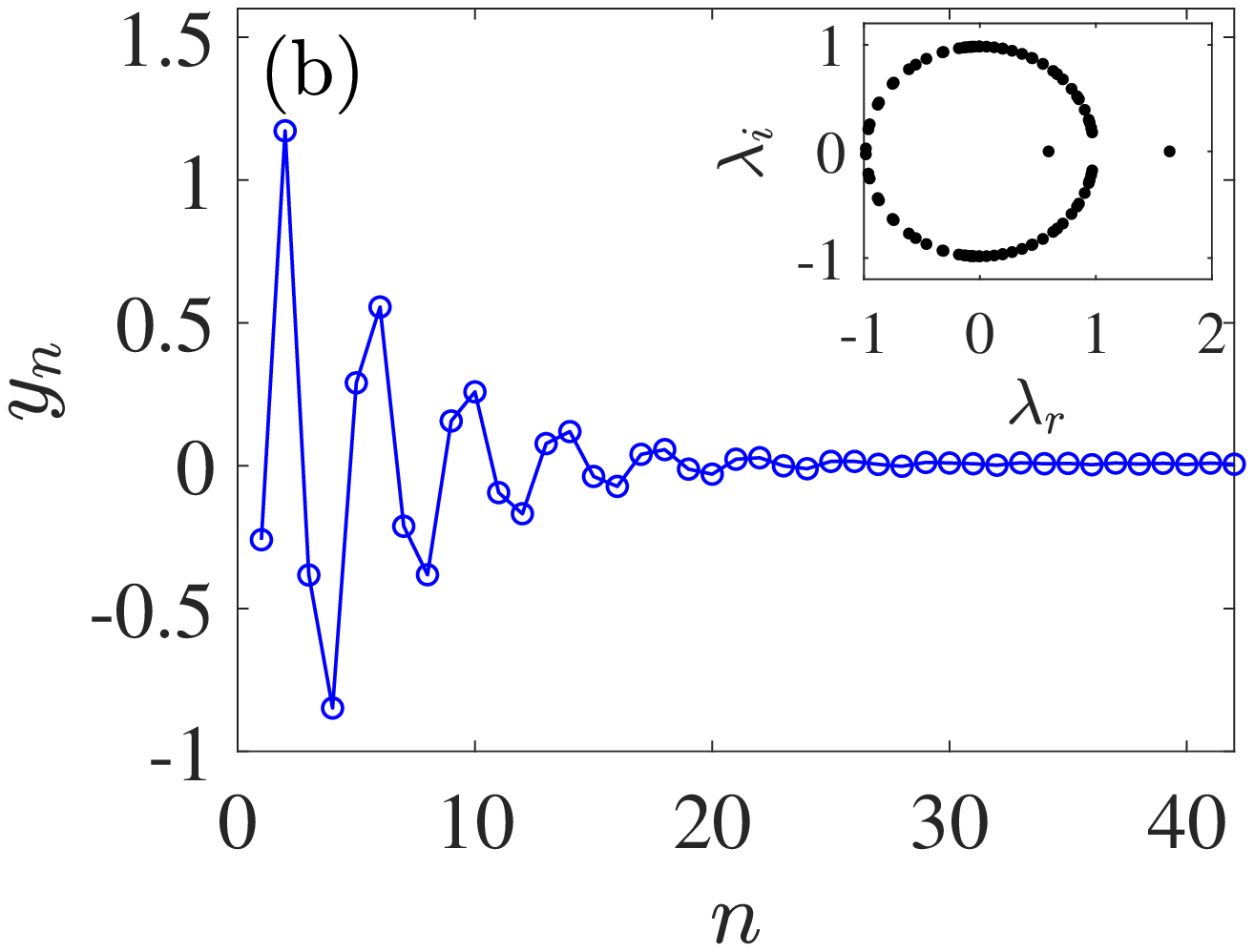}
\includegraphics[height=.17\textheight, angle =0]{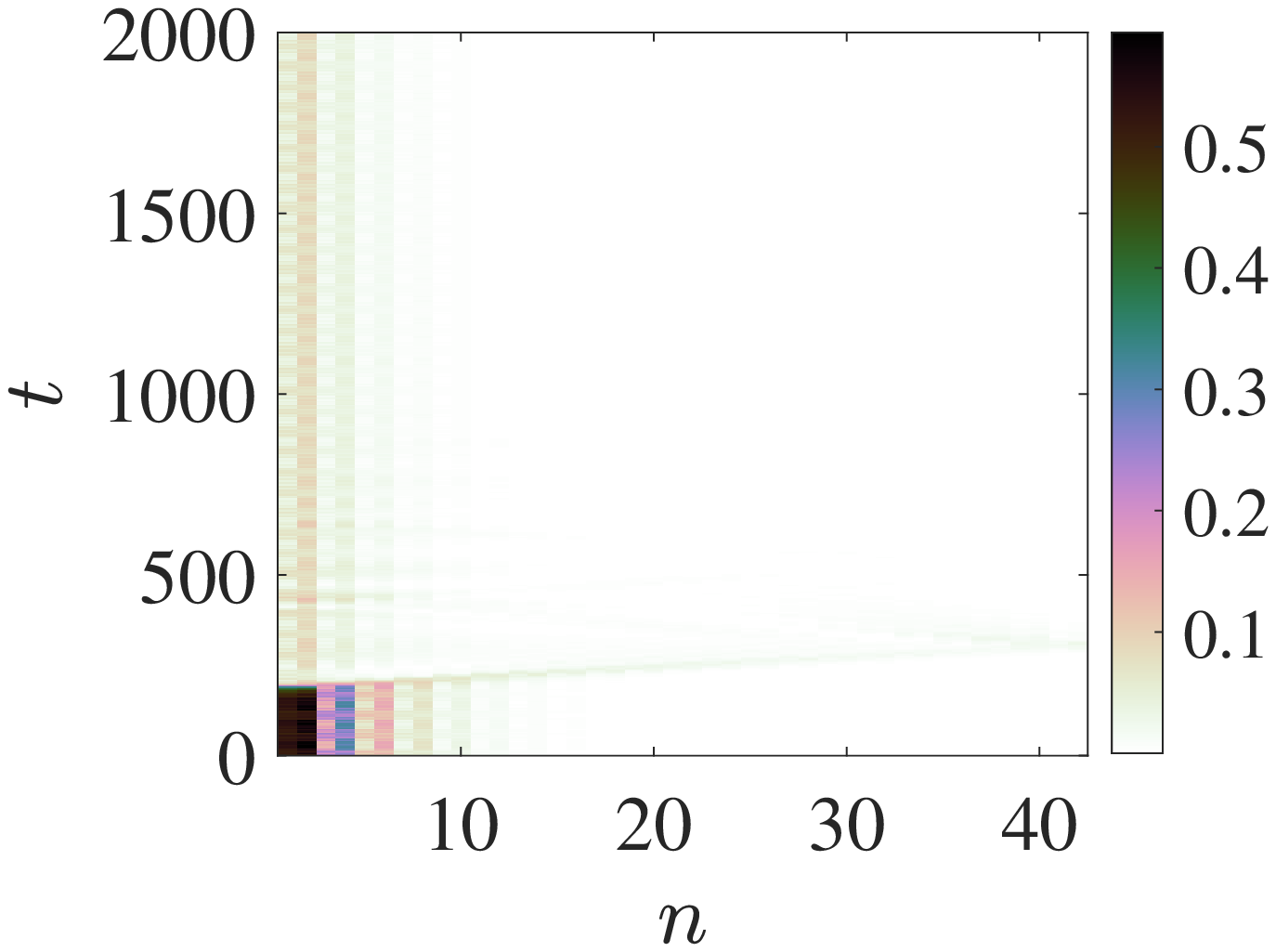}
\includegraphics[height=.17\textheight, angle =0]{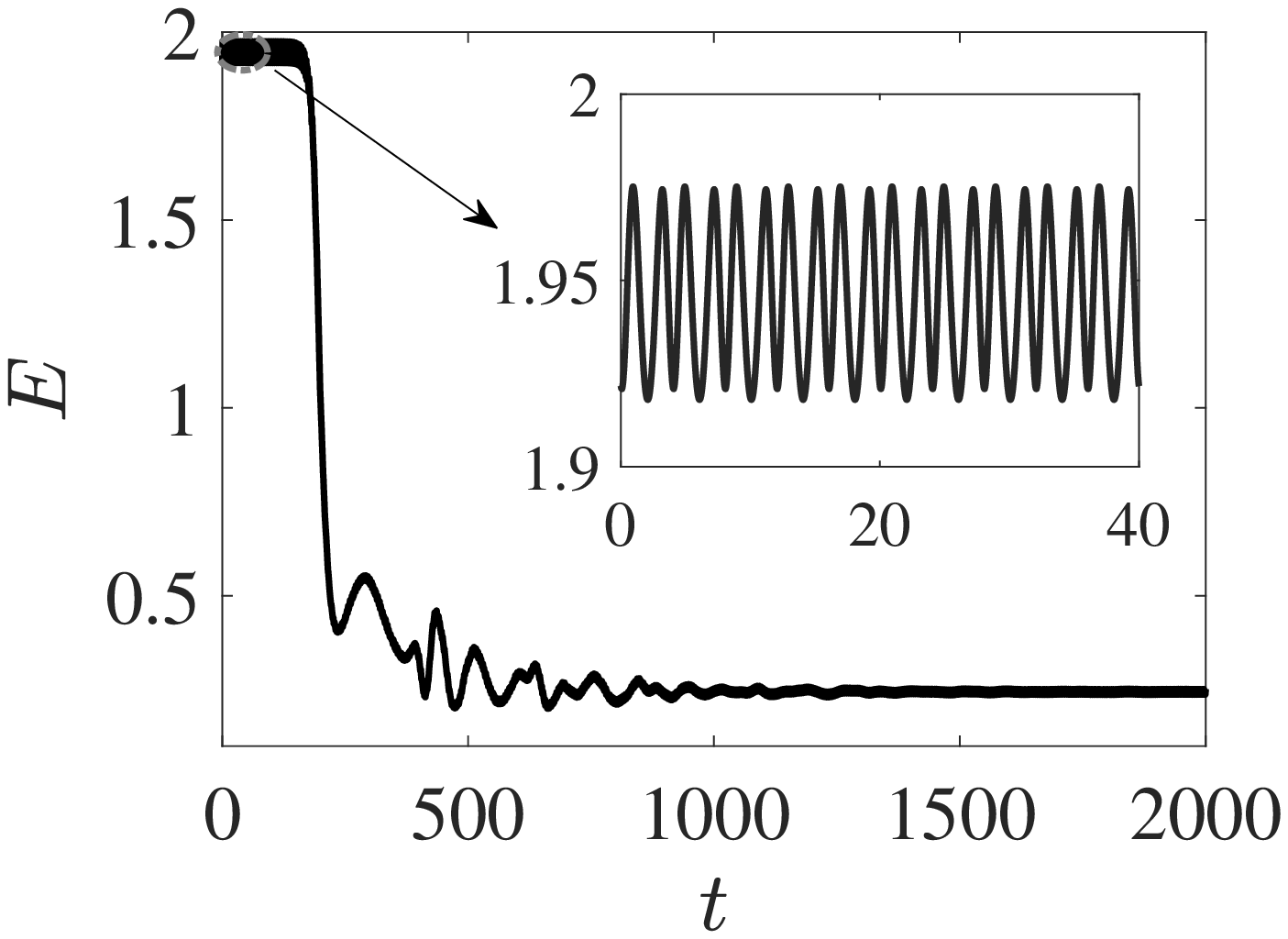}\\
\includegraphics[height=.17\textheight, angle =0]{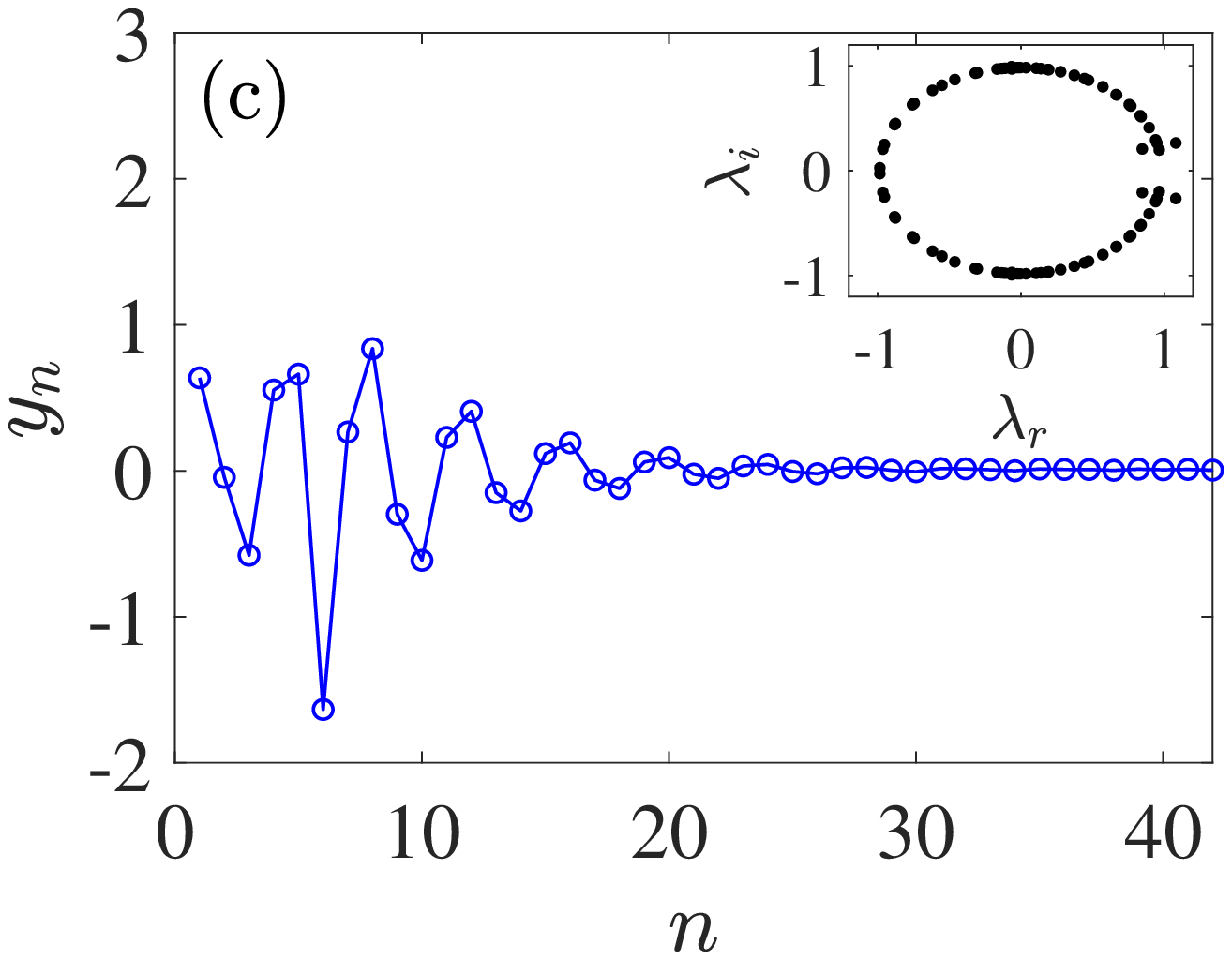}
\includegraphics[height=.17\textheight, angle =0]{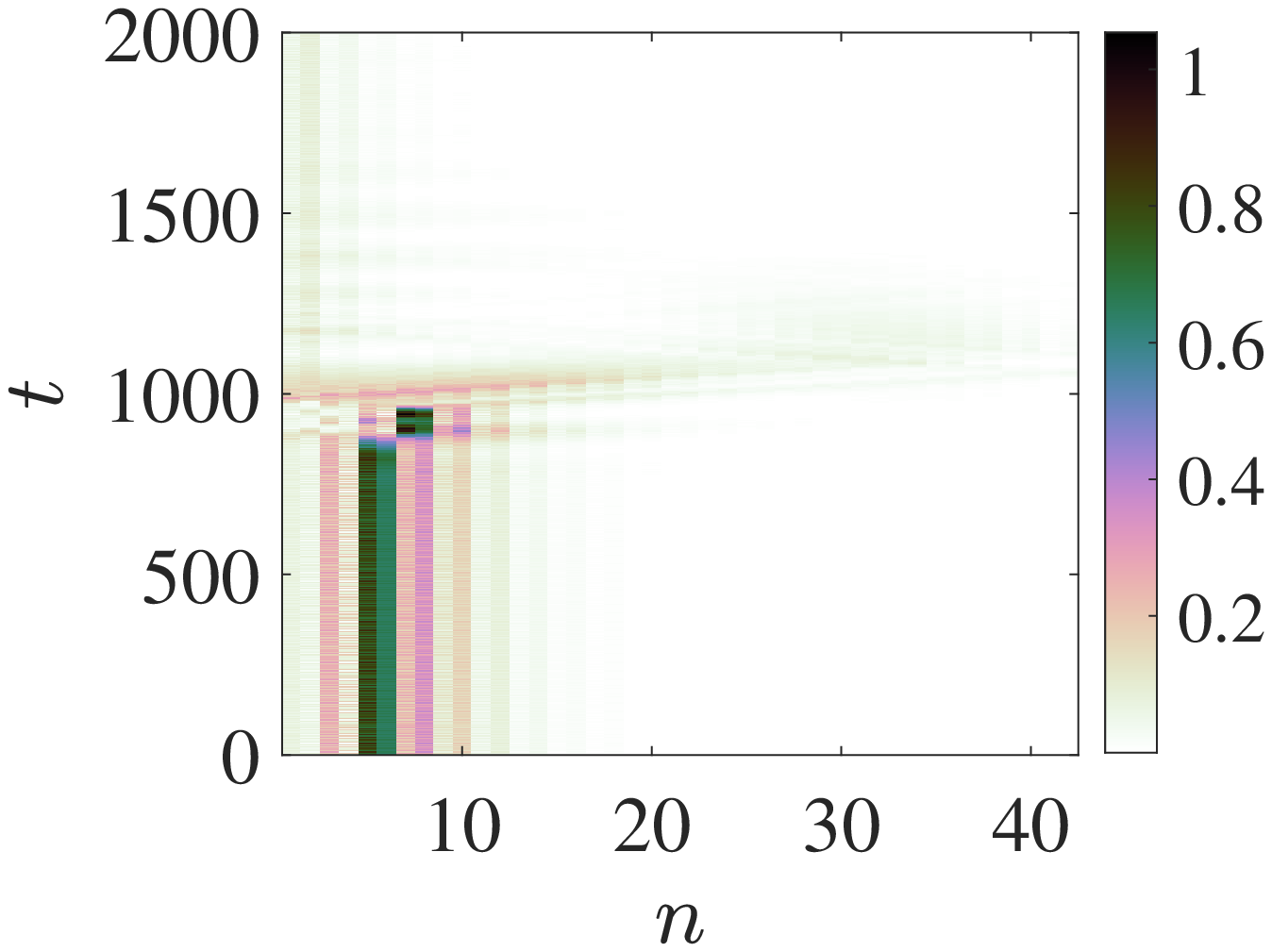}
\includegraphics[height=.17\textheight, angle =0]{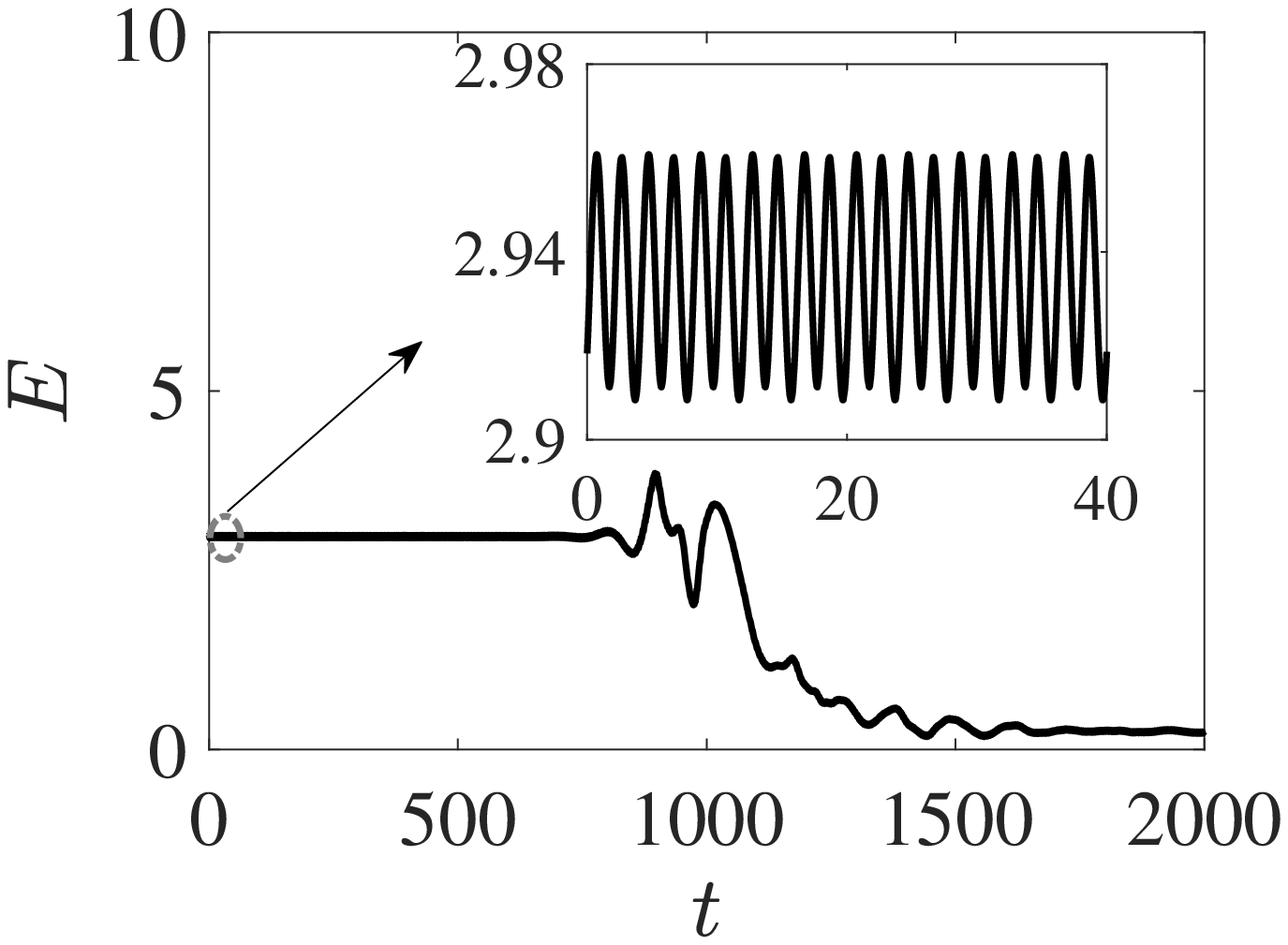}
\end{center}
\caption{(Color online)
Stable (top row) and unstable (middle and bottom rows) 
breather solutions for driving frequency $f_{d}=0.25$ and amplitude
$a_{d}=0.4$ associated with the labels of the right panel of Fig.~\ref{fig6}.
The left column shows the strain profiles as well as
Floquet spectra (see the insets therein). The middle column shows the spatio-temporal 
evolution of the energy density [cf. Eq.~\eqref{energ_dens}]. 
Note that the middle and bottom profiles are examples of an 
exponentially unstable solution (characterized by a purely real 
unstable eigenmode) and oscillatory one (described by an 
eigenvalue quartet), respectively. The right column monitors the 
total energy (see Eq.~\eqref{tot_E}) in the respective cases as 
a function of time $t$. 
\label{fig:profiles}
}
\end{figure}

We now fix the frequency (within the gap), and vary the amplitude of the drive, 
as shown in Fig.~\ref{fig6}. We show the dependence of $\overline{E}$ as a function 
of $a_{d}$ for a number of cases. The left and middle panels of Fig.~\ref{fig6} 
correspond to the hardening and softening cases, respectively, both with $f_{d}=0.23$ 
where the well-known hysteretic bifurcation diagram is obtained; see, e.g.,~\cite{Nature11,hooge12}. 
It should be noted in passing that the numerical results depicted in these panels are shown 
for amplitudes of the drive of up to $a_{d}=0.25$ and $0.5$, respectively. The stability 
of the hysteretic bifurcation diagram is in line with earlier works~\cite{Nature11,hooge12}. 
The fold bifurcations in the left panel of the figure occur at $a_{d}\approx 0.1961$ and 
$a_{d}\approx 0.1329$. In the middle panel the bifurcation values are $a_{d}\approx0.2726$ 
and $a_{d}\approx 0.1767$. Notice, however, that there are distinctive features such as 
the short interval of oscillatory instability (for $a_{d}\approx [0.1768,0.1865])$) preceded by 
a very narrow segment (for $a_{d}\approx[0.1767,0.1768]$) of stable breathers in the middle panel 
of the softening case very near the bifurcation point. The right panel of the figure is a representative 
example for the alternating stiffness case (i.e., with nonlinearity exponents given by Eq.~\eqref{nonlin_pow}) 
with $f_{d}=0.25$ lying in the frequency gap. The bifurcation curve (that is shown for drive amplitudes 
of up to $a_{d}=0.7$) becomes far more elaborate. It features a cascade of turning points, the first of 
which (i.e., at $a_{d}\approx 0.6935$) represents a fold bifurcation. Note that the labels (a)-(c) in 
the figure are connected with the numerical results of Fig.~\ref{fig:profiles}. Importantly, many of the 
numerous intermediate branches are found to be unstable through real instabilities, although the branch 
eventually features an oscillatory instability for sufficiently larger amplitude (as before). This additional 
complexity of the bifurcation diagram is introduced by means of the interplay between strain-softening and 
strain-hardening dynamics.

In Fig.~\ref{fig:profiles} we show select examples of the dynamics of breather solutions (with no perturbation
added on top of the respective initial conditions) for $f_{d}=0.25$, which is in the spectral gap. Those solutions 
correspond to values of $a_{d}=0.4$ and are associated with the labels of the right panel of Fig.~\ref{fig6}. 
The left column of Fig.~\ref{fig:profiles} shows the strain profiles, i.e., $y_{n}=u_{n-1}-u_{n}$ where the 
insets depict the Floquet spectra. The solution in the top row is spectrally stable (and remains accordingly 
robust, although the energy quantity we measure oscillates due to the presence of damping and driving) whereas 
the ones in the middle and bottom are unstable. The one in the middle row is an exponentially unstable waveform 
since the instability is characterized by a real Floquet multiplier pair (with a dominant unstable mode of 
$\lambda_{r}\approx 1.63576$), and the bottom one is oscillatory unstable corresponding to a complex multiplier 
quartet (with a dominant instability of $\lambda\approx 1.07981\pm 0.265122$). The stability results are corroborated 
in the respective panels of the middle column of the figure which present the spatio-temporal evolution of the energy 
density [cf. Eq.~\eqref{energ_dens}]. It can be seen in the middle panels of the second and third rows of the figure 
that after an unstable breather gets destroyed, a wavepacket is created which propagates with decreasing amplitude 
toward the bulk of the lattice.  An additional part of the relevant energy is localized at the left edge of the 
computational domain and appears to localize forming a smaller amplitude breathing structure therein. At longer times, 
these structures gradually asymptote to the stable solution of the top left panel of the figure. This finding can be 
further explored by looking at the right column of the figure. In particular, we present the total energy $E$ of 
Eq.~\eqref{tot_E} as a function of time in the respective cases. In the top right panel of the figure, the total energy 
of the stable solution oscillates, as discussed above, but maintains a nearly constant average of about 0.245 for 
$1000$ periods.  On the other hand, in the middle and bottom panels, the energy experiences large deviations, highlighting 
the onset of the instability. Both unstable solutions eventually approach the stable one shown in the top row. This can 
be seen in the energy plots, where the average energy approaches 0.245 at $t\gtrapprox1600$ for the unstable breather 
with label (b) and at $t\gtrapprox 1800$ for the unstable breather with label (c). 

\begin{figure}[pt!]
\begin{center}
\includegraphics[height=.17\textheight, angle =0]{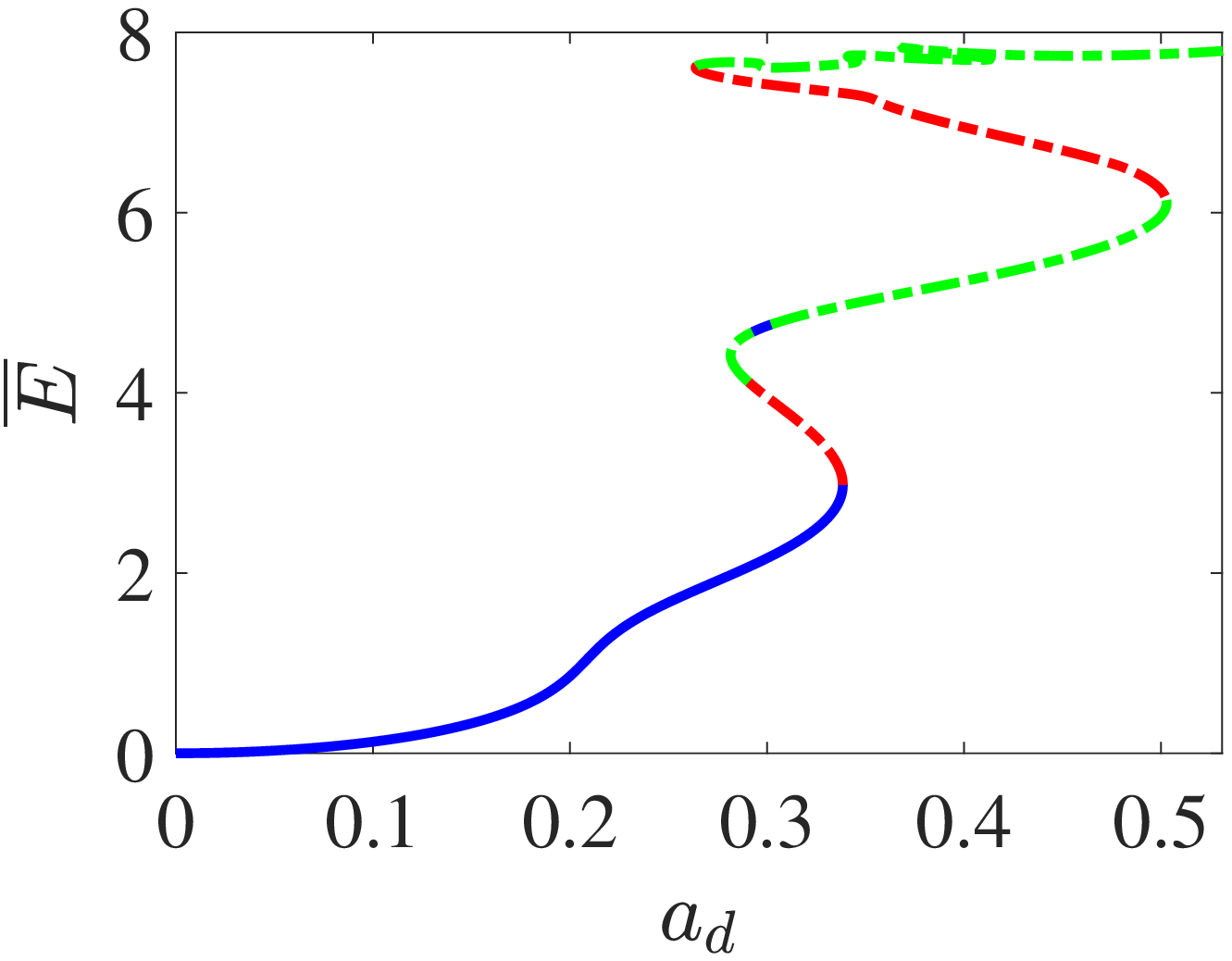}
\includegraphics[height=.17\textheight, angle =0]{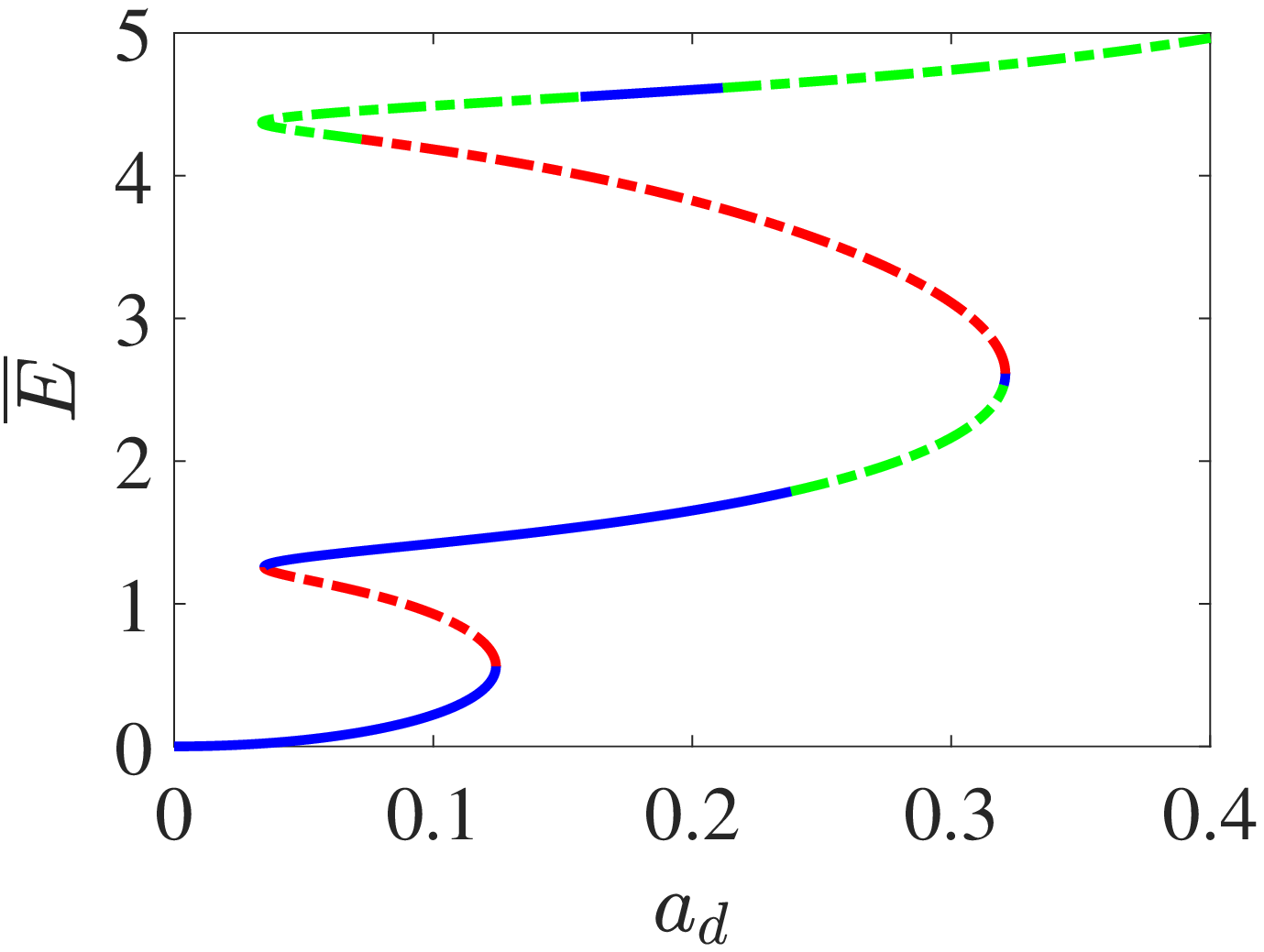}
\includegraphics[height=.17\textheight, angle =0]{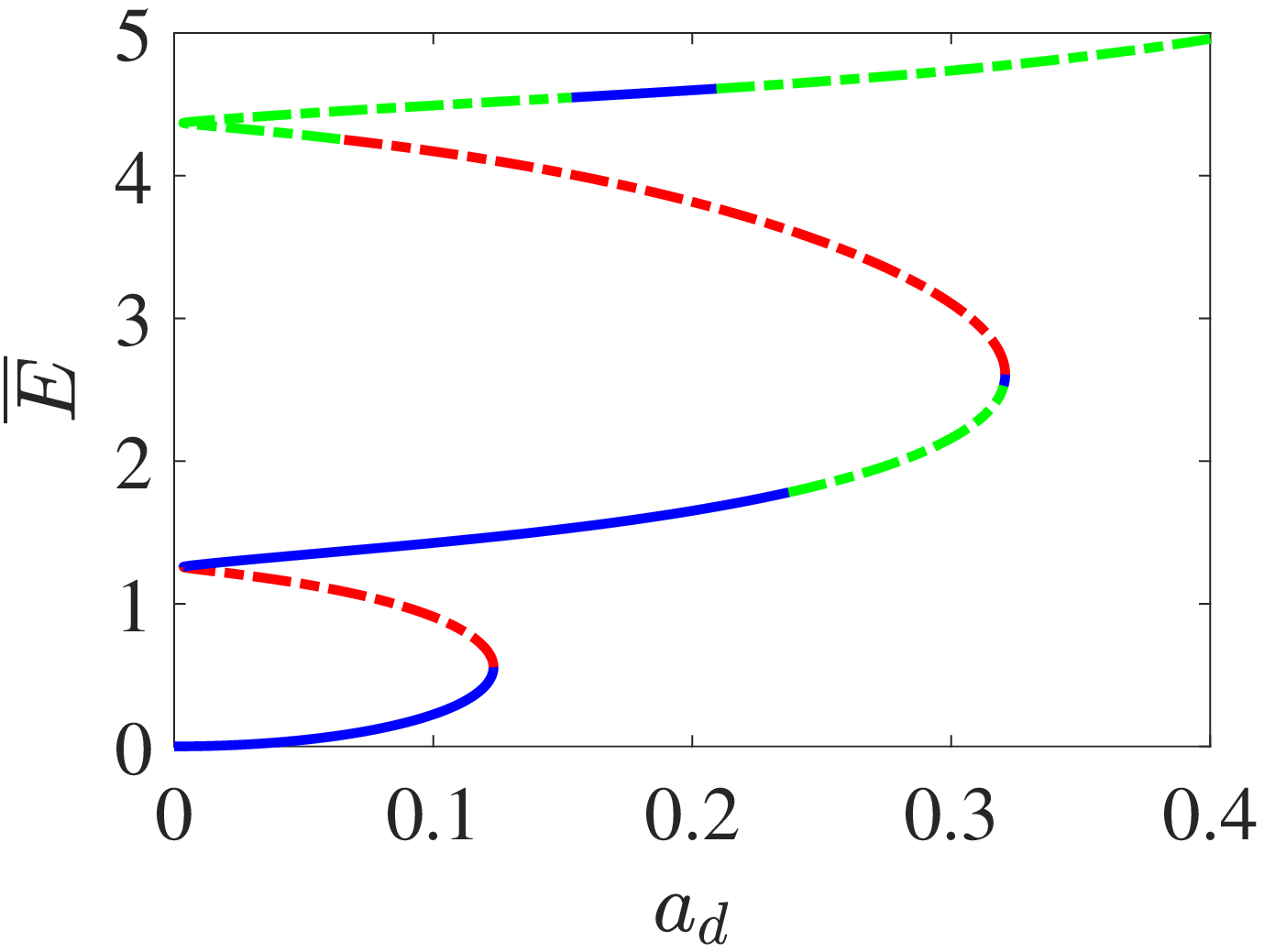}
\end{center}
\caption{(Color online)
Same as Fig.~\ref{fig6} but corresponding to the case with alternating 
nonlinearity and value of the driving frequency of $f_{d}\approx 0.2566$
(i.e., at the bottom edge of the optical band). Left, middle and right panels 
correspond to values of damping of $8\times 10^{-3}$, $8\times 10^{-4}$, and 
$\gamma=8\times 10^{-5}$, respectively. 
\label{fig:change_damping}
}
\end{figure}

In all the numerical results that we have presented so far, $\gamma=8\times 10^{-3}$ was considered 
in~Eq.~\eqref{gc_dimer_dim}. Next, we will study the effect of varying the dissipation parameter $\gamma$, 
where we gradually decrease it in order to approach the Hamiltonian limit. Moreover, this study will 
be beneficial while connecting the respective results presented herein with the ones of Sec.~\ref{sec:NLS} 
for $\gamma=0$. Let us start our discussion by considering Fig.~\ref{fig:change_damping} which corresponds 
to the value of driving frequency of $f_{d}\approx 0.2566$  
being itself the cut-off frequency of the bottom optical edge of the optical band (see the right panel of Fig.~\ref{fig1}).
From the left to the right panels of the figure, the dependence of $\overline{E}$ is shown as a function 
of $a_{d}$ for $\gamma=8\times 10^{-3}$, $8\times 10^{-4}$, and $\gamma=8\times 10^{-5}$, respectively.
We mention in passing the existence of a cascade of turning points in
the panels, with the first four happening
(as we trace the bifurcation curve from $a_{d}=0$ with increasing arclength) at 
$a_{d}\approx(0.33847, 0.28153, 0.50280, 0.26376)$ (left), 
$a_{d}\approx(0.12421, 0.03476, 0.32090, 0.03386)$ (middle), 
and $a_{d}\approx(0.12325, 0.00351, 0.32084, 0.00339)$ (right). 
A movie illustrating how the solution profile changes as one moves along
the bifurcation curve of the right panel is included in the supplement. 

As the value of the dissipation parameter decreases, when $a_d \rightarrow 0$, Eq.~\eqref{gc_dimer_dim} 
approaches its Hamiltonian sibling, i.e., when $(\gamma,a_d)\rightarrow (0,0)$. Indeed, the solutions at 
the turning points in the right panel with $a_{d}\ll 1$ (and $\gamma\ll 1$), aside from the lowest one 
introduced by the drive and vanishing in this limit, are expected to be proximal to Hamiltonian breathers, 
since they are large amplitude (energy) states that exist in the lattice with nearly zero drive and zero 
damping. Recall that in the above diagrams we have fixed the frequency at the lower optical band edge. 
Bifurcation diagrams for frequencies deeper in the gap generally follow the same trend, namely that the 
turning points move toward smaller drive amplitude as the damping is decreased. However, the overall bifurcation 
structure is more intricate, as we explore next. Moreover, we would like to report an intriguing feature 
that happens at the turning points of the above mentioned bifurcation diagrams. As we approach a turning point, 
two eigenvalues are moving (one from the first and the other one from the fourth quadrant inside the unit circle) 
toward the real axis, until they collide exactly at the turning point, and past that, a pair of real multipliers 
emerges. We note that for the Hamiltonian problem at hand, such a collision happens at $1+0\ii$ but for the
present damped-driven problem, the collisions happen slightly inside the unit circle. More concretely, in the 
middle panel of Fig.~\ref{fig:change_damping}, the eigenvalues at the first turning point (i.e., $a_{d}\approx0.1242$) 
are $\lambda_{1}\approx 0.9992$ and $\lambda_{2}\approx 0.9976$. Interestingly and in line with similar
earlier observations in~\cite{jesus}, they satisfy the relation $\lambda_{1}\lambda_{2}=e^{-\gamma T_{d}}$
which for the Hamiltonian case, i.e., $\gamma=0$, it gives $\lambda_{1}\lambda_{2}=1$. Indeed, we have 
$\lambda_{1}\lambda_{2}\approx 0.9968$ and $e^{-\gamma T_{d}}\approx 0.9968$. We further checked the 
relation $\lambda_{1}\lambda_{2}=e^{-\gamma T_{d}}$ at the other turning points and cases in $\gamma$ 
(right panel of the figure), and it holds indeed. It should also be noted that this implies that the stability 
change does not arise {\it at} the turning point but only nearby, given that the collision happens inside
of the unit circle, and hence further parametric tuning is needed for one of the (now real) multipliers to 
cross the $1+0\ii$ point.

\begin{figure}[pt!]
\begin{center}
\includegraphics[height=.17\textheight, angle =0]{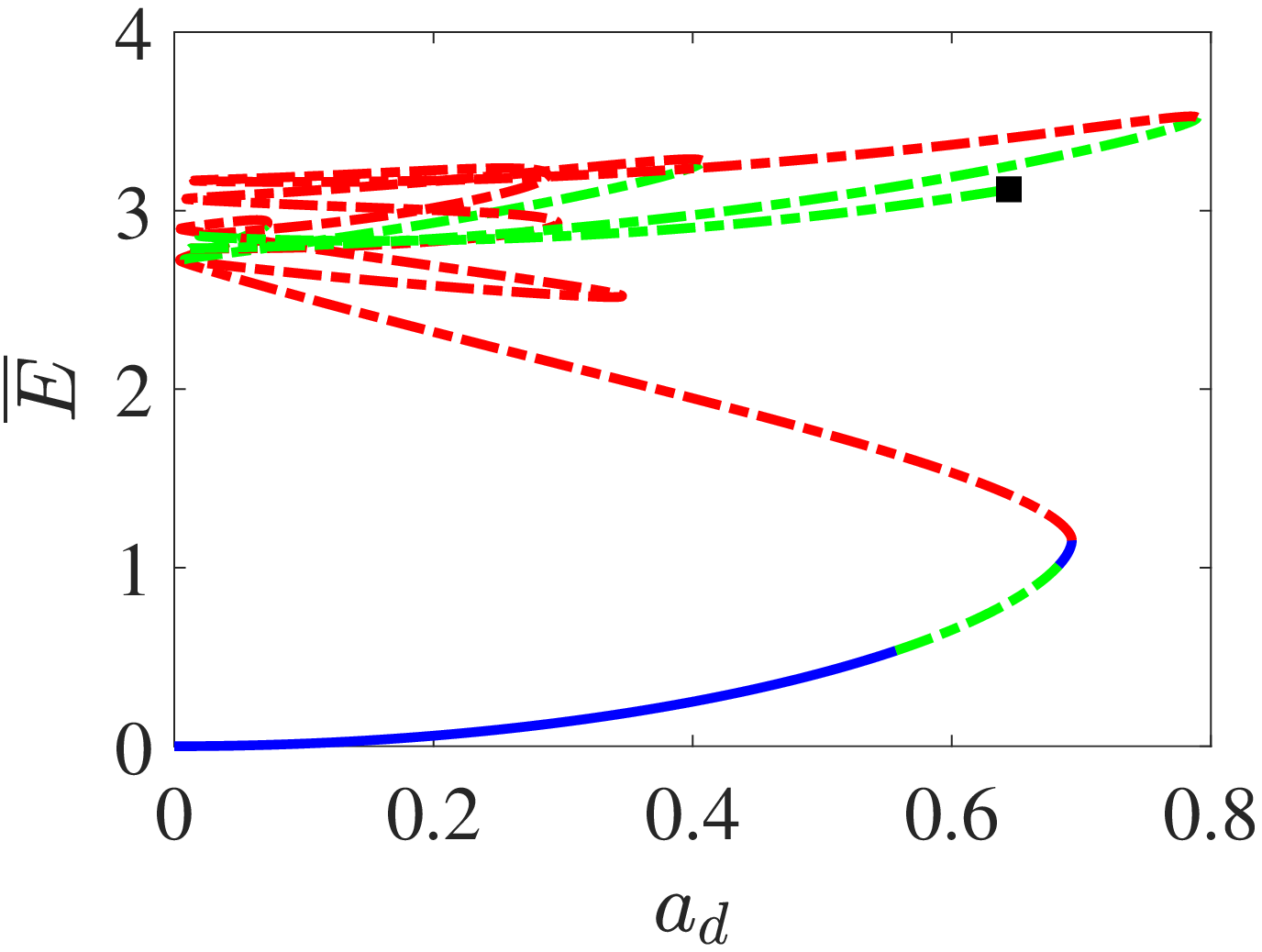}
\includegraphics[height=.17\textheight, angle =0]{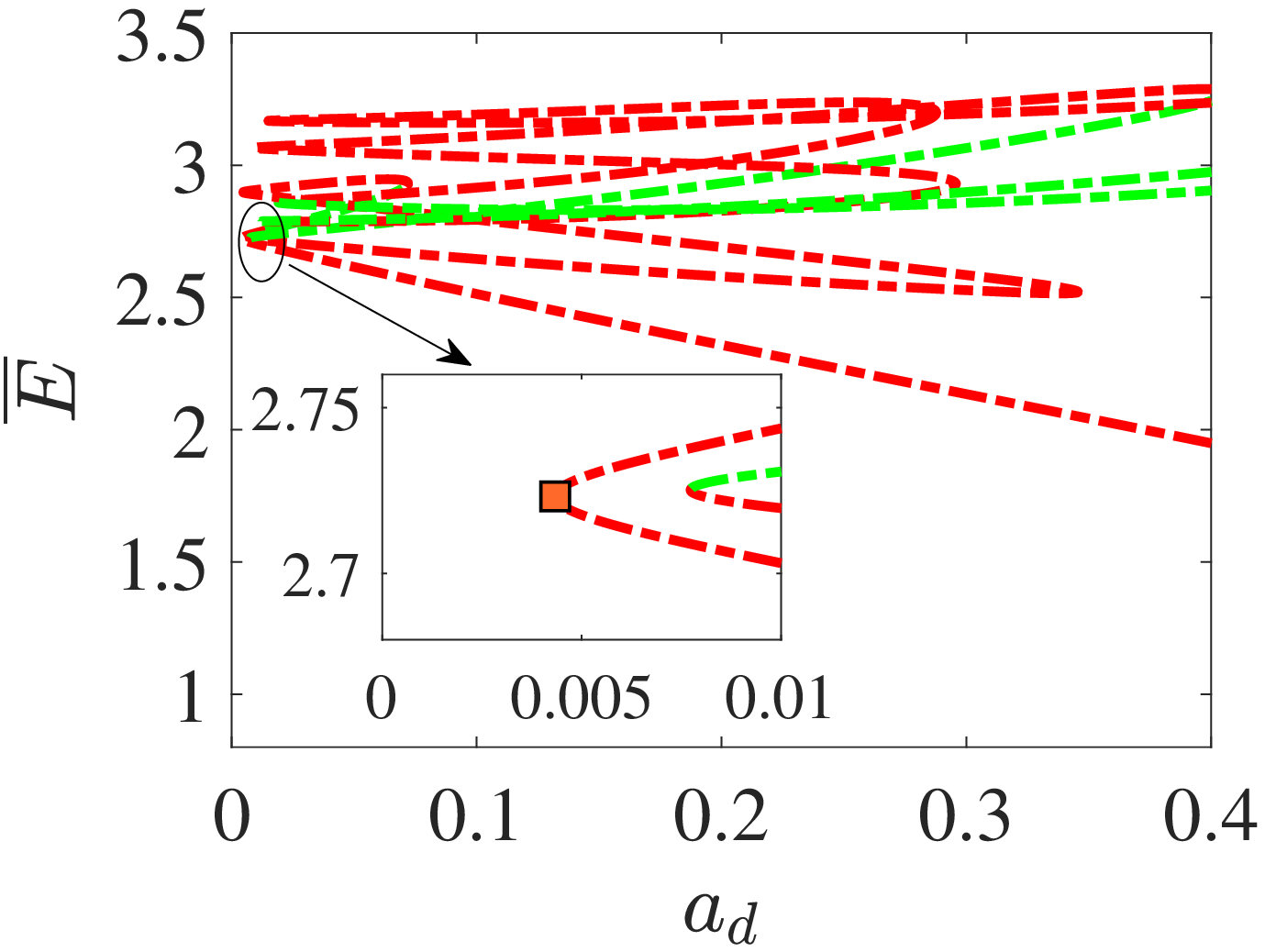}
\includegraphics[height=.17\textheight, angle =0]{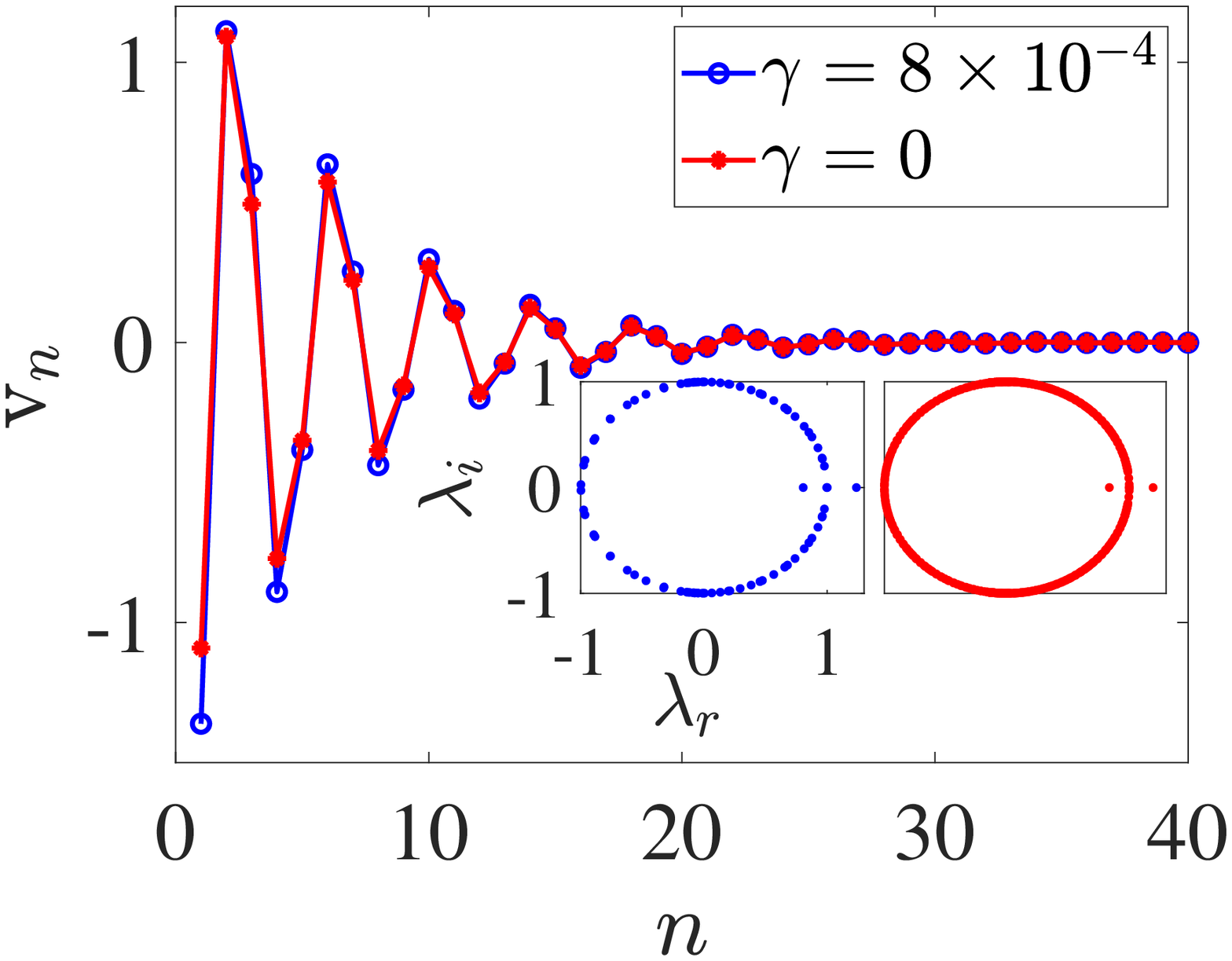}\\
\includegraphics[height=.17\textheight, angle =0]{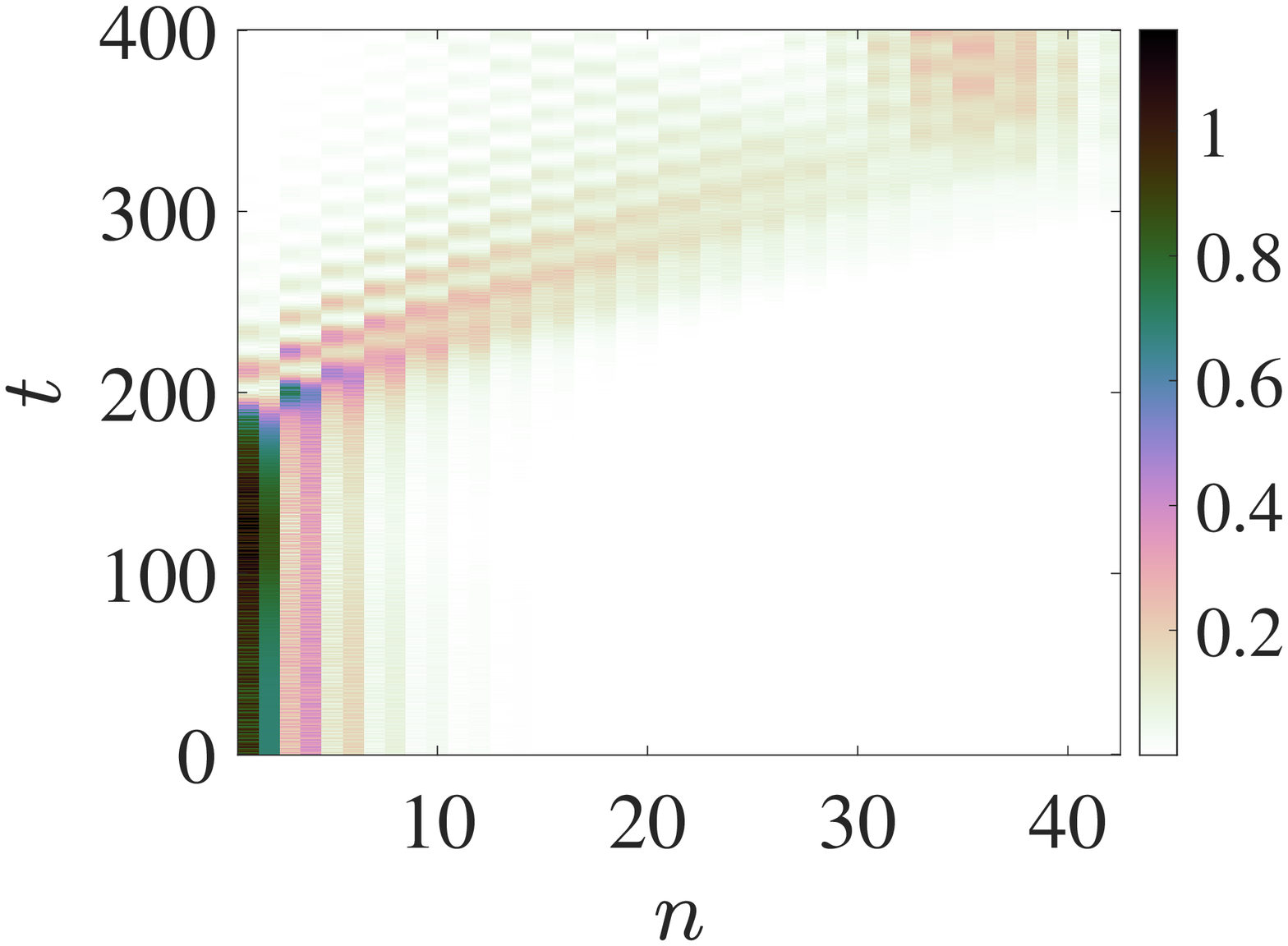}
\includegraphics[height=.17\textheight, angle =0]{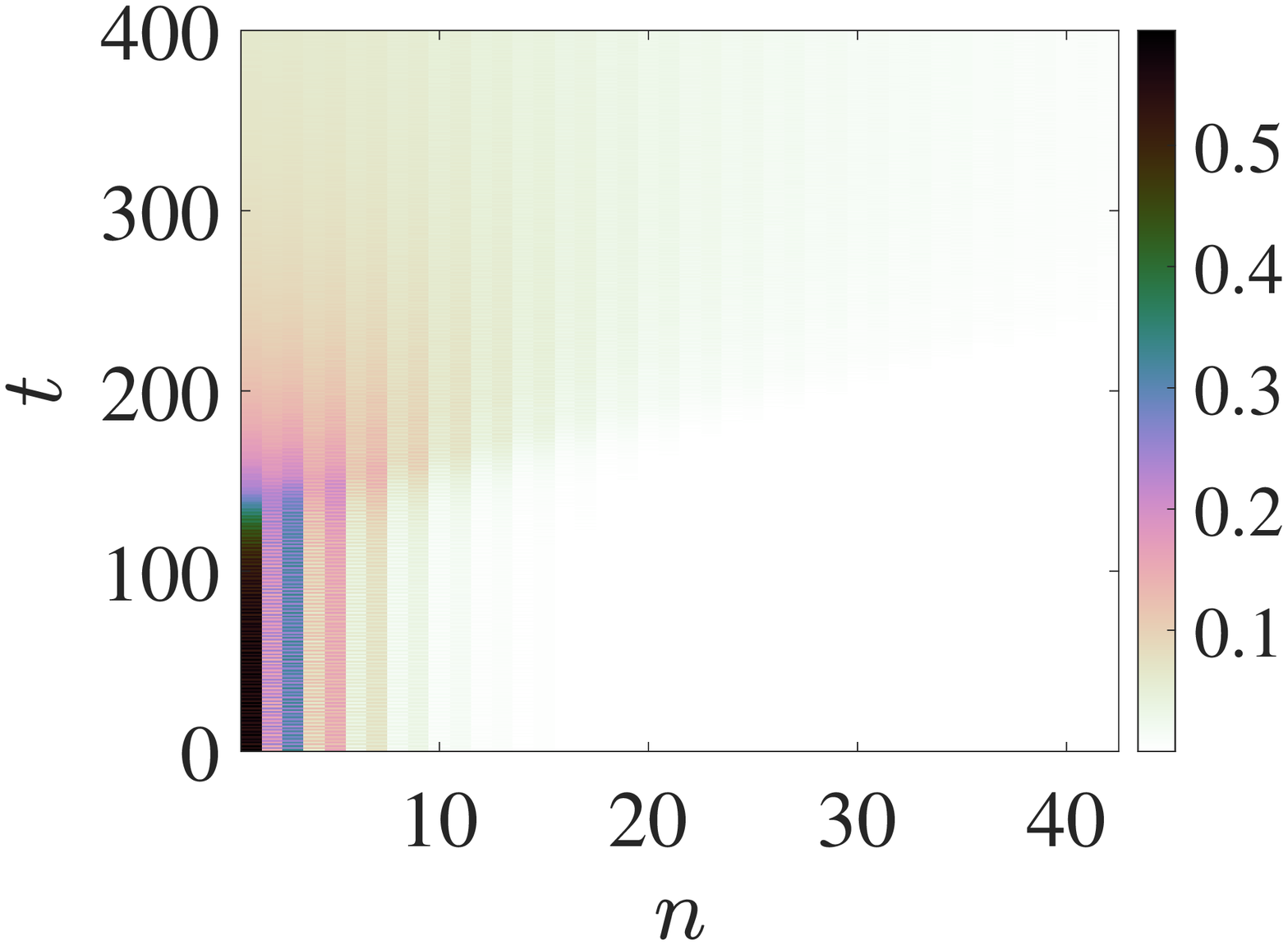}
\includegraphics[height=.17\textheight, angle =0]{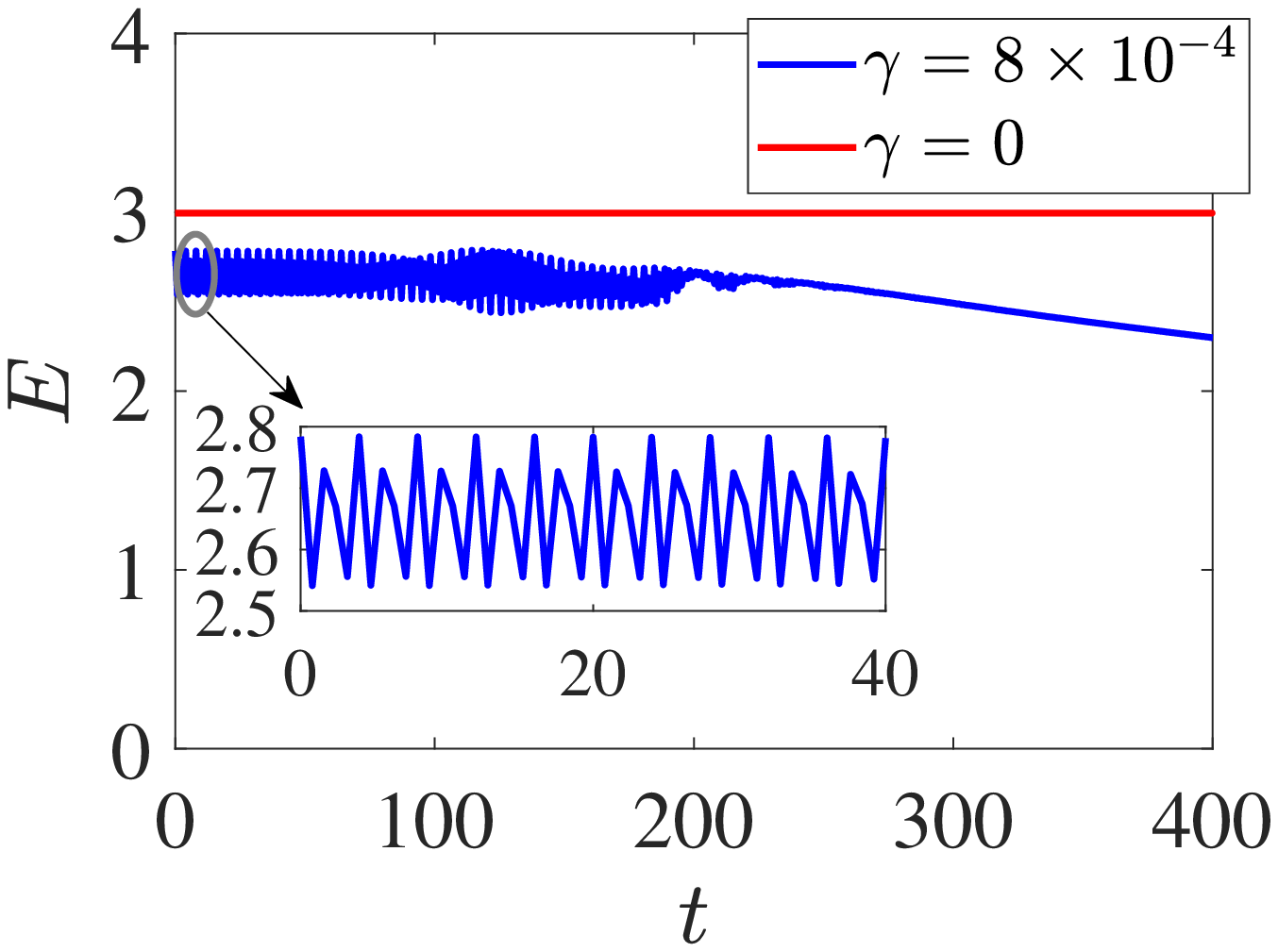}
\end{center}
\caption{
\label{fig9}
(Color online)
Summary of numerical results on breather solutions for $f_{d}=0.25$
and $\gamma=8\times 10^{-4}$. In particular, the top left panel highlights 
the dependence of the average energy density $\overline{E}$ as a function
of the driving amplitude whereas the top middle panel is a zoom in of the
(top) left one. The black square in the top left panel depicts the point 
in the parameter space where we stopped the continuation. The filled orange 
square marker in the inset of the top middle panel connects with the solution
shown in the top right panel whose velocity profile is shown in blue.
Therein, the driving amplitude of that solution is $a_{d}\approx 0.0043$. In the 
same panel, the associated Hamiltonian breather (i.e., $\gamma=0$) is shown in red. 
The bottom left and middle panels correspond to the spatio-temporal evolution 
of the energy density for breathers with $\gamma=8\times 10^{-4}$ and $\gamma=0$,
respectively, with the energy appearing to disperse upon the instability manifestation. 
Finally, the bottom right, showcases the dependence of the total energy $E$ on time 
$t$ (same coloring is used as in the top right panel). Note that in the Hamiltonian case, 
the total energy is conserved, as expected.
}
\end{figure}

The connection between the damped-driven and Hamiltonian breathers is explored 
in Fig.~\ref{fig9}.  We chose a frequency that is in the gap, namely $f_{d}=0.25$, 
and a small damping parameter $\gamma=8\times 10^{-4}$. At this frequency, the 
amplitude continuation (analogous to the one shown in Fig.~\ref{fig:change_damping})
yields a more intricate bifurcation diagram (see  the top left and middle panels in 
the figure). It is expected that a solution lying at a turning point with small driving 
amplitude (see the orange square marker in the inset of the top middle panel) will 
be proximal to an associated Hamiltonian breather, i.e., an intrinsic localized mode 
of Eq.~\eqref{gc_dimer_dim} with $\gamma=0$. The velocity profile of the damped-driven 
breather (corresponding to the orange square maker of the top middle panel) is shown 
in blue in the top right panel of the figure. The solution itself compares fairly well 
with the Hamiltonian breather, which is shown in red.  

In order to construct Hamiltonian breathers of Eq.~\eqref{gc_dimer_dim} numerically, we 
consider a lattice with $N=201$ beads and employ zero Dirichlet and free boundary conditions 
at the endpoints of the chain, respectively, i.e., $u_{0}=0$ and $u_{N+1}=u_{N}$. The larger 
lattice size was needed in order to minimize the effect of the boundary conditions (since the 
strain of the breather solution is nearly zero at the boundaries for large lattices). The initial 
guess to our solvers was provided by the associated linear mode at the bottom edge of the optical 
band of the Hamiltonian problem (i.e., when $\gamma=0$). The Hamiltonian breathers found in this 
way are centered at the middle of the lattice. Thus, in order to compare the obtained breather 
to the damped-driven one (which is localized at the source of the drive, namely the left 
boundary) we translate the Hamiltonian breather so that the largest amplitude is on the 
boundary. This is how the comparison shown the the top right panel of Fig.~\ref{fig9} was 
made, showcasing the proximity between the two.

\begin{figure}[pt!]
\begin{center}
\includegraphics[height=.17\textheight, angle =0]{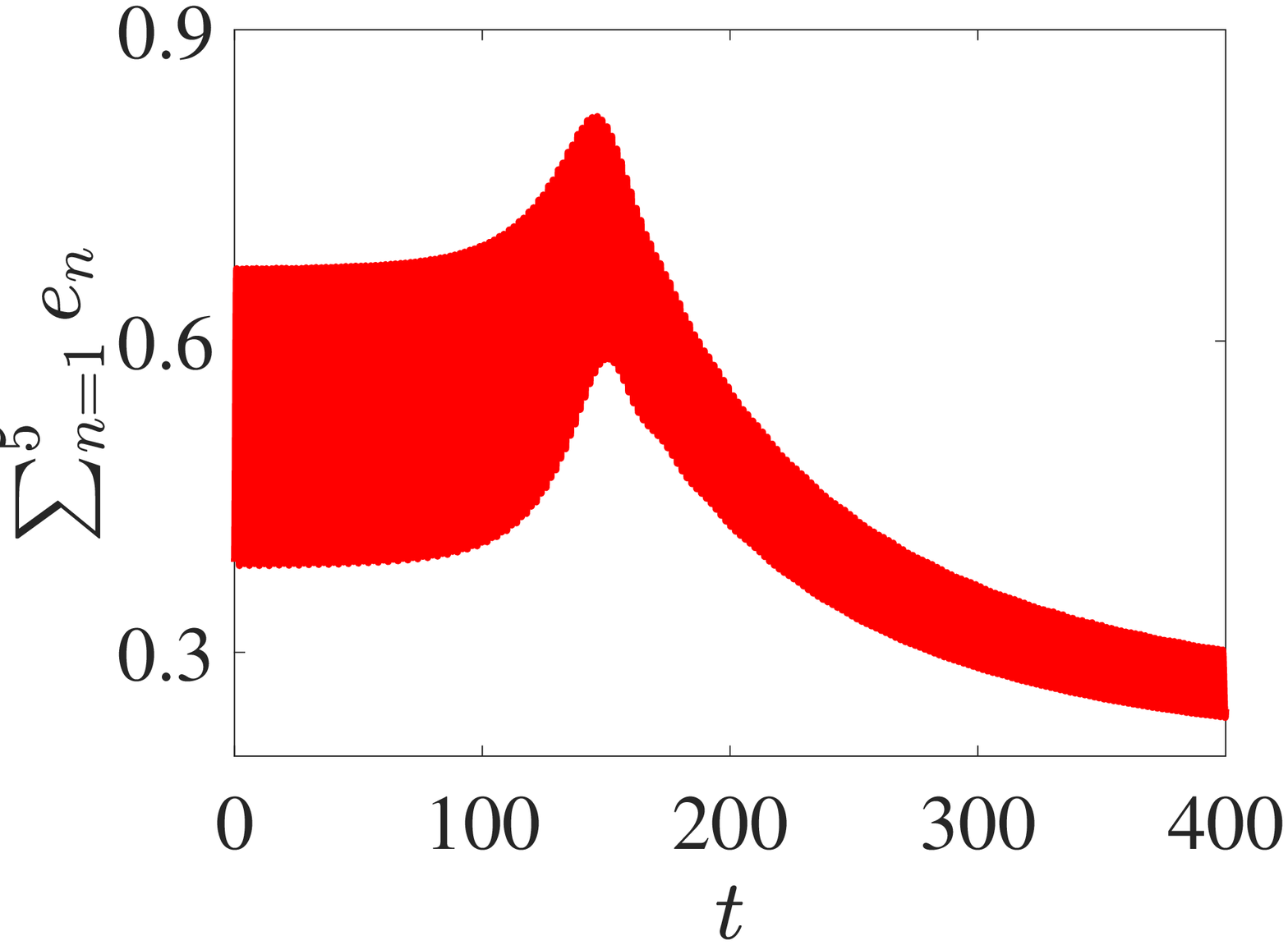}
\includegraphics[height=.17\textheight, angle =0]{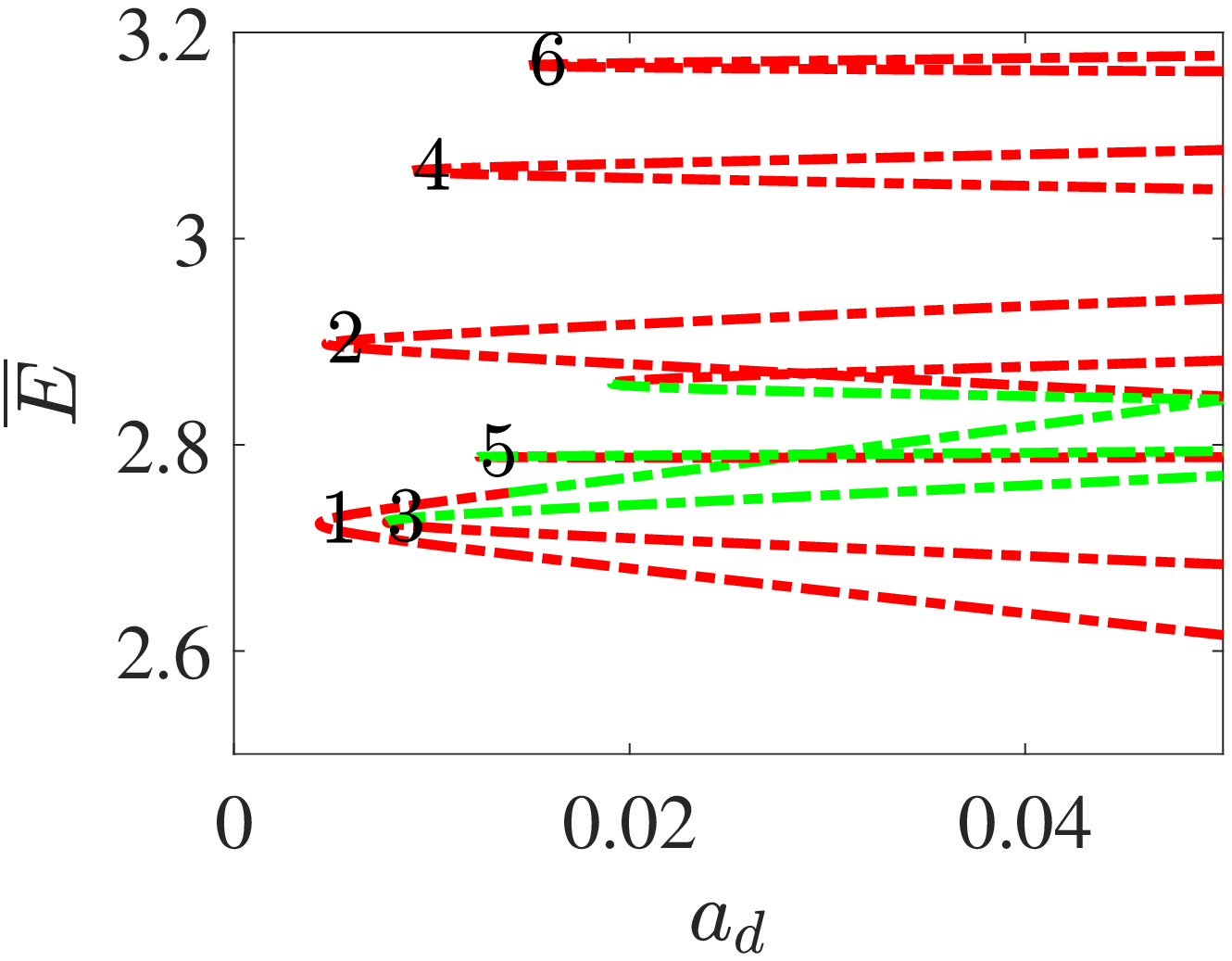}
\includegraphics[height=.17\textheight, angle =0]{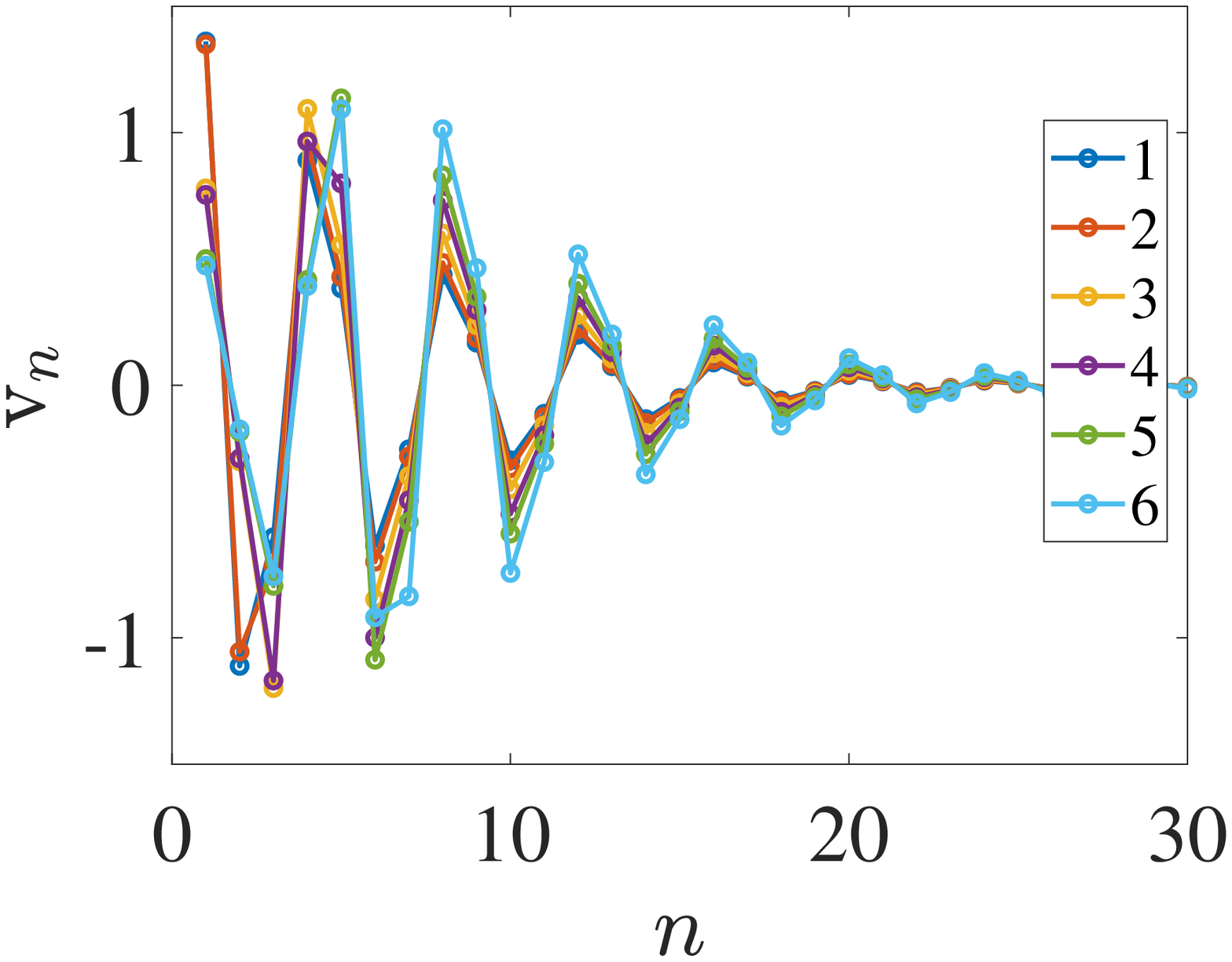}\\
\end{center}
\caption{
\label{zoo}
(Color online). {Total energy of the first 5 sites of the unstable Hamiltonian breather shown in 
the  bottom middle panel of Fig.~\ref{fig9}. The middle panel is a zoom of the turning points shown 
in the top left panel of Fig.~\ref{fig9}. Solution profiles for each of the numeric labels are 
shown in the right panel. Each solution is localized.}
}
\end{figure}

The insets of the top right panel of Fig.~\ref{fig9} indicate that the breathers of the 
damped-driven model and its Hamiltonian version are both exponentially unstable (with
dominant unstable modes of $\lambda_{r}\approx 1.23721$ and $\lambda_{r}\approx 1.19449$,
respectively), in addition to being extremely proximal to each other profile-wise. It 
should be reminded here that the difference in the density of Floquet multipliers populating 
the unit circle between the damped-driven (left) and Hamiltonian (right) insets stems from the
different corresponding size lattices. This stability analysis is corroborated in the bottom 
left and middle panels of Fig.~\ref{fig9} where the spatio-temporal evolution of the energy 
density of (randomly) perturbed breathers is shown therein. It can be discerned from the figure 
that the instability in the damped-driven case manifests itself slightly earlier than in the 
Hamiltonian one, in line with our observation of the slightly larger (instability) growth rate.
Moreover, the bottom right panel showcases the total energy as a function of time in both cases. 
Although the (averaged) energy in the damped-driven case (shown in blue therein) is bounded, it starts 
deviating at $t\approx 60$ from its original value signaling the onset of the instability. On the 
other hand, and as per the Hamiltonian nature of the breather, the energy is conserved (and shown 
in red), i.e., in this case, the energy is not a suitable diagnostic for discerning the instability,
yet its conservation is indicative of the accuracy of the numerical simulation. On the other hand,
if we probe the total energy of the first 5 sites, one can see that this local energy portion starts 
deviating around $t\approx 150$ from its initial (unstable equilibrium) value, thus signaling the 
onset of the instability, in line with the dynamical evolution of the left panel of Fig.~\ref{zoo}.
In both cases, over suitably long times, we observe that the resulting dynamical outcome is rather 
similar although the Hamiltonian system features longer-lived (and total-energy-preserving) dispersive
radiation. In the dissipative case, the relevant energy appears to decay away due to the damped nature 
of the lattice bulk. While here  we compared a Hamiltonian breather with damped-driven breather located 
at a single turning point in the bifurcation diagram, there is in fact a zoo of such solutions (one at 
each turning point in the bifurcation diagram), see the middle and right panels of Fig.~\ref{zoo}. A 
more systematic study of all the relevant waveforms may be an interesting topic for further study.

\section{NLS approximation of the Hamiltonian breathers} \label{sec:NLS}

We have just shown that when the damping is small, the Hamiltonian breathers
are close to breather solutions of the damped-driven lattice. In the former case 
we can analytically approximate breather solutions by deriving an NLS equation to 
describe slow modulations in time and space of an underlying carrier wave. The NLS 
equation has an exact soliton solution, and thus one can obtain an analytical approximation
in this way.  The derivation of the NLS equation we present is similar to the
one performed in \cite{Huang2} for a mass-dimer, but differs in a few key ways,
which is why we include the derivation here. 

At first, we re-write the equation of motion of Eq.~\eqref{gc_dimer_dim}
with $\gamma=0$ in terms of even ($v_{n}\coloneqq u_{2n}$) and odd 
($w_{n}\coloneqq u_{2n+1}$) particles as
\begin{subequations}
\begin{align}
\ddot{v}_{n}&=\left[1 + w_{n-1} - v_{n}\right]^{p_{2}}_{+} - %
\left[1 + v_{n} - w_{n}\right]^{p_{1}}_{+},
\label{v_non}
\\
\ddot{w}_{n}&=\left[1 + v_{n} - w_{n}\right]^{p_{1}}_{+} - %
\left[1 + w_{n} - v_{n+1}\right]^{p_{2}}_{+}.
\label{w_non}
\end{align}
\end{subequations}
Recall that $p_1=p-\delta$, while $p_2=p+\delta$ above and in what follows.
Upon assuming $|w_{n-1}-v_{n}|\ll 1$ and $|w_{n}-v_{n}|\ll 1$,
Eqs.~\eqref{v_non}-\eqref{w_non} reduce into
\begin{subequations}
\begin{align}
\ddot{v}_{n}&=M_{1}\left(w_{n}-v_{n}\right)-M_{2}\left(v_{n}-w_{n-1}\right),
\label{v_lin}
\\
\ddot{w}_{n}&=M_{2}\left(v_{n+1}-w_{n}\right)-M_{1}\left(w_{n}-v_{n}\right),
\label{w_lin}
\end{align}
\end{subequations}
where $M_{n}$ ($n=1,2$) is defined by 
\begin{align}
M_{n}(-x)=-\left[1-x\right]_{+}^{p_{n}}\approx -1 + p_{n} x -\frac{1}{2}p_{n}\left(p_{n}-1\right)x^{2}%
+\frac{1}{6}p_{n}\left(p_{n}-1\right)\left(p_{n}-2\right)x^{3} %
=J^{(1)}+J^{(2)}_{n}\,x+J^{(3)}_{n}\,x^{2}+J^{(4)}_{n}\,x^{3}\nonumber
\end{align}
with coefficients $J^{(1)}=-1$, $J^{(2)}_{n}=p_{n}$, $J^{(3)}_{n}=-p_{n}\left(p_{n}-1\right)/2$,
and $J^{(4)}_{n}=p_{n}\left(p_{n}-1\right)\left(p_{n}-2\right)/6$. 

Next, solutions in the form of fast oscillating yet small in amplitude 
patterns modulated by envelopes (that vary slowly in space and time) are 
sought by the following multiple scale Ans\"atze:
\begin{subequations} 
\begin{align} 
v_{n}(t)&=\sum_{s=1}^{3}\varepsilon^{s}\sum_{j=-s}^{s}U_{s,j}\left(X,T\right)\,E_{n}^{j}(t),
\label{v_multpl_ans}\\
w_{n}(t)&=\sum_{s=1}^{3}\varepsilon^{s}\sum_{j=-s}^{s}W_{s,j}\left(X,T\right)\,E_{n}^{j}(t),
\label{w_multpl_ans}
\end{align}
\end{subequations}
where $\varepsilon\ll 1$ as well as $X=\varepsilon\left(n-ct\right)$, $T=\varepsilon^{2} t$,
and $E_{n}(t)=e^{i\left(kn+\omega t\right)}$. Note that $U_{s,j}, W_{s,j}\in \mathbb{C}$,
and also $U_{s,-j}=\overline{U_{s,j}}$ and $W_{s,-j}=\overline{W_{s,j}}$ with the bar
standing for complex conjugation. Based on Eqs.~\eqref{v_multpl_ans}-\eqref{w_multpl_ans},
we also have that
\begin{subequations}
\begin{align}
v_{n\pm 1}(t)&=\sum_{s=1}^{3}\varepsilon^{s}\sum_{j=-s}^{s}U_{s,j}%
\left(X\pm\varepsilon,T\right)\,e^{\pm \ii kj}\,E_{n}^{j}(t),
\label{v_multpl_ans2}\\
w_{n\pm 1}(t)&=\sum_{s=1}^{3}\varepsilon^{s}\sum_{j=-s}^{s}W_{s,j}%
\left(X\pm\varepsilon,T\right)\,e^{\pm \ii kj}\,E_{n}^{j}(t),
\label{w_multpl_ans2}
\end{align}
\end{subequations}
together with the following Taylor expansions:
\begin{subequations}
\begin{align}
U_{s,j}\left(X\pm\varepsilon, T\right)&=U_{s,j}\left(X,T\right)%
\pm \partial_{X} U_{s,j}\left(X,T\right)\varepsilon %
+\frac{1}{2}\partial_{X}^{2} U_{s,j}\left(X,T\right)\varepsilon^{2}%
+\mathcal{O}\left(\varepsilon^{3}\right),
\label{U_Taylor}\\
W_{s,j}\left(X\pm\varepsilon, T\right)&=W_{s,j}\left(X,T\right)%
\pm \partial_{X} W_{s,j}\left(X,T\right)\varepsilon %
+\frac{1}{2}\partial_{X}^{2} W_{s,j}\left(X,T\right)\varepsilon^{2}%
+\mathcal{O}\left(\varepsilon^{3}\right).
\label{W_Taylor}
\end{align}
\end{subequations}

Since we are interested in standing breather solutions, we now
set $k=\pi$ and hence $c=0$. Thus, the two possible frequencies 
from which we perturb are $\omega_{+}(\pi)$, corresponding to the
the lower optical cut-off frequency, and $\omega_{-}(\pi)$, corresponding to
the upper acoustic cut-off frequency.

We now plug Eqs.~\eqref{v_multpl_ans}-\eqref{w_multpl_ans} into 
Eqs.~\eqref{v_lin}-\eqref{w_lin}, and collect coefficients
of each $\mathcal{O}\left(\varepsilon^{s}E_{n}^{j}(t)\right)$ term.
In particular, Eq.~\eqref{v_lin} (after suppressing the $t$ dependence 
for $E_{n}$) gives, up to the first two orders:
{\small
\begin{subequations}
\begin{align}
\varepsilon\,E_{n}^{0}:& \,\,\, 0=-2p U_{1,0}+2p W_{1,0}\Rightarrow U_{1,0}=W_{1,0}, 
\label{U10} \\
\varepsilon\,E_{n}^{1}:&\,\,\,  -\omega_{\pm}^{2}(\pi)U_{1,1}=%
-2 p U_{1,1} -2\delta W_{1,1}, 
\label{U11}     \\
\varepsilon^{2}\,E_{n}^{0}:&\,\,\,0=%
2\left[p\left(p-1\right)+\delta^{2}\right]%
(U_{1,1} \overline{W_{1,1}}+\overline{U_{1,1}} W_{1,1})%
+2p\left(W_{2,0}-U_{2,0}\right)-\left(p+\delta\right) \partial_{X}W_{1,0}
\nonumber \\
&+\delta (2p-1) \left[(U_{1,0}-W_{1,0})^2+2\left(|W_{1,1}|^{2}+|U_{1,1}|^{2}\right)%
\right],
\label{U20} \\
\varepsilon^{2}\,E_{n}^{1}:&\,\,\,-\omega^{2}_{\pm}(\pi)U_{2,1}=%
2\left(U_{1,0}-W_{1,0}\right)\left[\delta\left(2p-1\right)U_{1,1}+\left(p\left(p-1\right)+\delta^{2}\right)W_{1,1}\right]
-2pU_{2,1}-2\delta W_{2,1}+\left(p+\delta\right)\partial_{X}W_{1,1},
\label{U21} \\
\varepsilon^{2}\,E_{n}^{2}:&\,\,\, -4\omega^{2}_{\pm}(\pi)U_{2,2}= %
\delta\left(2p-1\right)U_{1,1}^{2}+2p\left(W_{2,2}-U_{2,2}\right)+%
\left[\left(p-\delta\right)\left(p-\delta-1\right)+\left(p+\delta\right)\left(p+\delta-1\right)\right]U_{1,1}W_{1,1}\nonumber \\
&+\frac{1}{2}\left[-\left(p-\delta\right)\left(p-\delta-1\right)+\left(p+\delta\right)\left(p+\delta-1\right)\right]W_{1,1}^{2}.
\label{U22}
\end{align}
\end{subequations}
}

Similarly, Eq.~\eqref{w_lin} gives:
{\small 
\begin{subequations}
\begin{align}
\varepsilon\,E_{n}^{0}:& \,\,\,0=2p U_{1,0}-2p W_{1,0}\Rightarrow U_{1,0}=W_{1,0},
\label{W10}\\
\varepsilon\,E_{n}^{1}:&\,\,\, -\omega_{\pm}^{2}(\pi)W_{1,1}=%
-2\delta U_{1,1}-2pW_{1,1}, 
\label{W11}     \\
\varepsilon^{2}\,E_{n}^{0}:&\,\,\,
0=-2\left[p\left(p-1\right)+\delta^2\right](U_{1,1}\overline{W_{1,1}}+\overline{U_{1,1}}W_{1,1})+%
\left(p+\delta\right)\partial_{X}U_{1,0}+2p (U_{2,0}-W_{2,0})\nonumber \\
&+\delta\left(1-2p\right)\Big[\left(U_{1,0}-W_{1,0}\right)^{2}+2\left(|U_{1,1}|^{2}+|W_{1,1}|^{2}\right)\Big],
\label{W20} \\
\varepsilon^{2}\,E_{n}^{1}:&\,\,\,-\omega^{2}_{\pm}(\pi)W_{2,1}=%
2\left(U_{1,0}-W_{1,0}\right)\left[\left(p\left(p-1\right)+\delta^{2}\right)U_{1,1}+\delta\left(2p-1\right)W_{1,1}\right]%
-2\delta U_{2,1}-2p W_{2,1}-\left(p+\delta\right)\partial_{X}U_{1,1},
\label{W21} \\ 
\varepsilon^{2}\,E_{n}^{2}:&\,\,\, -4\omega^{2}_{\pm}(\pi)W_{2,2}= %
\delta(1-2p)\left(U_{1,1}^{2}+W_{1,1}^{2}\right)+2p\left(U_{2,2}-W_{2,2}\right)%
-2\left[p\left(p-1\right)+\delta^{2}\right]U_{1,1}W_{1,1}.
\label{W22}
\end{align}
\end{subequations}
}

From Eqs.~\eqref{U11} and~\eqref{W11}, we can express the underlying 
system as $\mathcal{M}\left[U_{1,1}\,\,W_{1,1}\right]^{T}=0$ with
\begin{align}
\mathcal{M} =
\begin{bmatrix}
-2p+\omega^{2} & -2\delta \\
-2\delta       & -2p+\omega^{2}
\end{bmatrix},
\label{coeff_mat}
\end{align}
which has a non-trivial solution provided that $\mathcal{M}$ has a vanishing
determinant. It should be noted in passing that the matrix $\mathcal{M}$ 
is the same as the one of Eq.~\eqref{disp_mat_sys_dim} with $\gamma=0$.
The vanishing determinant of $\mathcal{M}$ yields the dispersion relation
\begin{align}
\omega_{\pm}^{2}(\pi)=2\left(p\pm \delta\right) \quad \left(\delta < p\right),
\label{rel_extra}
\end{align}
where the ``$+$'' subscript corresponds to the lower cut-off of the optical band,
and the ``$-$'' subscript corresponds to the upper cut-off of the acoustic band. In addition, 
from Eq.~\eqref{W11} (same argument can be deduced by using Eq.~\eqref{U11}), we obtain:
\begin{align}
W_{1,1}^{\left(\pm\right)}=\pm U_{1,1}^{\left(\pm\right)},
\label{rel1}
\end{align}
based on the dispersion relation. 

Upon utilizing Eqs.~\eqref{U10} and~\eqref{W10}, Eqs.~\eqref{U21} 
and~\eqref{W21} yield:
\begin{subequations}
\begin{align}
\left(\omega_{\pm}^{2}(\pi)-2p\right)U_{2,1}^{\left(\pm\right)}-2\delta W_{2,1}^{\left(\pm\right)}&=
-\left(p+\delta\right)\partial_{X}W_{1,1}^{\left(\pm\right)},\nonumber \\
-2\delta U_{2,1}^{\left(\pm\right)}+\left(\omega_{\pm}^{2}(\pi)-2p\right)W_{2,1}^{\left(\pm\right)}&=
\left(p+\delta\right)\partial_{X}U_{1,1}^{\left(\pm\right)},
\end{align}
\end{subequations}
which (with the aid of Eq.~\eqref{rel1}) can be cast into
\begin{align}
\mathcal{M}^{\left(\pm\right)}
\begin{bmatrix}
U_{2,1}\\
W_{2,1}
\end{bmatrix}
=
\mathcal{W}^{\left(\pm\right)}
\partial_{X}U_{1,1}^{\left(\pm\right)}, \quad 
\mathcal{W}^{\left(\pm\right)}=
\left(p+\delta\right)
\begin{bmatrix}
\mp 1\\
1
\end{bmatrix}.
\label{rel2}
\end{align}
Since the matrix $\mathcal{M}^{\left(\pm\right)}$ is singular (note that the superscripts emanate
from $\omega_{\pm}(\pi))$, it follows that $\left(\mathcal{M}^{\left(\pm\right)}\right)^{T}\left(=\mathcal{M}^{\left(\pm\right)}\right)$ 
(where $T$ stands for its transpose) has a non-trivial one-dimensional
kernel. Indeed, it is then a direct calculation to identify:
\begin{align}
\mathcal{W}^{\ast\left(\pm\right)}=
\begin{bmatrix}
2p-\omega_{\pm}^{2}\left(\pi\right)\\
-2\delta
\end{bmatrix},
\label{basis_ker}
\end{align}
such that $\left(\mathcal{M}^{\left(\pm\right)}\right)^{T}\,\mathcal{W}^{\ast\left(\pm\right)}=\mathbf{0}$ 
(upon using Eq.~\eqref{rel_extra} too). In addition, and upon considering
the Fredholm alternative for Eq.~\eqref{rel2}, we can show
that the latter indeed has a solution if and only if the right hand side of Eq.~\eqref{rel2} 
is orthogonal to the co-kernel of $\mathcal{M}^{\left(\pm\right)}$ (that is, the kernel of 
$\left(\mathcal{M}^{\left(\pm\right)}\right)^{T}$).

We now consider Eqs.~\eqref{U22} and~\eqref{W22}. With the aid of Eqs.~\eqref{rel_extra} 
and~\eqref{rel1}, we arrive at 
\begin{align} 
U_{2,2}^{\left(\pm\right)}=\alpha^{\left(\pm\right)}_{1}\left(U_{1,1}^{\left(\pm\right)}\right)^{2}, %
\quad W_{2,2}^{\left(\pm\right)}=-U_{2,2}^{\left(\pm\right)}, %
\quad 
\alpha^{\left(\pm\right)}_{1}=\pm\frac{\left(p\pm\delta\right)%
\left(1-p\mp\delta\right)}{2\left(p\pm2\delta\right)}.
\label{eqs_U22_W22}
\end{align}
Next we consider the $\varepsilon^2\,E^{0}$ equations, and in particular 
Eq.~\eqref{U20} which is solved with respect to $W_{2,0}$ (the exact same solution 
is obtained from Eq.~\eqref{W20}), thus yielding:
\begin{align}
W_{2,0}^{\left(\pm\right)}=U_{2,0}^{\left(\pm\right)}+\frac{1}{2p}\Bigg\lbrace{%
\left(p+\delta\right)\partial_{X}U_{1,0}^{\left(\pm\right)}%
-4\left[\delta\left(2p-1\right)\pm\left(p\left(p-1\right)+\delta^{2}\right)\right]|U_{1,1}^{\left(\pm\right)}|^{2}
\Bigg\rbrace}.
\label{rel4}
\end{align}

Next, we focus on the $\varepsilon^3\,E^{0}$ equations for $k=\pi$
(shown in the Appendix~\ref{append:e3E0eqs}) from which we will 
obtain an equation for $U_{1,0}$ that depends on $U_{1,1}$ and 
its (complex) conjugate. Upon utilizing Eqs.~\eqref{U10},~\eqref{W10}, 
and~\eqref{rel1}, one can find the $W_{3,0}^{\left(\pm\right)}$ component. 
Then, if one plugs the expression for $W_{2,0}^{\left(\pm\right)}$ 
[cf. Eq.~\eqref{rel4}] as well as $W_{3,0}^{\left(\pm\right)}$ into 
the $\varepsilon^3\,E^{0}$ equation for $v_{n}$, and solves subsequently 
with respect to $\partial_{X}^{2}U_{1,0}^{\left(\pm\right)}$, one 
obtains:
\begin{align}
\partial_{X}^{2}U_{1,0}^{\left(\pm\right)}&=%
\nu_{0}^{\left(\pm\right)}%
\partial_{X}|U_{1,1}^{\left(\pm\right)}|^{2},\quad
\nu_{0}^{\left(\pm\right)}=4\left(p\pm\delta-1\right).
\label{rel5}
\end{align}
Note that Eq.~\eqref{rel5} can be integrated once with respect to $X$, thus yielding
\begin{align}
\partial_{X}U_{1,0}^{\left(\pm\right)}=\nu_{0}^{\left(\pm\right)}|U_{1,1}^{\left(\pm\right)}|^{2}+f(T),
\label{rel5_sup}
\end{align}
where $f(T)$ is an arbitrary function of $T$. Considering our interest in structures that asymptotically 
vanish at $\infty$ we set $f(T)\equiv 0$. It should be noted in passing that when $\delta\equiv0$ (i.e., 
the monomer case), the prefactor $\nu_{0}^{\left(\pm\right)}=4\left(p-1\right)$ is exactly the same as 
the one in~\cite{granularBook} (see, Eq.~(A19) therein).

Finally, we focus on the $\varepsilon^3\,E^{1}$ equations for both $v$ and $w$, where we will find the 
NLS equation. At first, we express $W_{2,1}^{\left(\pm\right)}$ in terms of $U_{2,1}^{\left(\pm\right)}$:
\begin{align}
W_{2,1}^{\left(\pm\right)}=\pm\left[U_{2,1}^{\left(\pm\right)}%
+\frac{\left(p+\delta\right)}{2\delta}\partial_{X}U_{1,1}^{\left(\pm\right)}\right],\quad\textrm{and thus}
\quad
\partial_{X}W_{2,1}^{\left(\pm\right)}=\pm\left[\partial_{X}U_{2,1}^{\left(\pm\right)}%
+\frac{\left(p+\delta\right)}{2\delta}\partial_{X}^{2}U_{1,1}^{\left(\pm\right)}\right],
\nonumber 
\end{align}
which emanates from the (singular) system of Eq.~\eqref{rel2}. Next, we utilize 
Eqs.~\eqref{U10},~\eqref{W10},~\eqref{rel1},~\eqref{eqs_U22_W22}, and 
Eq.~\eqref{rel4}. This way, we obtain the following system at 
$\mathcal{O}\left(\varepsilon^3\,E^{1}\right)$:
\begin{subequations}
\begin{alignat}{2}
&-\omega_{\pm}^{2}\left(\pi\right)U_{3,1}^{\left(\pm\right)}+2\ii\omega_{\pm}\left(\pi\right)\partial_{T}U_{1,1}^{\left(\pm\right)}%
&=-2\delta W_{3,1}^{\left(\pm\right)}-2pU_{3,1}^{\left(\pm\right)}\pm\left(p+\delta\right)\partial_{X}U_{2,1}^{\left(\pm\right)}%
\pm\frac{p\left(p+\delta\right)}{2\delta}\partial_{X}^{2}U_{1,1}^{\left(\pm\right)}\nonumber\\
&&\pm\frac{2\left(p\pm\delta\right)^{2}\left[\delta\pm\left(p-1\right)\right]\left[3\delta\pm\left(p+1\right)\right]}{2\delta\pm p}%
|U_{1,1}^{\left(\pm\right)}|^{2}U_{1,1}^{\left(\pm\right)},\\
&-\omega_{\pm}^{2}\left(\pi\right)W_{3,1}^{\left(\pm\right)}+2\ii\omega_{\pm}\left(\pi\right)\partial_{T}W_{1,1}^{\left(\pm\right)}%
&=-2p W_{3,1}^{\left(\pm\right)}-2\delta U_{3,1}^{\left(\pm\right)}-\left(p+\delta\right)\partial_{X}U_{2,1}^{\left(\pm\right)}%
-\frac{1}{2}\left(p+\delta\right)\partial_{X}^{2}U_{1,1}^{\left(\pm\right)}\nonumber\\
&&+\frac{2\left(p\pm\delta\right)^{2}\left[\delta\pm\left(p-1\right)\right]\left[3\delta\pm\left(p+1\right)\right]}{2\delta\pm p}%
|U_{1,1}^{\left(\pm\right)}|^{2}U_{1,1}^{\left(\pm\right)},
\end{alignat}
\end{subequations}
which can be re-cast into
\begin{align}
\mathcal{M}^{\left(\pm\right)}%
\begin{bmatrix}
U_{3,1}^{\left(\pm\right)}\\
W_{3,1}^{\left(\pm\right)}
\end{bmatrix}
=W^{\left(\pm\right)}\partial_{X}U_{2,1}^{\left(\pm\right)}+G^{\left(\pm\right)},
\label{next_to_the_last}
\end{align}
where the residual vector $G^{\left(\pm\right)}$ is given by
\begin{align}
G^{\left(\pm\right)}=
\begin{bmatrix}
2\ii\omega_{\pm}\left(\pi\right)\partial_{T}U_{1,1}^{\left(\pm\right)}\mp\frac{p\left(p+\delta\right)}{2\delta}\partial_{X}^{2}U_{1,1}^{\left(\pm\right)}%
\mp\frac{2\left(p\pm\delta\right)^{2}\left[\delta\pm\left(p-1\right)\right]\left[3\delta\pm\left(p+1\right)\right]}{2\delta\pm p}%
|U_{1,1}^{\left(\pm\right)}|^{2}U_{1,1}^{\left(\pm\right)}\\
2\ii\omega_{\pm}\left(\pi\right)\partial_{T}W_{1,1}^{\left(\pm\right)}+\frac{1}{2}\left(p+\delta\right)\partial_{X}^{2}U_{1,1}^{\left(\pm\right)}%
-\frac{2\left(p\pm \delta\right)^{2}\left[\delta\pm\left(p-1\right)\right]\left[3\delta\pm\left(p+1\right)\right]}{2\delta\pm p}%
|U_{1,1}^{\left(\pm\right)}|^{2}U_{1,1}^{\left(\pm\right)}
\end{bmatrix}.
\end{align}
However, and due to the Fredholm alternative, the vector $G^{\left(\pm\right)}$ is orthogonal to 
$W^{\ast\left(\pm\right)}$, i.e., $W^{\ast\left(\pm\right)}\cdot G^{\left(\pm\right)}=0$, thus this 
compatibility condition yields the modulation equation for $U_{1,1}^{\left(\pm\right)}$, i.e., the 
standard NLS equation:
\begin{align}
\ii\partial_{T}U_{1,1}^{\left(\pm\right)}+\nu_{1}^{\left(\pm\right)}\partial_{X}^{2}U_{1,1}^{\left(\pm\right)}%
+\nu_{2}^{\left(\pm\right)}|U_{1,1}^{\left(\pm\right)}|^{2}U_{1,1}^{\left(\pm\right)}=0,
\label{NLS_dimer}
\end{align}
where
\begin{subequations}
\begin{align}
\nu_{1}^{\left(\pm\right)}&=\mp\frac{p^{2}-\delta^{2}}{8\delta\omega_{\pm}(\pi)},
\label{eq_nu1}\\
\nu_{2}^{\left(\pm\right)}&=\mp\frac{\left(p\pm\delta\right)^{2}\left[\delta\pm\left(p-1\right)\right]%
\left[3\delta\pm\left(p+1\right)\right]}{\omega_{\pm}(\pi)\left(2\delta\pm p\right)}.
\label{eq_nu2}
\end{align}
\end{subequations}
In the focusing case, i.e., when $\nu_{1}\nu_{2}>0$, 
the bright-soliton solution of the NLS [cf. Eq.~\eqref{NLS_dimer}] is given by
\begin{align}
U_{1,1}(X,T) = \sqrt{-\frac{2\mu}{\nu_{2}}}\sech{\left[\sqrt{-\frac{\mu}{\nu_{1}}}\,X\right]}e^{-\ii \mu T},
\label{bright_soliton}
\end{align}
where $\mu$ is an arbitrary constant that has the opposite sign of $\nu_{1}$. For 
simplicity sake, we set $|\mu| = 1$.  With the parameter values used throughout the 
manuscript, i.e., $p=1$ and $\delta=0.3$, we have that $\nu_{1}^{(-)} > 1$ and in this 
case $\mu=-1$. We also have that $\nu_{1}^{(+)} < 1$ and in this case $\mu=1$. Note that
$\nu_{1}^{(+)}\nu_{2}^{(+)}>0$ and $\nu_{1}^{(-)}\nu_{2}^{(-)}>0$, namely the NLS is 
focusing at both the top of the acoustic band and the bottom of the optical band. This 
is in contrast to classical mass-dimer chains, where only one edge typically leads to 
a focusing NLS equation \cite{granularBook}. Keeping the two lowest order terms of 
Eqs.~\eqref{v_multpl_ans} with the soliton solution for $U_{1,1}^{(\pm)}$ and the subsequent 
relation Eq.~\eqref{rel5_sup}  to compute $U_{0,1}^{(\pm)}$, we can finally write down 
an approximate breather solution.
For the breather near the top edge of the acoustic band $\omega_-(\pi)$, we have
\begin{eqnarray}
v_n &=& \frac{2 \nu_0^{(-)} \sqrt{\nu_1^{(-)}}}{\nu_2^{(-)}} %
\tanh{\left[ \sqrt{\frac{1}{\nu_1^{(-)}}} \varepsilon n    \right]  } +
\left(
\varepsilon \sqrt{\frac{2}{\nu_{2}^{(-)} }}\sech{\left[\sqrt{\frac{1}{\nu_{1}^{(-)} }}%
\, ( \varepsilon n  ) \right]}e^{\ii \varepsilon^2 t} e^{i \pi n + \omega_{-}(\pi) t} + c.c.  \right) + \mathcal{O}(\varepsilon^2)  \\
 &=&   \frac{2 \nu_0^{(-)} \sqrt{\nu_1^{(-)}}}{\nu_2^{(-)}} \tanh{\left[ \sqrt{\frac{1}{\nu_1^{(-)}}} \varepsilon n    \right]  } %
 +  (-1)^n \, 2 \varepsilon \sqrt{\frac{2}{\nu_{2}^{(-)}}}\sech{\left[\sqrt{\frac{1}{\nu_{1}^{(-)}}}\, %
 ( \varepsilon n  ) \right]} \cos( (\omega_{-}(\pi) + \varepsilon^2 )t ) + \mathcal{O}(\varepsilon^2),
\end{eqnarray}
and by virtue of Eqs.~\eqref{rel1} and \eqref{W10}, a similar expression for 
$w_n$ is obtained. This is a spatially localized solution with temporal frequency
$\omega_{-}(\pi) + \varepsilon^2$, which is slightly above the edge of the acoustic 
band $\omega_{-}(\pi)$ that sits on top of a stationary kink background. The acoustic 
breather would be out-of-phase if one subtracts off the stationary kink background 
(due to Eq.~\eqref{rel1}).
For the breather near the bottom edge of the optical band $\omega_+(\pi)$, we have
\begin{eqnarray}
v_n &=& -\frac{2 \nu_0^{(+)} \sqrt{-\nu_1^{(+)}}}{\nu_2^{(+)}} %
\tanh{\left[ \sqrt{\frac{-1}{\nu_1^{(+)}}} \varepsilon n    \right]  }  + (-1)^n  \, 2 \varepsilon \sqrt{\frac{-2}{\nu_{2}^{(+)}}}\sech{\left[\sqrt{\frac{-1}{\nu_{1}^{(+)}}}\, ( \varepsilon n  ) \right]} \cos( (\omega_{{(+)}}(\pi) - \varepsilon^2 )t )  %
+ \mathcal{O}(\varepsilon^2).
\end{eqnarray}
A similar expression for $w_n$ is obtained. Note that, to leading order, we have 
that $w_n =  v_n$. This implies the optical breather is in-phase.  The frequency 
of the oscillation is $\omega_{+}(\pi) - \varepsilon^2$, which is slightly below 
the edge of the optical band $\omega_{+}(\pi)$.

\begin{figure}[pt!]
\begin{center}
\includegraphics[height=.25\textheight, angle =0]{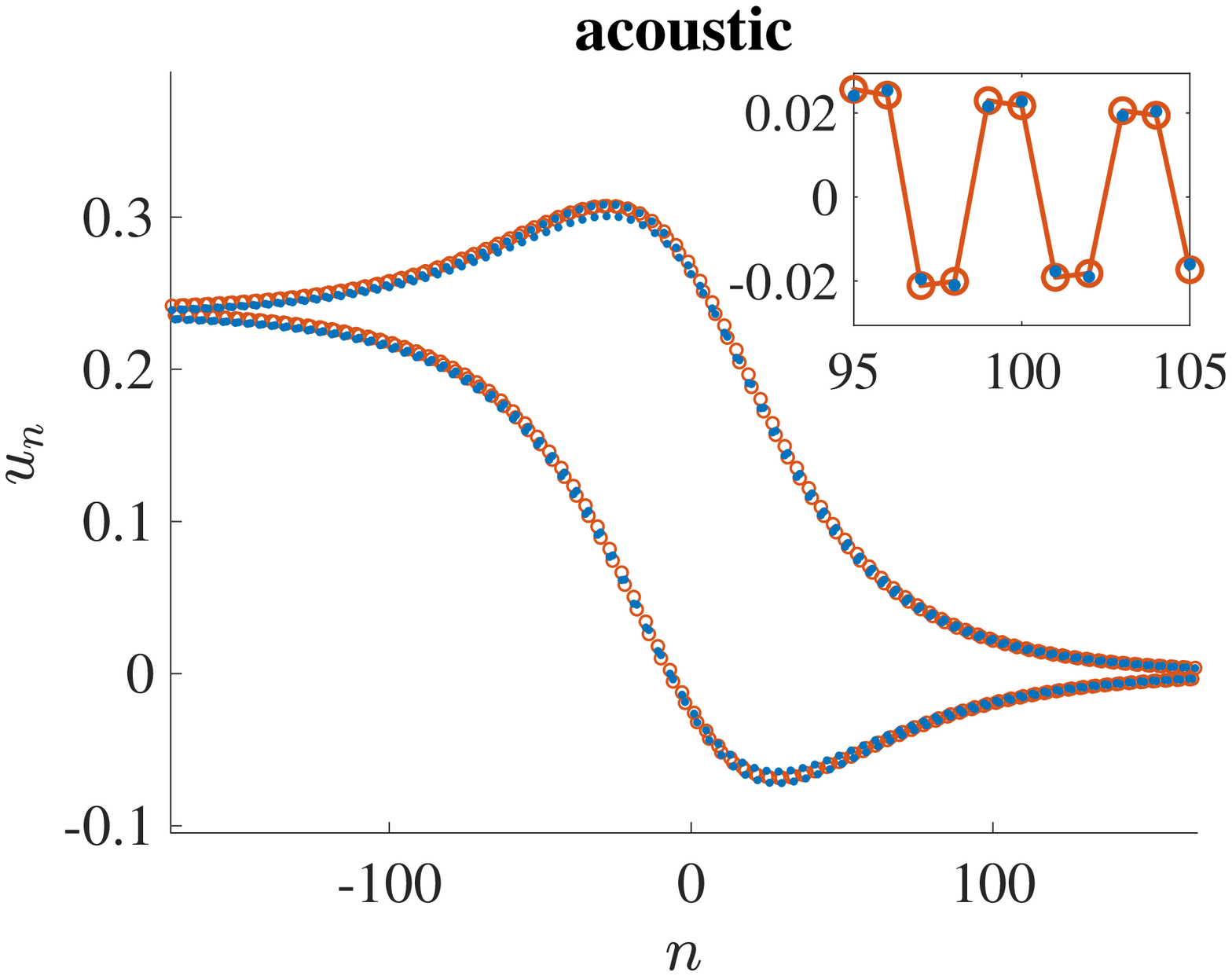}
\includegraphics[height=.25\textheight, angle =0]{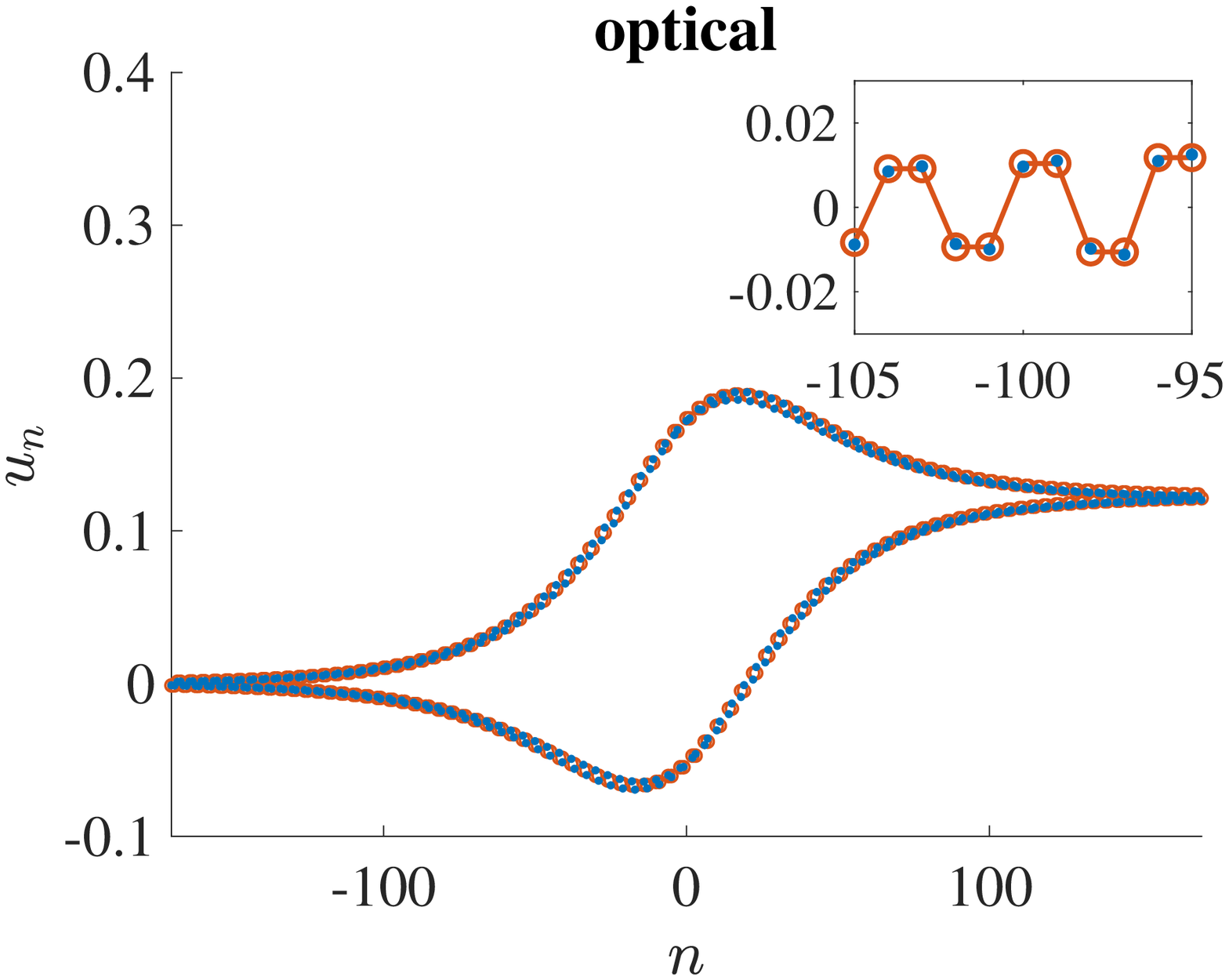}
\end{center}
\caption{(Color online) Plot of an acoustic (left) and optical (right) 
breather for $\varepsilon = 0.3$. The corresponding values of the frequencies 
are $f = 0.1885$ and $f = 0.2565$, respectively. The approximate solution, 
based on the NLS equation, is shown as red circles, and the numerical 
solution, obtained via Newton iterations, is shown as solid blue points. 
The insets show a zoom of the solution to better see the difference between 
approximation and numerical solution and to better see the phase structure. 
\label{fig:compare_NLS}
}
\end{figure}

\begin{figure}[pt!]
\begin{center}
\includegraphics[height=.22\textheight, angle =0]{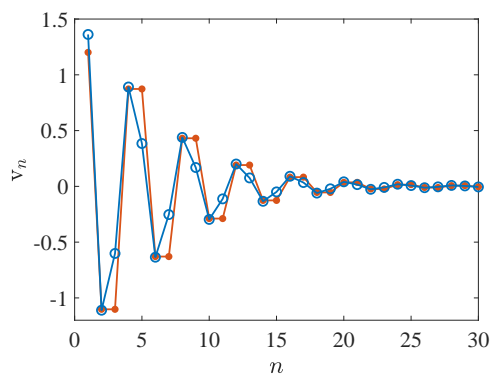}
\end{center}
\caption{(Color online) Plot of an optical breather for $f_{d}=0.25$. 
The approximate solution, based on the NLS equation, is shown as red filled circles, 
and the numerical solution with $\gamma=8\times 10^{-4}$, obtained via Newton iterations, 
is shown as blue open circles. 
\label{fig:compare_NLS_dissi}
}
\end{figure}

Both the acoustic and optical breathers have essentially the same structure, with 
the major difference being the sign of the leading order kink background. These 
approximate breather solutions compare well to those computed numerically for small 
values of $\varepsilon$, see Fig.~\ref{fig:compare_NLS} for example. Note that these 
formulas predict that the breathers will become larger, and more localized, as the 
frequency goes deeper into the spectral gap (and no bifurcation is predicted). 
However, the  approximation becomes less accurate as frequencies deviate from the 
spectral edges, since $\varepsilon$ is becoming larger. In particular, breathers of 
the full lattice system are expected to terminate or experience other bifurcations, 
like those reported in~\cite{dark}. Finally, we conclude our discussion here by mentioning 
the connection between the approximate breathers and the ones computed numerically for 
the dissipative problem with $\gamma=8\times 10^{-4}$ and $f_{d}=0.25$ (see, the profile
in the top right panel of Fig.~\ref{fig9}). The relevant comparison is shown in 
Fig.~\ref{fig:compare_NLS_dissi} where a proximity between the two profiles is observed. 
This confirms the fact that as we approach the point of small dissipation and drive amplitudes, 
the waveforms of interest essentially are the ones bifurcating from the Hamiltonian limit
that our above NLS multiple scales analysis allows us to accurately approximate.

\section{Conclusions and Future Challenges} \label{sec:theend}

We have proposed a new model that involves a mixture of strain-hardening 
and strain-softening interactions, and studied time-periodic and breather 
solutions therein. The nonlinear resonant peaks of the damped-driven system 
bend toward the frequency gap, but they do so in different ways in the case 
of the acoustic and optical gaps herein. The high-energy states that enter 
the frequency gap are spatially localized (e.g. breathers) and are found to 
compare well to Hamiltonian breathers when the damping and drive amplitude 
are small. The complex bifurcation diagrams associated with the damped-driven 
chain has been obtained and their fold bifurcations, as well as the ones 
leading to oscillatory instabilities have been elucidated. The dynamics of the 
unstable waveforms have also been probed, and it has been observed that they 
repartition the energy, possibly through weaker localization at the boundary and 
dispersion of wavepackets towards the bulk of the chain. In the Hamiltonian lattice 
limit, we derived an NLS equation and from there, we were able to construct approximate 
acoustic and optical breathers, both of which agree well with numerically computed 
breathers in the vicinity of the corresponding band edges.

While breathers are fundamental in the study of nonlinear lattices, other 
structures, such as solitary waves or dispersive shocks merit further exploration 
in lattices with alternating stiffness. Some indications towards the potential 
relevance of such structures have also appeared herein in settings where the 
energy was shown to somewhat coherently propagate through such lattices. It would 
be interesting to parallel relevant studies to the well-established picture of
resonances and anti-resonances in standard granular (mass) dimers~\cite{staros1}.
Similarly to the latter setting~\cite{JKdimer}, it is reasonable to
expect that such dimers can be implemented also in the present setting.
Moreover, the study of granular chains with nonlinearity exponents
close to unity is of extensive mathematical interest, as illustrated
by the work of~\cite{pelingj} and the numerous ones that followed
along the related direction of the continuum log-KdV model. Although
we did not pursue such a connection here, it would be a natural
one to also consider in the setting proposed herein. Extensions 
to higher spatial dimension are also worthy of additional exploration,
especially since the existence of breathers in the latter setting is
far less explored in the granular case; see, e.g.,~\cite{Chong_2021}
for a recent discussion. Finally, the proposed model seems well
within current experimental capabilities, and such studies would not 
only test the viability of breather solutions in such lattices, but may 
also offer insights for potential applications of lattices with alternating 
stiffness, such as in shock absorption or energy harvesting. Such studies 
are currently in progress and will be reported in future publications.

\begin{acknowledgments}
This material is based upon work supported by the Research, Scholarly \& 
Creative Activities Program awarded by the Cal Poly division of Research, 
Economic Development \& Graduate Education (MML, EGC, and SX)
and the US National Science Foundation under Grant Nos. DMS-2107945 (CC)
and DMS-1809074 (PGK). EGC expresses his gratitude to A.~Vainchtein 
(University of Pittsburgh) for discussions related to the derivation of the NLS.
\end{acknowledgments}

\appendix

\section{Newton's method and spectral stability}
We first compute time-periodic orbits of Eq.~\eqref{gc_dimer_dim} with 
period $T_d = 1/f_d$ with high precision by finding roots of the map 
$F := \mathbf{x}(T_d) - \mathbf{x}(0)$, where $\mathbf{x}(T_d)$ is the 
solution of Eq.~\eqref{gc_dimer_dim} at time $T_d$ with initial condition 
$\mathbf{x}(0)$.  Roots of this map (and hence time-periodic solutions of 
Eq.~\eqref{gc_dimer_dim}) are found via Newton iterations. This requires the 
Jacobian of $F$, which is of the form $V(T_d)-I$, where $I$ is the identity 
matrix, $V$ is the solution to the $N^2$ variational equations $\dot{V} = DF\cdot V$ 
with initial condition $V(0) = I,$ and  $DF$ is the Jacobian of the equations 
of motion evaluated at a given state vector of the form of $\mathbf{X}=[\mathbf{u} \, ,  \dot{\mathbf{u}}]^{T}$
where $\mathbf{u}=\left[u_{1}\, , u_{2}\, , \dots \,, u_{N} \right]^{T}$. Note 
that the solution frequency and drive frequency are both $f_d$ by construction. 
To investigate the dynamical stability of the obtained states, a Floquet analysis 
is used to compute the multipliers associated with the solutions. The Floquet multipliers 
for a solution are obtained by computing the eigenvalues of the monodromy matrix 
(which is $V(T_d)$ upon convergence of the Newton scheme). If a solution has all 
Floquet multipliers within or on the unit circle, the solution is called (spectrally) 
stable. An instability that results from a multiplier on the (positive) real line is 
called a real (exponential in nature) instability. However, there can also be 
oscillatory instabilities, which correspond to  complex-conjugate pairs of  Floquet 
multipliers  lying outside  the unit circle (in the complex plane). 

\section{The \protect{$\varepsilon^{3}E^{0}$} equations}
\label{append:e3E0eqs}
In this Appendix, we present the $\varepsilon^{3}E^{0}$ equations for 
the variables $v_{n}$ and $w_{n}$ ($k=\pi$ and $c=0$). We obtain 
these equations by utilizing $W_{1,0}^{(\pm)}=U_{1,0}^{(\pm)}$ together 
with Eq.~\eqref{rel1}. This way, the $\varepsilon^{3}E^{0}$ equations for 
the lower cut-off of the optical band read
\begin{subequations}
\begin{align} 
v_{n}:\,\,\,0&=4\left(p+\delta\right)\left(p+\delta-1\right)%
\left[ U_{1,1}^{(+)}\overline{U_{2,1}^{(+)}}+U_{1,1}^{(+)}%
\overline{W_{2,1}^{(+)}}+\overline{U_{1,1}^{(+)}}U_{2,1}^{(+)}%
+\overline{U_{1,1}^{(+)}}W_{2,1}^{(+)}\right]\nonumber \\
&+\left(p+\delta\right)\partial_{X}^{2}U_{1,0}^{(+)}-2\left(p+\delta\right)\partial_{X}W_{2,0}^{(+)}%
+4p\left(W_{3,0}^{(+)}-U_{3,0}^{(+)}\right)%
-4\left(p+\delta\right)\left(p+\delta-1\right)\partial_{X}|U_{1,1}^{(+)}|^{2},
\label{eq:e3E0vnopt}\\
w_{n}:\,\,\,0&=-4\left(p+\delta\right)\left(p+\delta-1\right)\left[U_{1,1}^{(+)}\overline{U_{2,1}^{(+)}}%
+U_{1,1}\overline{W_{2,1}^{(+)}}+\overline{U_{1,1}^{(+)}} U_{2,1}^{(+)}%
+\overline{U_{1,1}^{(+)}} W_{2,1}^{(+)}\right]\nonumber \\
&+\left(p+\delta\right)\partial_{X}^{2}U_{1,0}^{(+)}+2\left(p+\delta\right)\partial_{X}U_{2,0}^{(+)}%
+4p\left(U_{3,0}^{(+)}-W_{3,0}^{(+)}\right)%
-4\left(p+\delta\right)\left(p+\delta-1\right)\partial_{X}|U_{1,1}^{(+)}|^{2}.
\label{eq:e3E0wnopt}
\end{align}
\end{subequations}
As it was already mentioned in the text, we solve
Eq.~\eqref{eq:e3E0wnopt} with
respect to
$W_{3,0}^{(+)}$ first, and plug the resulting expression in Eq.~\eqref{eq:e3E0vnopt}
afterwards. Then, and upon using Eq.~\eqref{rel4} and its derivative (to eliminate 
the $W_{2,0}^{(+)}$ terms), we thus arrive at Eq.~\eqref{rel5} for the $U_{1,0}^{(+)}$
component.

In a similar vein, the $\varepsilon^{3}E^{0}$ equations for the upper cut-off 
of the optical band are given by
{\small 
\begin{subequations}
\begin{align} 
v_{n}:\,\,\,0&=-4\left(p-\delta\right)\left(p-\delta-1\right)%
\left[U_{1,1}^{(-)}\overline{U_{2,1}^{(-)}}-U_{1,1}^{(-)}\overline{W_{2,1}^{(-)}}%
+\overline{U_{1,1}^{(-)}} U_{2,1}^{(-)}-\overline{U_{1,1}^{(-)}} W_{2,1}^{(-)}\right]\nonumber \\
&+\left(p+\delta\right)\partial_{X}^{2}U_{1,0}^{(-)}-2\left(p+\delta\right)\partial_{X}W_{2,0}^{(-)}%
+4 p\left(W_{3,0}^{(-)}-U_{3,0}^{(-)}\right),
\label{eq:e3E0vnacou}
\\
w_{n}:\,\,\,0&=4\left(p-\delta\right)\left(p-\delta-1\right)\left[U_{1,1}^{(-)}%
\overline{U_{2,1}^{(-)}}-U_{1,1}^{(-)}\overline{W_{2,1}^{(-)}}+\overline{U_{1,1}^{(-)}} U_{2,1}^{(-)}%
-\overline{U_{1,1}^{(-)}} W_{2,1}^{(-)}\right]\nonumber \\
&+\left(p+\delta\right)\partial_{X}^{2}U_{1,0}^{(-)}%
+2\left(p+\delta\right)\partial_{X}U_{2,0}^{(-)}+4p\left(U_{3,0}^{(-)}-W_{3,0}^{(-)}\right),
\label{eq:e3E0wnacou}
\end{align}
\end{subequations}
}
and upon using a similar manipulation, i.e., extracting the $W_{3,0}^{(-)}$ component 
from Eq.~\eqref{eq:e3E0wnacou}, and plugging it into Eq.~\eqref{eq:e3E0vnacou}, we obtain
(with the aid of Eq.~\eqref{rel4} for the $W_{2,0}^{(-)}$) Eq.~\eqref{rel5} for the 
$U_{1,0}^{(-)}$ component.

\bibliographystyle{unsrt}
\bibliography{Chong}

\begin{thebibliography}{10}
\bibitem{isola_strategy}
This isola was obtained by performing a frequency continuation in {{AUTO}}.
  {{T}}he initial seed that was fed to the solver was extracted from the branch
  shown in the rightmost panel of {{F}}ig.~\ref{fig6} for $a_{d}=0.15$ (and
  $f_{d}=0.25$).
\end{thebibliography}

\end{document}